\documentclass[aps,prx,twocolumn,groupedaddress,showpacs,preprintnumbers,amsmath]{revtex4-1}
\usepackage{epsfig,amsmath,amssymb,graphics,color,calc,hyperref}
\usepackage{array}
\usepackage{mathtools}

\newcommand{\im}{\mbox{Im}}

\newcommand{\avg}[1]{\left\langle #1 \right\rangle}

\begin{document}

\title{Corrections to the Bethe lattice solution of Anderson localization}

\author{Matilde Baroni\textsuperscript{1}, Giulia Garcia Lorenzana\textsuperscript{2,3}, Tommaso Rizzo\textsuperscript{4,5}, and Marco Tarzia\textsuperscript{6,7}}

\affiliation{\textsuperscript{1}\mbox{LIP6, CNRS, Sorbonne Université, 4 place Jussieu, F-75005 Paris, France}\\
\textsuperscript{2}\mbox{LPENS, ENS, Université PSL, CNRS, Sorbonne Université, Université de Paris, F-75005 Paris, France}\\
	\textsuperscript{3} \mbox{Laboratoire Matière et Systèmes Complexes (MSC), Université de Paris, CNRS, 75013 Paris, France}\\
	\textsuperscript{4}\mbox{Dip. Fisica, Universit\`a ``Sapienza'', Piazzale A. Moro 2, I–00185, Rome, Italy}\\
	\textsuperscript{5}\mbox{ISC-CNR, UOS Rome, Universit\`a ``Sapienza'', Piazzale A. Moro 2, I-00185, Rome, Italy}\\
	\textsuperscript{6} \mbox{LPTMC, CNRS-UMR 7600, Sorbonne Universit\'e, 4 Pl. Jussieu, F-75005 Paris, France}\\
	\textsuperscript{7} \mbox{Institut  Universitaire  de  France,  1  rue  Descartes,  75231  Paris  Cedex  05,  France}}

\begin{abstract}

We study numerically Anderson localization on lattices that are tree-like except for the presence of one loop of varying length $L$.  The resulting expressions allow us to compute corrections to the Bethe lattice solution on i) Random-Regular-Graph (RRG) of finite size $N$ and ii) euclidean lattices in finite dimension. In the first case we show that the $1/N$ corrections to to the average values of observables such as the typical density of states and the inverse participation ratio have prefactors that diverge  exponentially approaching the critical point, which explains the puzzling observation that the numerical simulations on finite RRGs deviate spectacularly from the expected asymptotic behavior. In the second case our results, combined with the $M$-layer expansion, predict that corrections destroy the exotic critical behavior of the Bethe lattice solution in any finite dimension, strengthening the suggestion that the upper critical dimension of Anderson localization is infinity. This approach opens the way to the computation of non-mean-field critical exponents by resumming the series of diverging diagrams through the same recipes of the field-theoretical perturbative expansion.
\end{abstract}

\maketitle 

\section{Introduction}

Anderson localization (AL)~\cite{Anderson} is one of the central paradigm of condensed matter theory. It plays a central role in many areas of science, such as transport in disordered quantum systems, random matrices, and quantum chaos, and despite almost 60 years of research, its study continues to reveal new facets and subtleties~\cite{50years,lee,evers08}.

The critical properties of AL are well-established in two opposite limits: In low dimensional systems  the scaling arguments~\cite{gangoffour} allowed to identify in $d_L = 2$ the lower critical dimension~\cite{mott,weakloc} (for system with orthogonal symmetry), as later confirmed by the field-theory description of the transition in terms of a non-linear $\sigma$ model (NL$\sigma$M)~\cite{wegner,efetov}, for which the $2 + \epsilon$ renormalization group (RG) treatment~\cite{5loops} provides a quantitative ground for the scaling ideas. These advances culminated in a functional RG analysis of the NL$\sigma$M, which yields the multifractal spectra of wave-function amplitudes at the AL critical point in $d = 2 + \epsilon$~\cite{foster}.

There are also analytical results in the infinite dimensional limit, represented by AL on the Bethe lattice (BL)~\cite{Abou-Chacra} (i.e., an infinite random-regular graphs (RRG) of fixed connectivity~\cite{rrg}, a class of random lattices that have locally a tree-like structure but have very large loops and no boundaries, see below for a precise definition). This model allows for an exact solution, making it possible to establish the transition point and the corresponding critical behavior~\cite{Abou-Chacra,efetov_bethe,efetov_bethe1,zirn,mirlin,mirlin1,verba,noi,mirlintikhonov,tikhonov_critical,large_deviations,aizenmann,semerjian,parisi,victor}, which turns out to be very peculiar and to exhibit a few important differences with respect to finite $d$: The BL criticality displays exponential instead of power-law singularities (when the localization transition is approached from the delocalized regime)~\cite{efetov_bethe,efetov_bethe1,zirn,mirlin,mirlin1,verba,noi,mirlintikhonov,tikhonov_critical,large_deviations}, and the Inverse Participation Ratio (IPR), defined as $I_2 = \langle \sum_{i=1}^N \vert \psi_i \vert^4 \rangle$ ($\psi_i$ being the value of the wave-function on site $i$, and $N$ the number of nodes of the lattice), has a discontinuous jump at the transition from $O(1)$ in the insulating phase toward a $1/N$ scaling in the extended phase,  in contrast with the $d$-dimensional case in which $I_2$ vanishes smoothly at the critical disorder.  Yet, as discussed in Ref.~\cite{mirlin94}, such apparent discrepancies~\cite{efetov_bethe1,zirn} between the exotic BL criticality and the predictions of the scaling hypothesis can be rationalized in terms of the ``pathological'' space structure of the BL, which is inconsistent with the euclidean structure at any finite $d$: On the BL the volume of a finite portion of the tree of radius $L$ is exponential in $L$, $V_{\rm BL} (L) \propto k^L$ ($k$ being the branching ratio of the graph), instead of $V_d (L) \propto L^d$ for an euclidean lattice. For this reason, the BL criticality has been argued to be incorrect in any finite $d$, since the true critical behavior should correspond to  power-law singularities with $d$-dependent critical exponents~\cite{mirlin94}. In this sense, the case $d = \infty$ has been argued to be a singular point for AL, and to play the role of the upper critical dimension~\cite{mirlin94,noilarged,garcia,castellani,dobro}. 

AL on the BL has attracted a renewed interest in the last few years because of its connection to Many Body Localization (MBL)~\cite{BAA}, a new kind of quantum out-of-equilibrium dynamical phase transition between an ergodic metal at low disorder and a non-ergodic insulator at strong disorder in which the (interacting) system is unable to self-equilibrate~\cite{BAA,Gornyi,reviewMBL,reviewMBL2,reviewMBL3,reviewMBL4,reviewMBL5}. The analogy of this problem with single-particle AL was put forward in the seminal work of Ref.~\cite{dot}, where the quasi-particles decay in the Fock space of many-body states is mapped onto an appropriate non-interacting tight binding model on a disordered tree-like graph (see also Refs.~\cite{Gornyi,BetheProxy1,BetheProxy2,scardicchioMB,roylogan,tikhonov_mirlin_21,biroli_tarzia_17,gabrielKT}). Hence, beside its own fundamental importance, understanding the critical behavior of AL in large $d$ could provide valuable insights to sharpen many important questions related to the MBL criticality.

In this context, motivated by the connection with MBL, over the past ten years, numerous numerical investigations have been conducted on the Anderson model on RRGs~\cite{noi_nonergo,scardicchio1,ioffe1,ioffe3,refael2019,pinorrg,bera2018,detomasi2020,mirlinRRG,Bethe,lemarie}. The results of these studies have been interpreted by some authors as the manifestation of the possible existence of an intermediate delocalized but non-ergodic phase characterized by multifractal eigenfunctions  covering a wide range of disorder values before the localization transition occurs~\cite{noi_nonergo,scardicchio1,ioffe1,ioffe3,refael2019,pinorrg}. The supporting arguments for this hypothesis primarily rely on extrapolations from numerical results obtained through Exact Diagonalization (ED) of substantial yet finite size samples. 

While the possibility of this multifractal delocalized phase is undeniably fascinating, particularly in its connection to MBL~\cite{dot}, it contradicts the analytical predictions based on the supersymmetric approach for the Anderson model on sparse random graphs~\cite{mirlintikhonov}, and has been heavily debated in recent years. Several numerical investigations, based on finite-size scaling analysis of spectral and wave-function statistics, have provided compelling evidence countering the existence of a truly intermediate non-ergodic extended phase in the Anderson model on RRGs and similar sparse random lattices~\cite{mirlinRRG,Bethe,lemarie,levy,large_deviations}. The findings indicate a non-monotonic trend in the observables as the system size varies on the delocalized side of the transition, which can be explained by two factors. Firstly, there is the presence of a characteristic scale that grows exponentially fast as the transition point is approached, which becomes very large even at significant distances from the transition~\cite{lemarie, levy,Bethe,large_deviations,mirlintikhonov,mirlinRRG}. Secondly, the critical point itself tends to become localized in the limit of infinite dimensions~\cite{efetov_bethe,efetov_bethe1,zirn,mirlin,mirlin1,verba,noilarged}. The combination of these factors leads to significant and intricate finite-size effects, even at considerable distances from the critical point, characterized by a pronounced non-ergodic behavior within a crossover region where the correlation volume surpasses the accessible system sizes. (In contrast, it is widely accepted that the eigenstates of the Anderson model on loop-less Cayley trees are genuinely multifractal in the whole delocalized regime~\cite{mirlinCT,garel,noiCT,khaymovichCT}.)

In this paper we develop a perturbative loop expansion which provides for the first time corrections to the BL solution of AL, focusing on the metallic phase. We apply this loop expansion to study two complementary (and crucial) aspects of the problem:  On the one hand, we obtain the $1/N$ corrections to the average value of relevant observables on RRGs of finite size $N$. On the other hand, within the framework of the $M$-layer construction~\cite{mlayer,mrfim,mboot,mkcm,msg1,msg2,vonto,mori} --- a novel technique recently introduced to treat problems in which mean-field theory is only available on the BL --- we analyze finite-dimensional corrections to the BL solution of AL on euclidean lattices. In both cases we find that the corrections are huge, as they diverge exponentially fast upon approaching the critical point. This behavior is in striking contrast with conventional phase transitions, which are instead characterized by a less dramatic algebraic divergence of corrections, and leads to deep physical consequences: In the latter case our results imply that the exotic critical behavior of the BL solution is destroyed by corrections in any finite dimension, providing a quantitative ground to the insights of Ref.~\cite{mirlin94}, and opening the way for the computation of finite $d$ corrections in a systematic and rigorous framework. In fact, the hypothesis that the upper critical dimension of AL is infinity is decades-old~\cite{mirlin94,noilarged,garcia,castellani,dobro} and our computation yields the first concrete and quantitative evidence for its validity. In the former case, our approach clarifies the physical origin of the spectacular finite-size effects observed on RRGs of finite size, which resides in the fact that the $1/N$ corrections to the pertinent observables have prefactors that diverge exponentially fast approaching the critical point.

The paper is organized as follows: In Sec.~\ref{sec:Mlayer} we briefly present the main ingredients of the loop expansion and of the $M$-layer approach; In Sec.~\ref{sec:model} we introduce the Anderson tight-binding model and briefly recall the key features of its BL solution; In Sec.~\ref{sec:ginzburg} we apply the Ginzburg criterion to determine the region of validity of the mean-field critical behavior; In Sec.~\ref{sec:LDOS} we study the $1$-loop corrections to (the logarithm of) the typical density of states (DoS), which can be thought as a proxy for the order parameter of AL, focusing both on euclidean lattice in finite dimensions and on RRGs of finite size; In Sec.~\ref{sec:IPR} we discuss the $1$-loop corrections to the IPR in the metallic phase on large but finite RRGs; Finally, in Sec.~\ref{sec:conclusions} we provide a few concluding remarks along with some perspectives for future investigations. In the main text we only present the key results of our analysis, while the technical details are reported at length in the appendix sections: In App.~\ref{app:perco} we illustrate the implementation of the $M$-layer expansion at the $1$-loop level for the simplest statistical mechanics model with a second order phase transition, i.e., random percolation. In App.~\ref{app:Anderson} we provide more details and numerical results related to several points discussed in the main text concerning the loop expansion for the Anderson model.

\begin{figure}
	\includegraphics[width=0.4\textwidth]{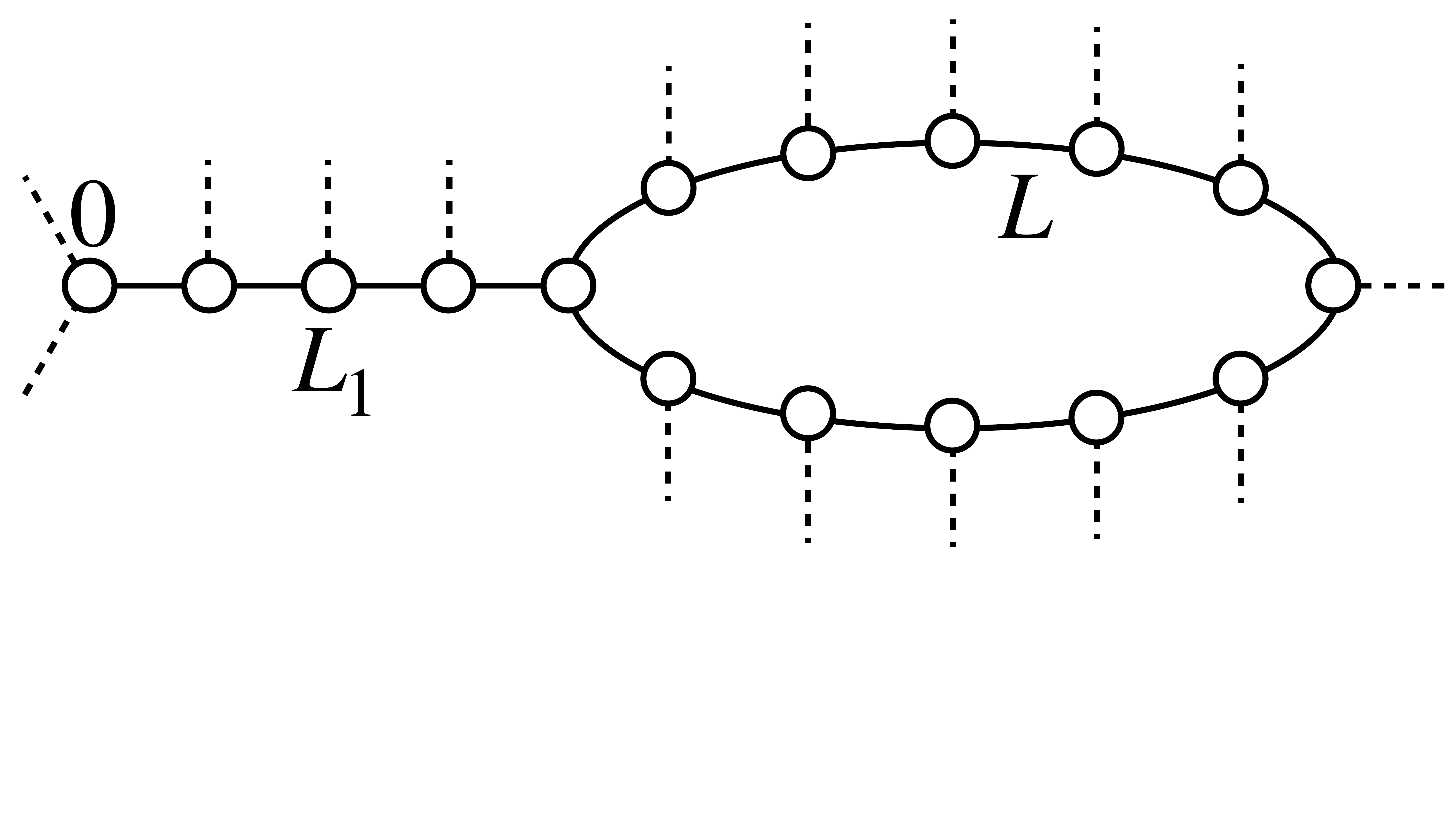}
	\vspace{-1.5cm}
	\caption{\label{fig:1loopdiagram} 1-loop topological diagrams contributing  to the corrections of one point observables (with $L=8$, $L_1=4$, and $k=2$). $L$ and $L_1$  are the lengths of the lines of the diagrams, and $0$ denotes the site on which we measure the corrections to the pertinent observable. The dashed lines correspond to  the semi-infinite branches of the loop-less BL attached to each node to ensure that the local connectivity is strictly equal to $k+1$.}
\end{figure}

\section{Loop expansion around the Bethe lattice solution} \label{sec:Mlayer}

For some physical problems, including AL, the mean-field theory is only available on the BL. In these cases improving the predictions of the BL solution systematically by taking into account the presence of loops in the underlying lattice  would be extremely valuable.

In principle, as shown in Ref.~\cite{mlayer}, {\it any} observable on {\it any} lattice can be written as the sum of the BL result plus the contributions of the loops. To be more specific, the average value of any $1$-point observable on any given lattice can be expanded (at the $1$-loop level) as~\cite{mlayer}:
\begin{equation} \label{eq:expansion}
\avg{O} = \avg{O}_{\rm BL} + \sum_{L,L_1} {\cal N}_G (L,L_1) \, \delta[O(L,L_1)] + \ldots \, ,
\end{equation}
where $\avg{O}_{\rm BL}$ is the average of the observable on the infinite loop-less BL, ${\cal N}_G (L,L_1)$ is the number of $1$-loop diagrams for the specific  graph considered, which for $1$-point observables have the structure shown in Fig.~\ref{fig:1loopdiagram},  and $\delta[O(L,L_1)] = \avg{O}_{(L,L_1)\textrm{-loop}} - \avg{O}_{\rm BL}$ is the difference between the average of the observable computed in presence and in absence of the loop. The dots in the right hand side represent the contributions coming from diagrams with more than one loop. This expansion can be generalized to any $n$-point observables and hence to correlation functions (see Sec. IV of Ref.~\cite{mlayer}). Note that if the chosen lattice is random (as in the cases considered below), an extra average must be performed in Eq.~\eqref{eq:expansion} over the ensemble of all possible realizations of the graph.

Although the diagrams of Fig.~\ref{fig:1loopdiagram} look similar to standard Feynman diagrams, in our case the loops have a specific geometrical meaning.  To evaluate the contribution of a given diagram  one has to consider the graph as a portion of the original lattice, replacing the internal lines with appropriate one-dimensional chains, and attaching to the internal points the appropriate number of branches of infinite-size BL to restore the correct local connectivity of the original model~\cite{mlayer}. 

Eq.~\eqref{eq:expansion} holds in general, for every model on every lattice. Yet, for a generic graph, this expansion is non-perturbative in the sense that all terms in the expansion need to be evaluated because they have the same order of magnitude. For a generic $d$-dimensional euclidean lattice, for instance, there are many short loops and therefore their contribution is important and gives very strong corrections. One should indeed at least consider all the loops shorter than the correlation length to have a reliable result and this expansion is in general useless.

There are, however, two classes of (random) lattice for which the loop expansion can be made perturbative and used to make accurate quantitative predictions. The first is represented by (finite size) RRGs (and other similar kind of sparse random graphs). RRGs are a class of random lattices that have locally a tree-like structure but do not have boundaries. More precisely, a $(k+1)$-RRG is a lattice chosen uniformly at random among all possible graphs of $N$ vertices where each of the sites has fixed degree $k+1$. The properties of such random graphs have been extensively studied (see Ref.~\cite{rrg} for a review). A RRG can be essentially thought as a finite portion of a tree wrapped onto itself. It is known in particular that for large number of vertices any finite portion of such a graph is a tree with a probability going to one as $N \to \infty$: RRGs have loops of all size but  short loops are rare and their typical length is of order $\ln N/\ln k$~\cite{rrg}. Thanks to these properties the expansion in loops on the RRG is well defined: at variance with $d$-dimensional euclidean lattices, the probability of finding a loop starting from a given node decrease as $1/N$~\cite{rrg,guilhem,remi} and the topological loop expansion is naturally ordered in terms of that small parameter.

More specifically, going back to Eq.~\eqref{eq:expansion}, for a RRG of $N$ nodes and degree $k+1$ the average number of $1$-loop diagrams having the structure shown in  Fig.~\ref{fig:1loopdiagram}, $\avg{{\cal N}_G (L,L_1)}$, can be computed explicitly as explained below. The starting point of this computation is the precise combinatorial estimation of the average number of closed circuits of length $L$ in a RRG (in the large $N$ limit) carried out in Refs.~\cite{rrg,guilhem,remi} yielding:
\[
\avg{{\cal C} (L)} = \frac{k^L}{2 L} \, 
\]
(for $L \ll \ln N/ \ln k$). The average number of nodes which are on a circuit of length $L$ is thus  $L \avg{{\cal C} (L)} = k^L/2$. The $1$-loop diagrams of Fig.~\ref{fig:1loopdiagram} are obtained by attaching an external leg of length $L_1$ to one of the nodes of the circuit of length $L$. The average number of these diagrams  originating from a given node (labeled as $0$) of the RRG is then finally obtained by dividing the total number of nodes at distance $L_1$ from the loop, $k^{L_1} L \avg{{\cal C} (L)} = k^{L+L_1}/2$, by the total number of nodes $N$:
\begin{equation} \label{eq:BLpaths}
\avg{{\cal N}_G (L,L_1)} = \frac{k^{L+L_1}}{2 N} \, .
\end{equation}
The loop expansion~\eqref{eq:expansion} can thus be used to compute the $1/N$ corrections of physically relevant observables to the BL solutions on RRGs of finite size. As discussed in the introduction, this is a key issue, since AL on RRGs (and similar sparse random networks) has attracted a considerable amount of interest in the last decade due to its deep connection with MBL~\cite{BAA,BetheProxy1,BetheProxy2,scardicchioMB,roylogan,tikhonov_mirlin_21,biroli_tarzia_17,gabrielKT}, and the puzzling behavior of its finite-size corrections compared to the infinite BL asymptotic behavior has been at the core of an intense debate~\cite{noi_nonergo,scardicchio1,ioffe1,ioffe3,refael2019,pinorrg,bera2018,detomasi2020,mirlinRRG,Bethe,lemarie,levy}.

\begin{figure*}
\includegraphics[width=0.31\textwidth]{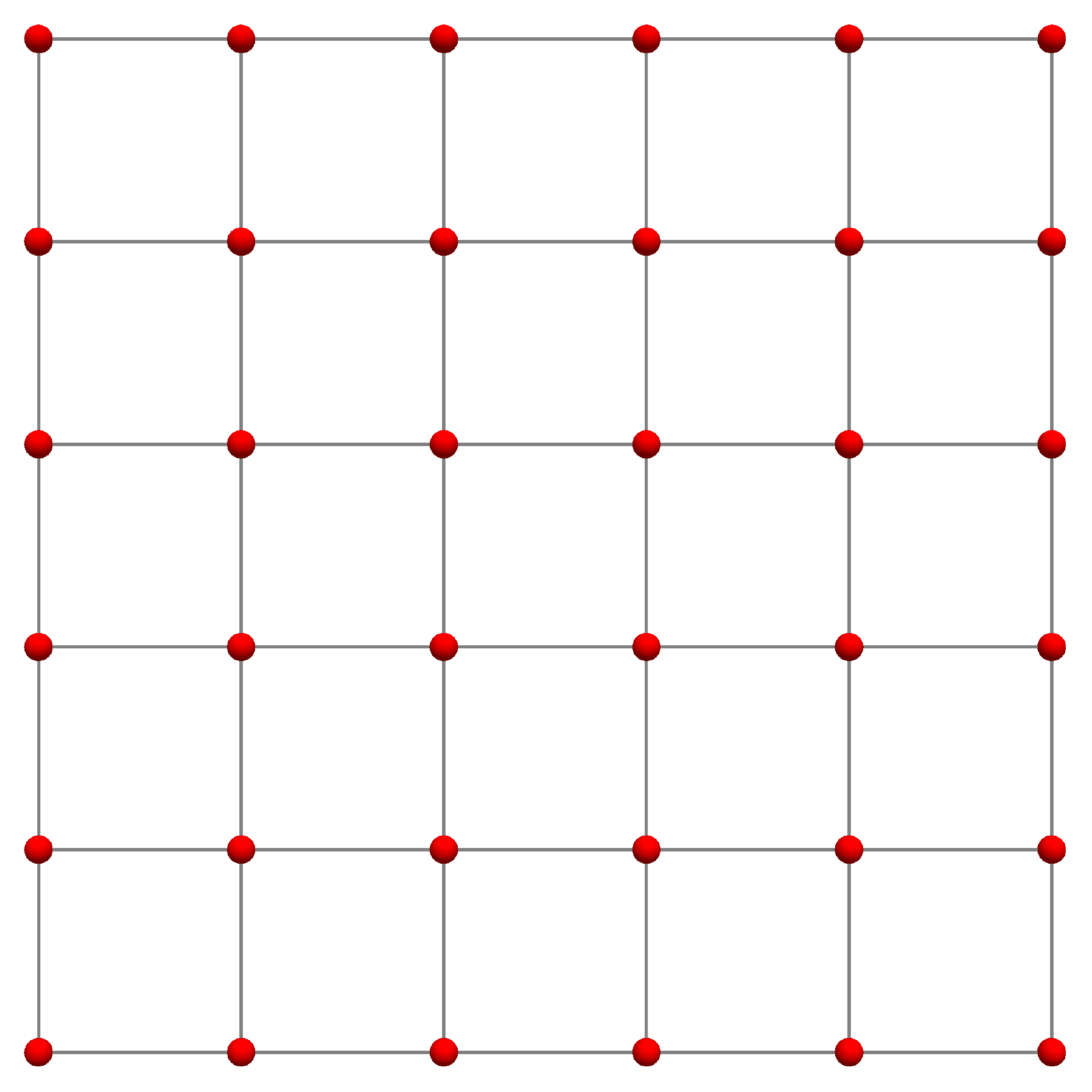} \hspace{+0.25cm} \includegraphics[width=0.31\textwidth]{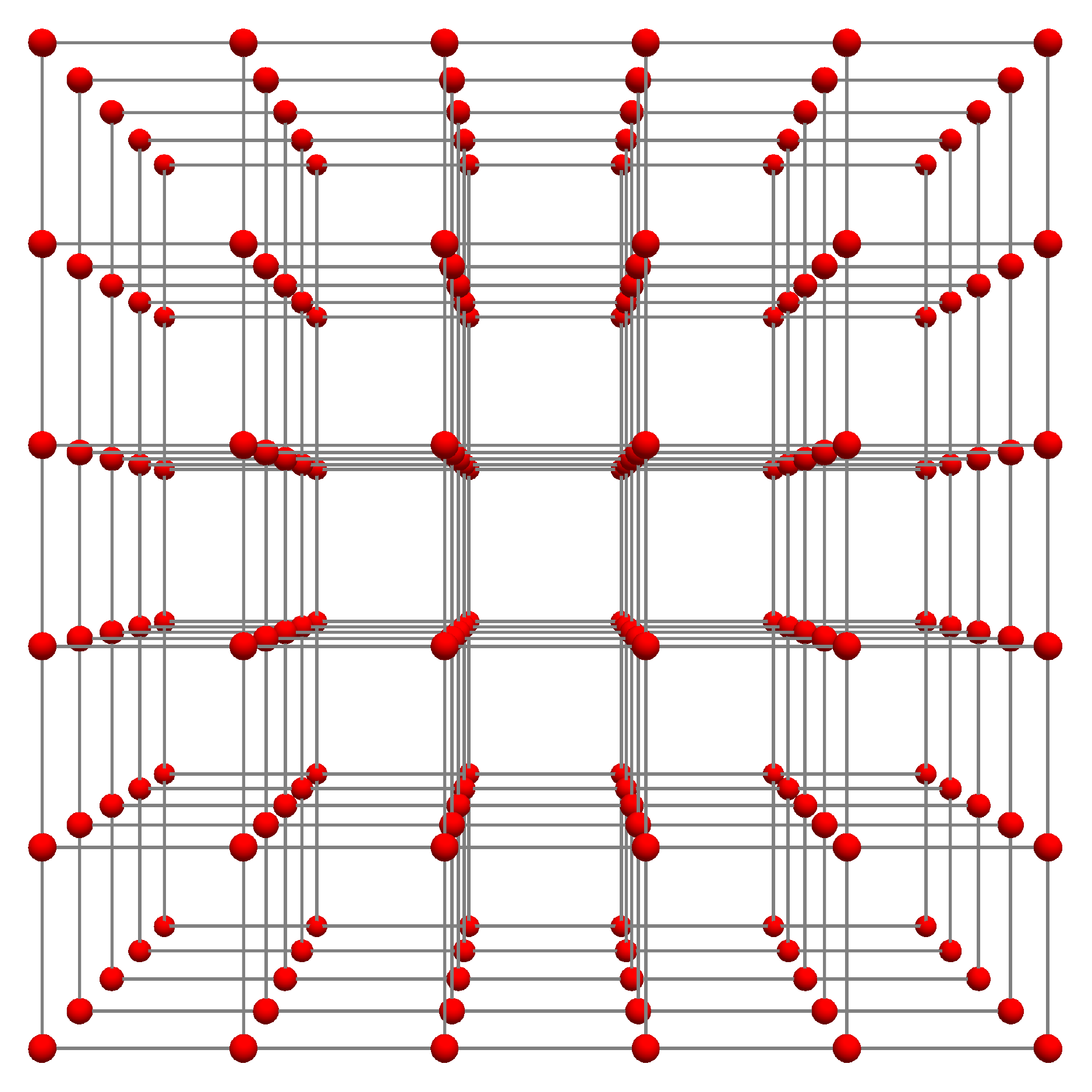}
\hspace{+0.25cm} \includegraphics[width=0.31\textwidth]{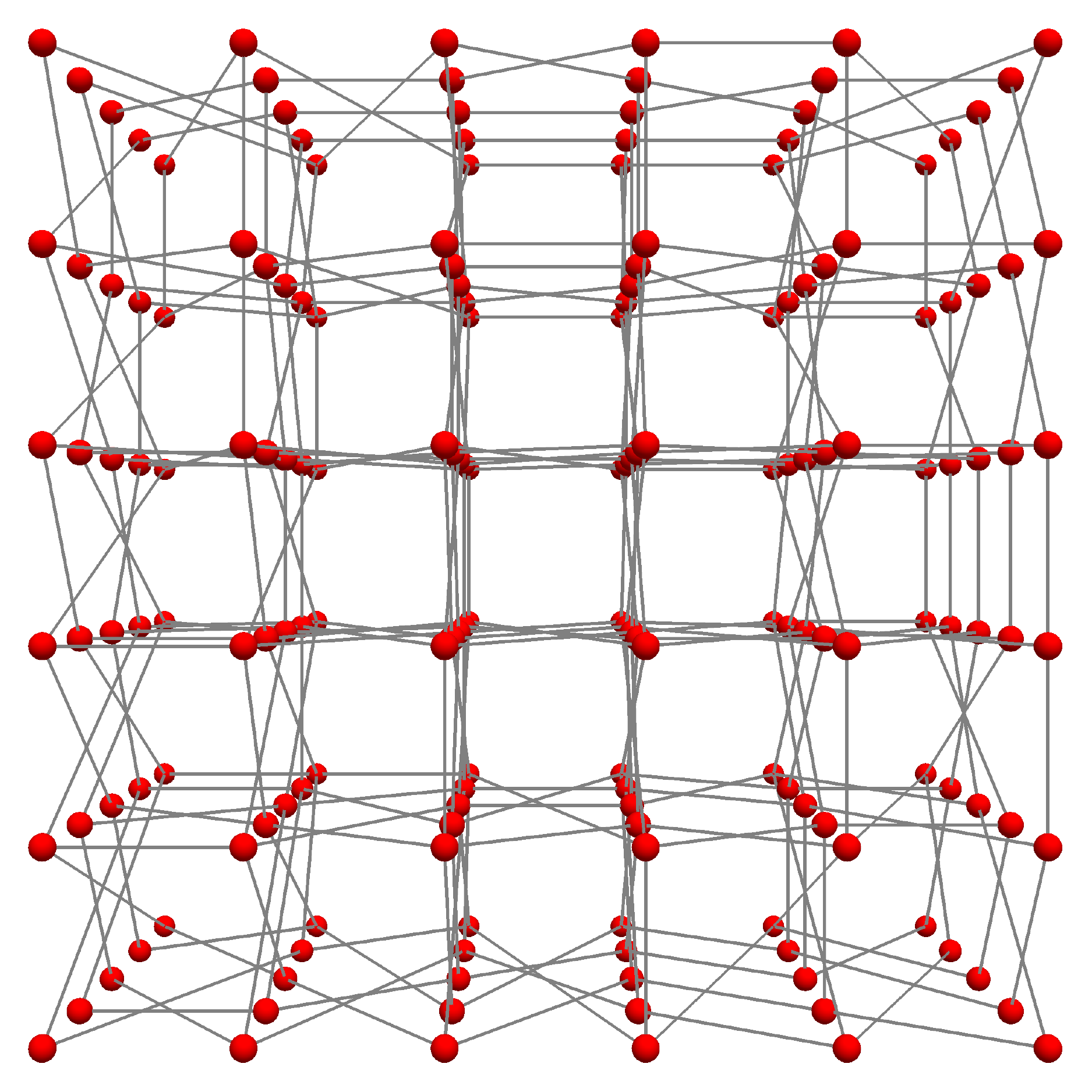} 
	\caption{\label{fig:mlayer} Illustration of the $M$-layer construction for a $2d$ square lattice of four sites with $M = 5$. The original lattice (left) is replicated $M$ times (middle); For {\it each} edge in the original graph, a random rewiring of the $M$ links is considered (right).   }
\end{figure*}

The other case in which the loop expansion~\eqref{eq:expansion} can be made perturbative in terms of a small parameter is the so-called $M$-layer construction. This is a technique that has been put forward recently precisely to build a perturbative series in topological diagrams around the Bethe solution for any physical model defined on a $d$-dimensional euclidean lattice. A general and detailed description of this technique can be found in Ref.~\cite{mlayer} (see also Refs.~\cite{vonto,mori} for the formulation of the same  approach in the context of computer science). For the interested reader in App.~\ref{app:perco} we also illustrate the $M$-layer expansion (at the 1-loop level) for the simplest statistical mechanics model with a second order phase transition, namely random percolation. 

The starting point of the $M$-layer approach consists in constructing $M$ copies of the model defined on a $d$-dimensional lattice, as sketched in the middle drawing of Fig.~\ref{fig:mlayer}. For {\it each} of the original edges one chooses uniformly and independently at random a permutation  of the set of vertices (as sketched in the right drawing of Fig.~\ref{fig:mlayer}), creating inter-layer links (the average over all possible rewirings has to be taken at the end of the computation). In the $M \to \infty$ limit the probability to have a loop goes to 0 as $1/M$ and the Bethe approximation becomes asymptotically exact, while the original model in $d$ dimensions is recovered for $M = 1$. This approach has been recently applied to several problem for which the field-theoretical formulation is missing, either because the transition in the fully-connected model is absent (as for the $k$-core percolation~\cite{mboot} or the spin-glass in an external field~\cite{msg1,msg2}), or very different from the true one (as for the random field Ising model~\cite{mrfim}).

For the specific case of the Anderson model the $M$-layer construction is formally analogous to the classical Wegner's $n$-orbital model~\cite{wegner,wegner1,efetoflarkin}, the only difference being the fact that each electron state has a single non-vanishing transition rate with each of its neighbors. Thanks to universality, for any finite $M$ the critical properties on the $M$-layered replicated lattice are those of the $d$ dimensional model. In this respect the $M$-layer approach is tightly related to Efetov’s effective medium approximation~\cite{ema} (see App.~\ref{app:ema}) and also similar in spirit to the construction proposed in Ref.~\cite{FMS}. 

Performing the average over all possible permutations of the edges, the average number of $1$-loop diagrams of Fig.~\ref{fig:1loopdiagram} in the $M$-layer replicated lattice is given by the number of such diagrams on the original euclidean lattice, ${\cal N} (L,L_1)$, divided by the number of layers, $M$:
\begin{equation} \label{eq:NgavgMlayer}
\avg{{\cal N}_G (L,L_1)} = \frac{{\cal N} (L,L_1)}{M} \, .
\end{equation}
This clearly illustrates why the $M$-layer approach becomes useful on $d$-dimensional lattices, as it allows to suppress by powers of $1/M$ the number of loops in the replicated lattice, thereby allowing one to rewrite the loop expansion~\eqref{eq:expansion} in a perturbative way, in terms of the small parameter $1/M$. The asymptotic expressions giving the number of $1$-loop diagrams in the original $d$ dimensional lattice (when the length of the lines is large) is 
\begin{equation} \label{eq:loops}
	{\cal N} (L,L_1) \simeq \frac{k^{(L+L_1)}}{L^{d/2}} \, .
\end{equation}
In this context, the loop expansion can be applied to compute finite-dimensional corrections to the BL solution of AL.
In principle a full computation of the singular behavior of  the loop corrections (including multi-point observables and higher order diagrams) should allow one to compute  the non-mean-field critical exponents by resumming the series through the same recipes of the field theoretical expansions~\cite{SFT,lebellac,ZJ,cardy}.

\section{The model} \label{sec:model}

For concreteness we consider the simplest  model for AL, which consists in non-interacting spin-less fermions in a disordered potential:
\begin{equation} \label{eq:H}
	{\cal H} = - t \sum_{\langle i, j \rangle} \left (  c^\dagger_i c_j  + c^\dagger_j c_i \right) - \sum_{i=1}^N \epsilon_i c^\dagger_i c_i \, .
\end{equation}
The second sum runs over all $N$ sites of the lattice, and the first sum runs over all pairs of nearest neighbors sites; $c^\dagger_i$ and $ c_i$  are creation and annihilation operators, and $t$ is the hopping kinetic energy scale, which we take equal to 1 throughout. The on-site energies $\epsilon_i$ are independent random variables uniformly distributed in the interval $[-W/2, W/2]$, $W$ being the disorder strength. Localization begins from the band edges~\cite{noi,edges}, therefore to see if all states are localized it is sufficient to look at the band center, $E=0$. 

AL can be cast in the framework of spontaneous symmetry breaking, with an order parameter function  related to the probability distribution of the local density of states (LDoS) at energy $E$~\cite{mirlin94}, 
\begin{equation} \label{eq:LDoS}
\rho_i  \equiv \sum_\alpha |\psi_\alpha (i)|^2 \delta(E-E_\alpha ) \, .
\end{equation}
In the insulating phase the LDoS vanishes while in the metallic phase it is finite with probability density $P(\rho)$.

The statistics of the LDoS is encoded in the statistics of the (diagonal) elements of the resolvent matrix, $G_{ij} = ({\rm i} \eta {\cal I} - {\cal H})^{-1}_{ij}$, where ${\cal I}$ is the identity matrix, ${\cal H}$ is the Hamiltonian~\eqref{eq:H}, and $\eta$ is an infinitesimal imaginary regulator that softens the pole singularities in the denominator. On the infinite BL the diagonal elements of $G$ verify an exact self-consistent recursion relation~\cite{Abou-Chacra} (see App.~\ref{app:cavity}). Its fixed point yields the probability distribution of the LDoS $P(\rho)$, as well as the IPR (which is essentially proportional to the second moment of $P(\rho)$):
\begin{eqnarray}
\label{eq:ldos} \rho_i & = &\frac{1}{\pi} \lim_{\eta \to 0^+} {\rm Im} G_{ii} \, , \\
\label{eq:I2} I_2 & = &\frac{1}{\pi \langle \rho \rangle } \lim_{\eta \to 0^+} \! \frac{1}{N} \sum_i \eta |G_{ii}|^2 = \lim_{\eta \to 0^+} \! \frac{\eta \avg{|G|^2}}{\avg{{\rm Im} G}} \, ,
\end{eqnarray}
Since the BL solution is the starting point of our perturbative loop expansion, it is useful to briefly recall here its key properties.
In the delocalized phase $P(\rho)$ converges to a stable non-singular distribution which becomes very broad and asymmetric upon approaching the critical disorder from below, and a (large) characteristic scale $\Lambda$ spontaneously emerges (see Fig.~\ref{fig:Prho} of App.~\ref{app:cavity}): The probability distribution has a sharp maximum for $\rho \sim \Lambda^{-1}$ followed by a power law decay $P(\rho) \sim \rho^{-3/2}$ with a cutoff at $\rho$ of order $\Lambda$~\cite{mirlin94,large_deviations}. Such $\Lambda$ is found to diverge exponentially at the critical point as~\cite{efetov_bethe,zirn,mirlin,mirlin1,verba,tikhonov_critical,mirlintikhonov,large_deviations}
\begin{equation} \label{eq:rhotypBL} 
\Lambda \propto \exp{ \left[ A (W_c-W)^{-\nu_{\rm del}} \right] } \, ,
\end{equation}
(with $\nu_{\rm del} = 1/2$), and can be interpreted as the {\it correlation volume} of typical eigenstates: for $W \lesssim W_c$ the wave-functions have $O(N/\Lambda)$ bumps localized in a small region of the BL where the amplitude is of order $\Lambda/N$ (to ensure normalization), separated by regions of radius $\ln \Lambda$ where the amplitude is very small~\cite{mirlin94}.

The typical value of the LDoS, $\rho_{\rm typ} = \exp[ \langle \ln \rho_i \rangle ] \propto \Lambda^{-1}$, can be taken as a proxy for the order parameter (the average DoS is instead analytic across the transition), providing another intuitive argument that allows one to interpret $\Lambda$ as the correlation volume: in order to be in a regime representative of the thermodynamic limit the typical number of states per unit of energy contributing to the LDoS on a given site, $N \rho_{\rm typ}$, must be much larger than $1$. Since $\rho_{\rm typ} \propto \Lambda^{-1}$ this condition is fulfilled only if $N \gg \Lambda$ (see also App.~\ref{app:2points} and the bottom panels of Fig.~\ref{fig:G0L2}). 

In this paper we focus on the delocalized side of the transition only. The localized regime will be studied in a forthcoming work. 

\section{The Ginzburg criterion} \label{sec:ginzburg}

The Ginzburg criterion consists in requiring that finite-dimensional fluctuations do not modify the mean-field critical behavior. To this aim, we apply the $M$-layer construction discussed in Sec.~\ref{sec:Mlayer} to the Anderson model on a generic $d$-dimensional euclidean lattice (see Fig.~\ref{fig:mlayer}) and define a local distribution of LDoS $P_i(\rho)= (1/M) \sum_{\alpha=1}^M \delta(\rho-\rho_{i,\alpha})$ where $\rho_{i,\alpha}$ is the LDoS of site $i$ on the $\alpha$-th layer. In the large-$M$ limit the BL solution becomes {\it exact} in the sense that on all sites $i$ of the original lattice, $P_i(\rho)$ is essentially equal to $P(\rho)$ on the BL with small $O(1/M)$ fluctuations around the mean. Nonetheless according to the Ginzburg criterion we have to check that the prefactor of $1/M$ does not diverge at the transition in order for the mean-field predictions to be valid. Usually a local order parameter, say an Ising spin, has always large fluctuations, therefore to have an order parameter with small fluctuations with respect to the mean one has to consider a coarse-grained order parameter up to the scale of the correlation length. In order to apply the Ginzburg criterion it is thus customary to considers fluctuations on the scale of the correlation length (see e.g. Ref.~\cite{huang}). In the $M$-layer model even the local order parameter, being already the sum of $M$ terms, has small fluctuations and the Ginzburg criterion can be applied directly to it.
  
Let us therefore consider as a local observable $(1/M) \sum_{\alpha=1}^M \ln \rho_{i,\alpha}$. The typical DoS plays in fact the role of a proxy for the order parameter distribution, as $e^{\avg{\ln \rho}}$ vanishes proportionally to $\Lambda^{-1}$ when $W$ approaches $W_c$ from below and is identically equal to zero in the insulating phase. The average of $(1/M) \sum_{\alpha=1}^M \ln \rho_{i,\alpha}$ converges to the BL result $\ln \rho_{\rm typ}$ at large values of $M$. The fluctuations of its value at two lattice sites at distance $r$ on the original $d$-dimensional euclidean $M$-layered lattice can be written as~\cite{mlayer}:
\begin{equation} \label{eq:Cd}
C_d (r) = \frac{1}{M} \sum_{L=1}^{\infty} {\cal N} (r,L) \, C_{\rm BL} (L) \, .
\end{equation}
where ${\cal N} (r,L)$ is the number of non-backtracking paths of length $L$ on the original euclidean lattice between  two sites at distance $r$, that for large $r$ and $L$  obeys:
\begin{equation} \label{eq:NrL}
{\cal N} (r,L)  \simeq  k^L e^{-r^2/(4L)} L^{-d/2} \, .
\end{equation}
The second factor $C_{\rm BL} (L)$  is the correlation function computed on the infinite BL between two sites at distance $L$: 
\begin{equation} \label{eq:CBL}
C_{\rm BL} (L) = \langle \ln \rho_i \ln \rho_{i+L} \rangle - \langle \ln \rho_i \rangle \langle \ln \rho_{i+L} \rangle \, ,
\end{equation}
To the best of our knowledge $C_{\rm BL} (L)$ has never been studied before in the literature, and characterizing its asymptotic behavior is an interesting result {\it per se}. A  convenient way to compute it is to apply exact decimation relations~\cite{aoki,dobro,noilarged} which allows one to integrate out progressively all the intermediate vertices on the branch, as explained in details in App.~\ref{app:decimation}. A careful analysis of $C_{\rm BL} (L)$ in the limit of large $L$ and close to the critical disorder is presented in App~\ref{app:ginzburg}, thereby providing the behavior of $C_d(r)$ through Eq.~\eqref{eq:Cd}.

\begin{figure}
\includegraphics[width=0.469\textwidth]{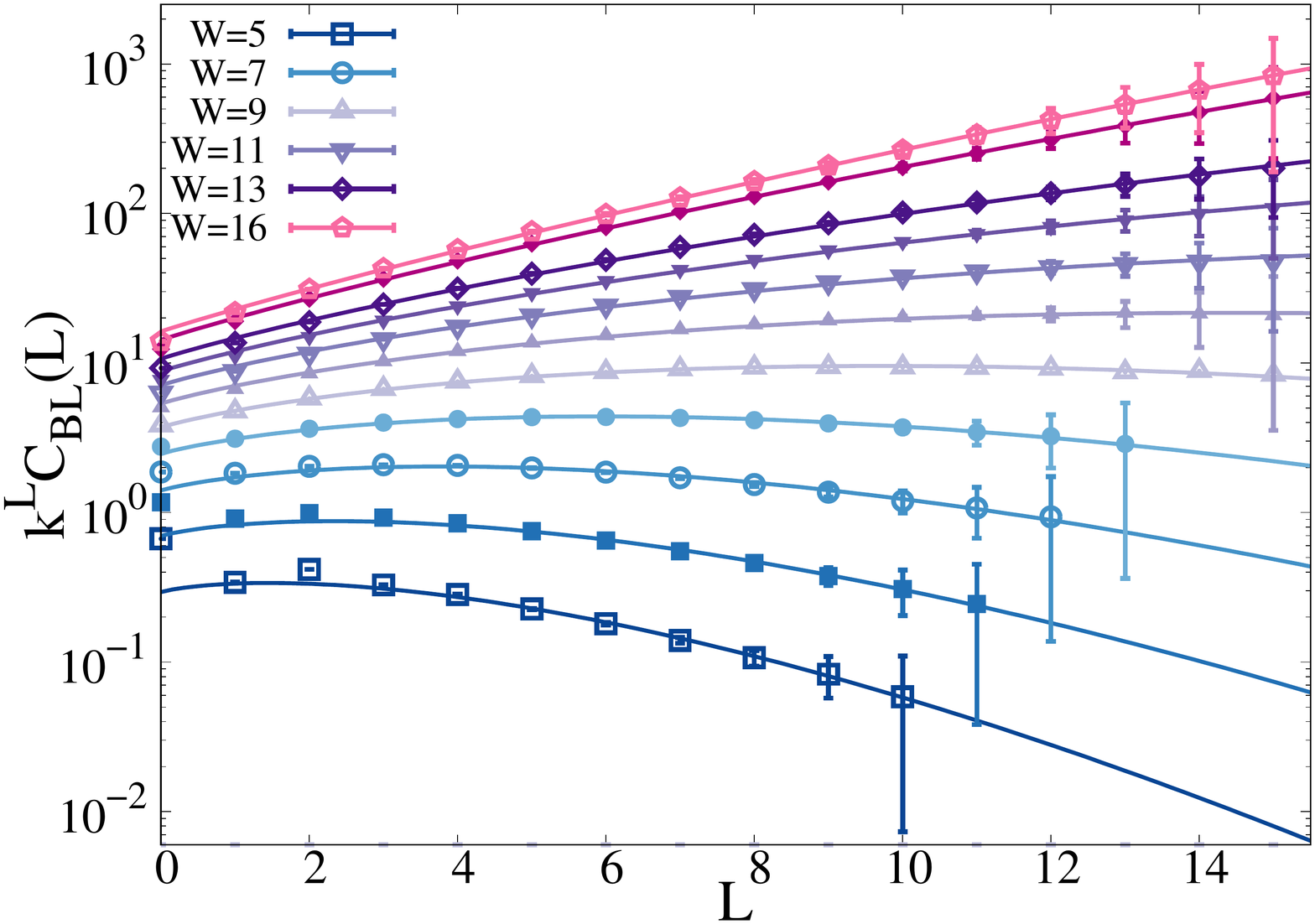} 
\caption{$k^L C_{\rm BL} (L)$ as a function of $L$ for different values of the disorder from $W=5$ to $W=16$ for $k=2$. The continuous lines correspond to the fits of the data according to Eq.~\eqref{eq:fit} (see App.~\ref{app:ginzburg}). The critical value of the disorder for $k=2$ is $W_c \approx 18.2$~\cite{tikhonov_critical}.
\label{fig:ginzburg}}
\end{figure}

In Fig.~\ref{fig:ginzburg} we plot $C_{\rm BL}(L)$ multiplied by  $k^L$ (i.e. the volume of the sphere of radius $L$ on the BL which enters in the expression~\eqref{eq:NrL} of the number of paths between two sites at distance $r$ of the origina lattice) for $k=2$. The figure shows that $k^L C_{\rm BL} (L)$ first grows for $L$ small enough, and then decreases at larger $L$, after going through a maximum in $L_\star$. The position of the maximum moves to larger and larger values as $W$ is increased and its height grows very rapidly with $W$. As illustrated in App.~\ref{app:ginzburg} we find 
\begin{equation} \label{eq:Lstar}
\begin{aligned}
&L_\star  \approx a (\ln \Lambda)^\delta \, , \\
& k^{L_\star} C_{\rm BL} (L_\star) \approx e^{b  \ln \Lambda}  \, ,
\end{aligned}
\end{equation}
where $a$ and $b$ are constant of order $1$, and $\Lambda$ is the correlation volume introduced above.

Setting $r=0$ in Eq.~\eqref{eq:CBL} we obtain the fluctuations of $\ln \rho_{i,\alpha}$ at a given position $i$ of the $M$-layered euclidean lattice among different layers $\alpha$ at the order $1/M$. According to the Ginzburg criterion we have to check that the prefactor of the fluctuations of order $1/M$ does not diverge at the transition in order for the mean-field predictions to be valid. The sum over $L$ in the expression above is dominated by the maximum of $k^L C_{\rm BL} (L)$ at $L=L_\star$, which grows exponentially fast as the disorder is increased. We thus obtain that at the saddle-point level the fluctuations of $\ln \rho$ at a given position in the $d$-dimensional space between different layers of the lattice  behave as (see App.~\ref{app:ginzburg} for more details):
\begin{equation} \label{eq:Cd0}
C_d (0) \propto \frac{1}{M}\frac{e^{c \ln \Lambda}}{(\ln \Lambda)^{d \delta/2}} \propto \frac{1}{M}(W_c - W)^{\frac{d \delta}{4}} e^{ {\rm cst}/\sqrt{W_c-W}}
\, ,
\end{equation}
(with $\delta \approx 1.5$), which grows exponentially fast close to $W_c$ in any dimension. Conversely, the square of the order parameter only diverges algebraically, as $\langle \ln \rho \rangle^2 \propto (\ln \Lambda)^2 \propto (W_c-W)^{-1}$, therefore {\it in any dimension}, no matter how large $M$ is, there will always be a region of disorder values close to $W_c$, the critical disorder on the BL, where the local order parameter $(1/M) \sum_{\alpha=1}^M \ln \rho_{i,\alpha}$ has fluctuations much larger than its mean. Therefore the BL critical behavior should not be trusted in any finite dimension $d$, confirming the idea that the upper critical dimension of the Anderson transition is $d_U \to \infty$~\cite{noilarged,garcia,mirlin94,castellani,dobro}.
On the contrary in more conventional second-order phase transitions, like ferromagnetism or percolation, instead of an exponential divergence there is an algebraic divergence that can be compensated by the first factor at sufficiently large values of $d$ leading to a finite upper critical dimension (see  App.~\ref{app:GP} for a detailed explanation of the Ginzburg criterion within the $M$-layer construction for the case of the percolation transition, for which $d_U=6$).

\section{1-loop corrections to the typical DoS} \label{sec:LDOS}
Next we consider the corrections to the critical behavior of $\langle \ln \rho \rangle$ at the 1-loop level within the $M$-layer construction. The 1-loop diagrams for one-point observables are shown in Fig.~\ref{fig:1loopdiagram} where the node on which we compute the LDoS is labeled as $0$. The asymptotic expressions giving the number of such diagrams in $d$ dimensions, ${\cal N} (L,L_1)$, when the length of the lines is large is given in Eq.~\eqref{eq:loops}. As explained in Sec.~\ref{sec:Mlayer}, the average number of $1$-loop diagrams on the $M$-layer replicated lattice is simply given by dividing ${\cal N} (L,L_1)$ by the number of layers $M$. The 1-loop contribution to the correction to $\ln \rho_{\rm typ}$ thus reads~\cite{mlayer} (see Sec.~\ref{sec:Mlayer} and Eqs.~\eqref{eq:expansion} and~\eqref{eq:NgavgMlayer}):
\begin{equation} \label{eq:rhotyp1loop}
\begin{aligned}
\Delta \! \avg{\ln \rho}_{\rm 1loop} &= \frac{1}{M} \sum_{L_1=0}^\infty \sum_{L=3}^\infty {\cal N} (L,L_1) \, \delta [\ln \rho (L,L_1)] \, ,
\end{aligned}
\end{equation}
where $\delta [\ln \rho (L,L_1)]$ is the line connected value of $\langle \ln \rho \rangle$   defined as the difference between the average of $\ln \rho_0$ computed in presence and in absence of the loop: 
\begin{equation} \label{eq:deltalnrho}
\begin{aligned}
\delta [\ln \rho (L,L_1)] & \equiv \langle \ln \rho_0 \rangle_{(L,L_1)\textrm{-loop}} - \langle \ln \rho_0\rangle_{\rm BL}  \, . 
\end{aligned}
\end{equation}
A thorough numerical analysis of the asymptotic behavior of $\delta [\ln \rho (L,L_1)]$ is performed in App.~\ref{app:deltalnrho}. We report below only the main results. Similarly to the case of the percolation transition discussed in App.~\ref{app:DeltaP}, we find that the dependence on $L$ and $L_1$ completely factorizes (see  Fig.~\ref{fig:loop1LA} of App.~\ref{app:deltalnrho}). Such factorization property can be understood by realizing that $\delta [\ln \rho (L,L_1)]$ is essentially a response function: It measures the variation of $\ln \rho$ on site $0$ due to the variation of the  LDoS on a site at distance $L_1$ from $0$ due to the presence of a loop of length $L$ originating from it. The length of the loop sets the amplitude of the perturbation. Since a small perturbation of the value of one of the cavity propagators in the right hand side of the BL iteration relations, Eq.~\eqref{eq:Gcav}, propagates linearly along a chain of the tree, $\delta [\ln \rho (L,L_1)]$ is given by the product of the amplitude of the perturbation (which depends only on $L$) times the response function (which depends only on $L_1$).

As shown in Fig.~\ref{fig:loop} of App.~\ref{app:deltalnrho}, sufficiently close to the localization transition and at large enough $L$ the asymptotic dependence of $\delta [\ln \rho (L,L_1)]$ at fixed $L_1$ (e.g. for $L_1=0$) is roughly proportional (apart from a minus sign) to the connected correlation function of $\ln \rho_i$ studied in the previous section, $\delta [\ln \rho (L,0)] \propto - \, C_{\rm BL} (L)$. This behavior can be rationalized as follows: If the values of the LDoS on two nodes at distance $L$ placed at the two ends of a chain embedded in the BL are correlated, then if one connects the two nodes to the same site $0$ producing a loop, the LDoS on site $0$ will be modified compared to its BL counterpart (i.e. the case in which the loop is absent). Conversely, if the the LDoS on the two nodes at the two ends of the chain are uncorrelated, then if one connects the two nodes to the same site $0$, the LDoS on site $0$ will be on average the same as on the infinite BL in which the loop  is absent and all the neighbors of $0$ are uncorrelated. The minus sign reflects the fact that finite-dimensional fluctuations reduce the value of the critical disorder. 

The dependence of  $\delta [\ln \rho (L,L_1)]$ on $L_1$ at fixed $L$ is studied in Fig.~\ref{fig:loop1LA} of App.~\ref{app:deltalnrho}, indicating that at large enough $L_1$ the 1-loop corrections to $\ln \rho_{\rm typ}$ decrease exponentially as $\lambda^{-L_1}$. We then finally obtain that:
\begin{equation}
\delta [\ln \rho (L,L_1)] \propto - \, C_{\rm BL} (L) \, \lambda^{-L_1} \, 
\end{equation}
The decay rate $\lambda$ is larger than $k$ for $W < W_c$ (ensuring the convergence of the sum over $L_1$ in Eq.~\eqref{eq:rhotyp1loop}), and decreases smoothly as $W$ is increased, approaching $\lambda \to k$ for $W \to W_c$. Close enough to $W_c$ we find that $\lambda - k \propto (1 - W/W_c)^{\omega}$ with $\omega \approx 3/2$. Plugging this expression into Eq.~\eqref{eq:rhotyp1loop} and performing the sums over $L$ and $L_1$, one obtains that close to the critical points the 1-loop corrections to the logarithm of the typical DoS are given by:
\begin{equation} \label{eq:Deltaavglnrho}
\begin{aligned}
\Delta \! \avg{\ln \rho}_{\rm 1loop} \propto - \frac{1}{M}(W_c - W)^{\frac{\delta d}{4} - \omega} e^{ {\rm cst} / \sqrt{ (W_c - W)} } \, .
\end{aligned}
\end{equation}
Hence, at any finite $M$, the corrections to the BL result diverge exponentially fast (with a subleading algebraic term),  and are overwhelmingly larger than the order parameter itself in any dimension, in agreement with the outcome of the Ginzburg criterion.

\begin{figure}
\includegraphics[width=0.45\textwidth]{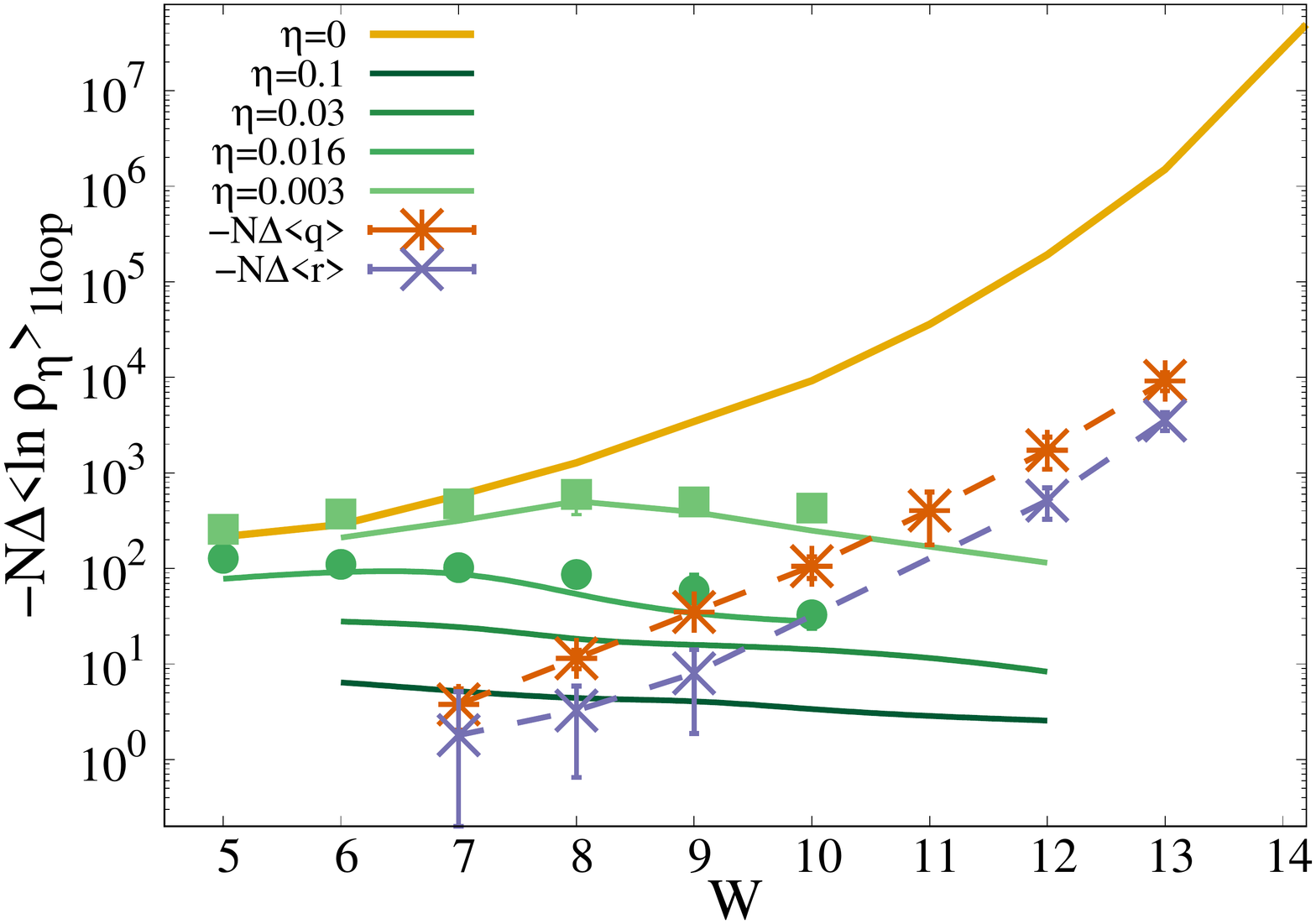}
\caption{$1$-loop corrections to the logarithm of the typical DoS (multiplied by $-1$) as a function of the disorder $W$ predicted within the $M$-layer construction, Eq.~\eqref{eq:Dlnrhoeta}, for finite $\eta$ (green) and in the $\eta \to 0^+$ limit (yellow). These predictions are compared with the $1/N$ corrections to $\avg{\ln \rho_\eta}$ obtained from EDs of RRGs of $N$ nodes (filled symbols) at the corresponding value of $\eta$, showing a good agreement. The dashed lines show the values of the $1/N$ corrections (multiplied by $-1$) to the average overlap between consecutive eigenvector, $-N \Delta \avg{q}$, and to the average gap ratio, $-N \Delta \avg{r}$, computed from EDs (see App.~\ref{app:Ncorrlnrho}).
\label{fig:BLcorr}}
\end{figure}

\subsection{$1/N$ corrections of the typical DoS on finite RRGs} \label{sec:deltalnrhoRRG}

As explained in Sec.~\ref{sec:Mlayer}, the knowledge of the asymptotic behavior of $\delta [\ln \rho (L,L_1)]$ at large $L$ and $L_1$ in the delocalized phase can be also exploited to estimate the finite-size corrections of order $1/N$ to the logarithm of the typical DoS on RRGs of large but finite sizes. As discussed in the introduction, this is an important issue, since a complete understanding of AL on the RRG has thus far proved to be notoriously elusive, mostly due to the alleged discrepancies between the extrapolations of numerical results obtained from EDs of large but finite samples and the expected asymptotic behavior of the supersymmetric solution. Due to the deep connection between AL on  sparse random networks and MBL~\cite{BAA,BetheProxy1,BetheProxy2,scardicchioMB,roylogan,tikhonov_mirlin_21,biroli_tarzia_17,gabrielKT}, these features have produced a strong debate in the last decade~\cite{noi_nonergo,scardicchio1,ioffe1,ioffe3,refael2019,pinorrg,bera2018,detomasi2020,mirlinRRG,Bethe,lemarie,levy}.

Plugging the average number of $1$-loop diagrams on RRGs of $N$ nodes, Eq.~\eqref{eq:BLpaths} into the expansion~\eqref{eq:expansion}, the 1-loop corrections of order $1/N$ to the logarithm of the typical DoS on finite RRGs at the order $1/N$ is given by:
\begin{equation} \label{eq:rhotyp1loopRRG}
\begin{aligned}
\Delta \! \avg{\ln \rho}_{\rm 1loop} &= \frac{1}{N}
\sum_{L_1=0}^\infty \sum_{L=3}^\infty \frac{k^{L+L_1}}{2} \, \delta [\ln \rho (L,L_1)] \, .
\end{aligned}
\end{equation}
On the other hand, the $1/N$ corrections to $\avg{\log \rho}$ can be, in principle, also computed numerically by exact full diagonalization of the Hamiltonian of the Anderson model on $(k+1)$-RRGs of large but finite sizes. Yet, there is a subtle point that needs to be discussed thoroughly at this point. The role of $\eta$ in the denominator of the resolvent is to regularise the distribution of the LDoS by introducing a broadening of the energy levels and transforming the Dirac $\delta$-functions in Cauchy distributions of width $\eta$. In the $N \to \infty$ limit the spectrum in the metallic phase becomes absolutely continuous, the sum over the eigenvalues converges to an integral. Consistently, one finds that the distribution of the LDoS is regular and becomes independent of $\eta$ (provided that $\eta$ is sufficiently small). However, the DoS of any finite system is, by definition, point-like. As a consequence, even deep in the metallic regime, although all eigenvectors are extended, if one takes the limit $\eta \to 0^+$ at $N$ fixed and finite the probability distribution of the LDoS will look completely different from the one found in the thermodynamic limit on the BL. Hence, to compare with ED on the RRG one has to replace the delta function in the definition $\rho_i$, Eq.~\eqref{eq:LDoS}, with a smooth function, $\delta(x) \rightarrow \eta (x^2+\eta^2)^{-1} \pi^{-1}$, depending on a parameter $\eta$. The fact that the system is in the delocalized phase manifests itself in the emergence of a broad interval of $\eta$ in which the probability distribution of the LDoS becomes essentially independent of $\eta$. This interval extends over a broader and broader range of $\eta$ as the system size is increased, and eventually diverges in the thermodynamic limit. These properties have been analyzed thoroughly in~\cite{Bethe}, where it was shown that for $W < W_c$ the typical DoS is essentially independent of $\eta$ for $\eta$ larger than the mean level spacing, $\Delta_N = (N \avg{\rho})^{-1}$, and smaller than a disorder-dependent value of the regulator above which the typical DoS starts to increase with $\eta$. The value of $\eta$ above which the observables start to depend on $\eta$ is, not surprisingly, of the order of the inverse of the correlation volume $\Lambda^{-1}$, which is finite for $W<W_c$ but becomes exponentially small upon approaching the critical disorder.

For these reasons the numerical evaluation of $\avg {\ln \rho}$ on finite RRGs must be done at $\eta$ small but finite. The largest system sizes for which we are able to invert exactly the matrix ${\cal H} - {\rm i} \eta {\cal I}$ and computes $\avg{\ln \rho}$ numerically is $N =47104$. Since the correlation volume grows exponentially fast as $W$ approaches $W_c$ and is very large already far away from the transition, we are not able to set $\eta$ in the interval $(N \rho)^{-1} \ll \eta \ll \Lambda^{-1}$, within which $\avg{\ln \rho}$ would be independent of $\eta$, and thus representative of the $\eta \to 0^+$ limit. Hence, the numerical values of $\avg{\ln \rho}$ that we measure on finite RRGs not only suffer from finite-size corrections with respect to their counterpart in the thermodynamic limit, but also of finite-$\eta$ corrections, due to the fact that we are constrained to consider values of $\eta$ larger than $\Lambda^{-1}$. As a consequence, in order to compare the numerical results obtained for samples of finite size at finite $\eta$ with the predictions of the 1-loop result, we need first to determine the effect of introducing a finite (but small) regulator on the asymptotic behavior of the $1$-loop line connected value of the average of the logarithm of the LDoS.

A thorough investigation of the effect of the imaginary regulator is reported in App.~\ref{app:Ncorrlnrho} (see in particular Fig.~\ref{fig:k2IratioTYP}). In practice we find that at finite $\eta$ the 1-loop corrections to $\avg{\ln \rho}$ develops  additional exponential decays of the form:
\begin{equation}
\delta [\ln \rho (L,L_1)] \propto - \, C_{\rm BL} (L) \, e^{- r_0 \eta L} \lambda^{-L_1} \, e^{- r_0^\prime \eta L_1} \, . 
\end{equation}
The rate of the exponential functions $r_0$ and $r_0^\prime$ depend very strongly on $W$ and are found to increase exponentially fast as the disorder is increased towards the localization transition, proportionally to the correlation volume: $r_0, r_0^\prime \propto \Lambda$, (see App.~\ref{app:Ncorrlnrho} for more details).

Plugging this expression into Eq.~\eqref{eq:rhotyp1loopRRG} and summing over $L$ and $L_1$, where ${\cal N} (L,L_1)$ is the number of $1$-loop diagrams on finite RRGs given in Eq.~\eqref{eq:BLpaths}, we obtain that the correction to the BL result is 
\begin{equation} \label{eq:Dlnrhoeta}
\Delta \! \avg{\ln \rho_\eta}_{\rm 1loop} \simeq - \frac{\varphi}{N} \,  \frac{e^{b \ln \Lambda  (1 - c_2 \eta \Lambda)}}{1 - \frac{k}{\lambda} \, e^{- \eta \Lambda^{c_3}}}  \, ,
\end{equation}
where $\varphi$, $b$, $c_2$, and $c_3$ are constants of order $1$ (whose numerical values are provided in Apps.~\ref{app:deltalnrho} and~\ref{app:Ncorrlnrho}) and $\lambda - k \propto (W_c - W)^\omega$. The prediction of Eq.~\eqref{eq:Dlnrhoeta} are shown in Fig.~\ref{fig:BLcorr}: if $\eta \ll \Lambda^{-1}$ the contribution coming from the regulator is negligible and the $1/N$ corrections to the logarithm of the typical DoS diverge exponentially fast close to $W_c$ as $-(W_c-W)^{-\omega} e^{{\rm cst}/\sqrt{W_c-W}}$ (yellow line); For $\eta$ finite, instead, $-\Delta \! \avg{\ln \rho_\eta}_{\rm 1loop}$ increases exponentially at small disorder (for $W$ such that $\eta \ll \Lambda^{-1}$), and then decreases at large disorder (for $W$ such that $\eta \Lambda \gtrsim 1$) due to the effect of the strong exponential cutoff produced by the regulator (green lines). This results in a non-monotonic behavior of the $1/N$ corrections to $\avg{\ln \rho_\eta}$ at finite-$\eta$. 

These theoretical predictions can be compared to the numerical results of the $1/N$ corrections to the logarithm of the typical DoS (for $\eta$ small but finite) on RRGs of large but finite sizes. As shown in App.~\ref{app:Ncorrlnrho}, one clearly observes a regime at large $N$ in which the finite-size corrections to the logarithm of the typical DoS are linear in $1/N$. By performing a linear fit of $\avg{\ln \rho_\eta (N)} - \avg{\ln \rho_\eta}_{\rm BL}$ as a function of $1/N$, one can evaluate numerically the $1/N$ corrections to the $N \to \infty$ value of $\avg{\ln \rho_\eta}_{\rm BL}$, which can be easily obtained from the solution of the self-consistent equations for the infinite BL. As shown in the figure the ED results agree well with the theoretical predictions of Eq.~\eqref{eq:Dlnrhoeta}.

We have also computed the $1/N$ corrections to the average value of the mutual overlap between two subsequent eigenvectors $\avg{q}$, and the average gap ratio $\avg{r}$, defined as:
\begin{equation} \label{eq:Q}
\begin{aligned}
\avg{q} &= \left \langle \sum_{i=1}^N \left | \psi_\alpha (i)  \right | \left | \psi_{\alpha+1} (i) \right | \right \rangle \, , \\  
\avg{r} &= \left \langle \frac{{\rm min} \{\lambda_{\alpha+2} - \lambda_{\alpha+1}, \lambda_{\alpha+1} - \lambda_{\alpha} \}}{{\rm max} \{ \lambda_{\alpha+2} - \lambda_{\alpha+1}, \lambda_{\alpha+1} - \lambda_{\alpha} \}} \right \rangle \, ,
\end{aligned}
\end{equation}
where $\lambda_\alpha$ and $\psi_\alpha$ are the (sorted) eigenvalues and the eigenvectors of the Anderson Hamiltonian. $\avg{q}$ and $\avg{r}$ are both  related to the level statistics, and take different universal values in the delocalized/GOE and in the localized/Poisson phases~\cite{huse,Bethe,large_deviations} (see App.~\ref{app:Ncorrlnrho} for more details). $\avg{r}$ and $\avg{q}$ can be directly measured from ED of RRGs of finite but large sizes and have the advantage of being defined directly in $\eta = 0^+$ limit.  We have computed numerically the $1/N$ corrections to  $\avg{r}$ and $\avg{q}$, defined as $\Delta \! \avg{r}$ and $\Delta \! \avg{q}$ respectively, (see in particular the right panel of Fig.~\ref{fig:BLcorrSI} of App.~\ref{app:Ncorrlnrho}), which turns out to be negative. The numerical results for $-N \Delta \! \avg{q}$ and $-N \Delta \! \avg{r}$ are shown as orange and violet dashed lines  in Fig.~\ref{fig:BLcorr}, and  follow the same exponential trend as (minus) the corrections to the logarithm of the typical DoS in the $\eta \to 0$ limit, $\avg{\ln \rho_{0^+}}$, in agreement with the numerical results of Refs.~\cite{mirlinRRG,large_deviations}.


\begin{figure}
\includegraphics[width=0.48\textwidth]{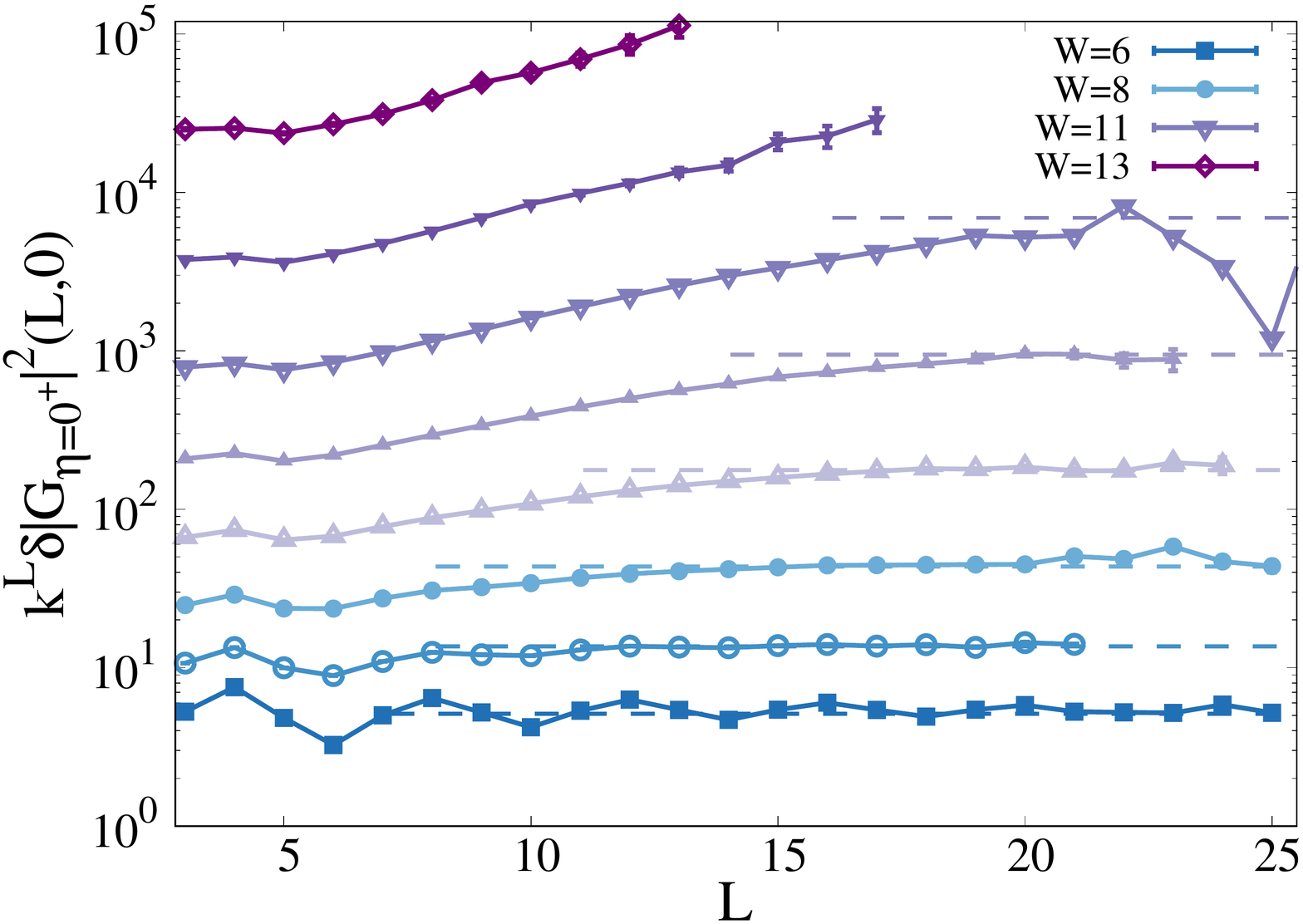} 
\caption{$k^L \delta |G_{\eta=0^+}|^2 (L, L_1=0)$ as a function of $L$ (for $L_1$ fixed and equal to $0$) for several values of the disorder from $W=6$ to $W=13$ across the metallic phase. The critical value of the disorder for $k=2$ is $W_c \approx 18.2$~\cite{tikhonov_critical}.
\label{fig:k2Iloop_main}}
\end{figure}

\section{Finite-size corrections to the IPR on the RRG in the delocalized phase} \label{sec:IPR}

Here we apply the same approach to determine the $1$-loop corrections to the IPR on RRGs of large but finite sizes in the delocalized phase. In fact in the $N \to \infty$ limit the IPR is identically equal to zero in the metallic regime. However on a RRG of $N$ nodes the IPR should scale as $1/N$ times a disorder-dependent constant which has been predicted to be proportional to the correlation volume within the supersymmetric formalism: $I_2 \propto \Lambda/N$~\cite{mirlintikhonov}. As we have seen in Sec.~\ref{sec:deltalnrhoRRG}, the $1/N$ corrections on the RRG are obtained by studying $1$-loop diagrams, and hence the $1/N$ prefactor of the IPR must be recovered considering the diagrams of Fig.~\ref{fig:1loopdiagram}. 

The spectral representation of the IPR in the thermodynamic limit is given in Eq.~\eqref{eq:I2}. Since $\avg{\rho}$ is a smooth decreasing function of $W$, which is non-singular across the localization transition, the 1-loop corrections to $\avg{\rho}$ are expected to be non-divergent at $W_c$. Hence, the $1/N$ prefactor of the IPR is given by the corrections to $\eta \langle |G_\eta|^2 \rangle$ at 1-loop (in the $\eta \to 0$ limit): 
\begin{equation} \label{eq:iprRRG}
\begin{aligned}
I_2 & \simeq \lim_{\eta \to 0^+} \frac{1}{N \pi \avg{\rho}} \sum_{L,L_1} \avg{{\cal N}_G (L,L_1)} \, \eta \delta |G_\eta|^2 (L, L_1) \, ,
\end{aligned}
\end{equation}
where $\avg{{\cal N}_G (L,L_1)}$ is the average number of $1$-loop diagrams originating from node $0$ on a RRG of $N$ nodes given by Eq.~\eqref{eq:BLpaths}. To evaluate this expression thus we need to determine the asymptotic behavior at large $L$ and $L_1$ of $\delta |G_\eta|^2 (L, L_1)$ at small $\eta$, which is the line connected value of $\langle |G_\eta|^2 \rangle$ at the 1-loop level, defined as the difference between $\langle |G_\eta|^2 \rangle$ computed on site $0$ of the diagram of Fig.~\ref{fig:1loopdiagram} in presence and in absence of the loop:
\begin{equation} \label{eq:deltaG2}
\delta |G_\eta|^2 (L, L_1) = \avg{|G_\eta|^2}_{(L, L_1)\textrm{-loop}} - \avg{|G_\eta|^2}_{\rm BL} \, .
\end{equation}
To perform this analysis it is again essential to consider the case in which the $\delta$-function is replaced by a smooth function and then taking the limit $\eta \rightarrow 0$. The computation is not trivial due to the presence of a term that is divergent at $\eta=0$ associated to the broken symmetry in the supersymmetric formalism \cite{mirlintikhonov}, as discussed below. 

Analogously to the 1-loop corrections to $\avg{\ln \rho}$ discussed above, we find that the dependence on $L$ and $L_1$ of $\delta |G_\eta|^2 (L, L_1)$ completely factorizes (see App.~\ref{app:deltaG2} for details). In Fig.~\ref{fig:k2Iloop_main} $L_1$ is fixed to $0$ (no external leg) and $k^L \delta |G_{\eta=0^+}|^2 (L, 0)$ is plotted as a function of the length of the loop $L$ when the imaginary regulator is set to $\eta=0$. This plot indicates that  for $L$ large enough (i.e. $L$ larger than a characteristic scale proportional to some power of $\ln \Lambda$) $\delta |G_{\eta=0^+}|^2$ behaves as $C k^{-L}$. The value of the disorder-dependent prefactor $C$ is found to grow very fast as the disorder is increased, roughly as the square of the correlation volume, $C \propto \Lambda^2$ (see App.~\ref{app:deltaG2}, in particular the right panel of Fig.~\ref{fig:k2Iratio}).

The dependence of  $\delta |G_{\eta=0^+}|^2(L,L_1)$ on the length of the external leg $L_1$ (for fixed $L$) is again a simple exponential decay for $L_1$ large, as $\lambda^{-L_1}$,  with $\lambda>k$, and $\lambda$ approaching $k$ algebraically when $W$ approaches $W_c$ from below. The values of the exponential rate $\lambda$ obtained from fitting the data  (see Fig.~\ref{fig:k2Iloop} of App.~\ref{app:deltaG2}) are compatible with the values of the rate describing the exponential decay of $\delta [\ln \rho (L,L_1)]$ at fixed $L$.

Putting all these results together we obtain the following asymptotic behavior at large $L$ and $L_1$ for $\eta=0$:
\begin{equation} \label{eq:dG2h0}
\delta |G_{\eta=0^+}|^2 (L, L_1) \approx C \, k^{-L} \lambda^{-L_1} \, ,
\end{equation}
where $\lambda - k \propto (W_c-W)^\omega$ with $\omega \approx 3/2$, and $C \propto \Lambda^{2}$. Hence, when multiplying $\delta |G_{\eta=0^+}|^2$ by the number of diagrams on RRGs of $N$ nodes, Eq.~\eqref{eq:BLpaths}, and summing over $L$ and $L_1$, the $1$-loop corrections to $\avg{|G_{\eta=0^+}|^2}$ diverge. Yet, since the IPR is proportional to $\eta \avg{|G_\eta|^2}$ for $\eta \to 0$, in order to obtain the 1-loop corrections to the IPR we need to study the behavior of $\delta |G_{\eta}|^2 (L, L_1)$ at small but {\it finite} $\eta$. 

A thorough investigation of the effect of the imaginary regulator is reported in App.~\ref{app:deltaG2} (see in particular Fig.~\ref{fig:k2Iratio}). We again find that at finite $\eta$ the 1-loop corrections to $\avg{|G_{\eta}|^2}$ develops additional exponential decays of the form:
\begin{equation}
\delta |G_{\eta}|^2 (L, L_1) \approx C \, k^{-L} e^{- \tilde{r}_0 \eta L} \lambda^{-L_1} e^{- \tilde{r}_0^\prime \eta L} \, . 
\end{equation}
The rate of the exponential functions $\tilde{r}_0$ and $\tilde{r}_0^\prime$ depend very strongly on $W$ and are found to increase exponentially fast as the disorder is increased towards the localization transition, proportionally to the correlation volume: $\tilde{r}_0, \tilde{r}_0^\prime \propto \Lambda$, (see the right panel of Fig.~\ref{fig:k2Iratio}).

\begin{figure}
\includegraphics[width=0.45\textwidth]{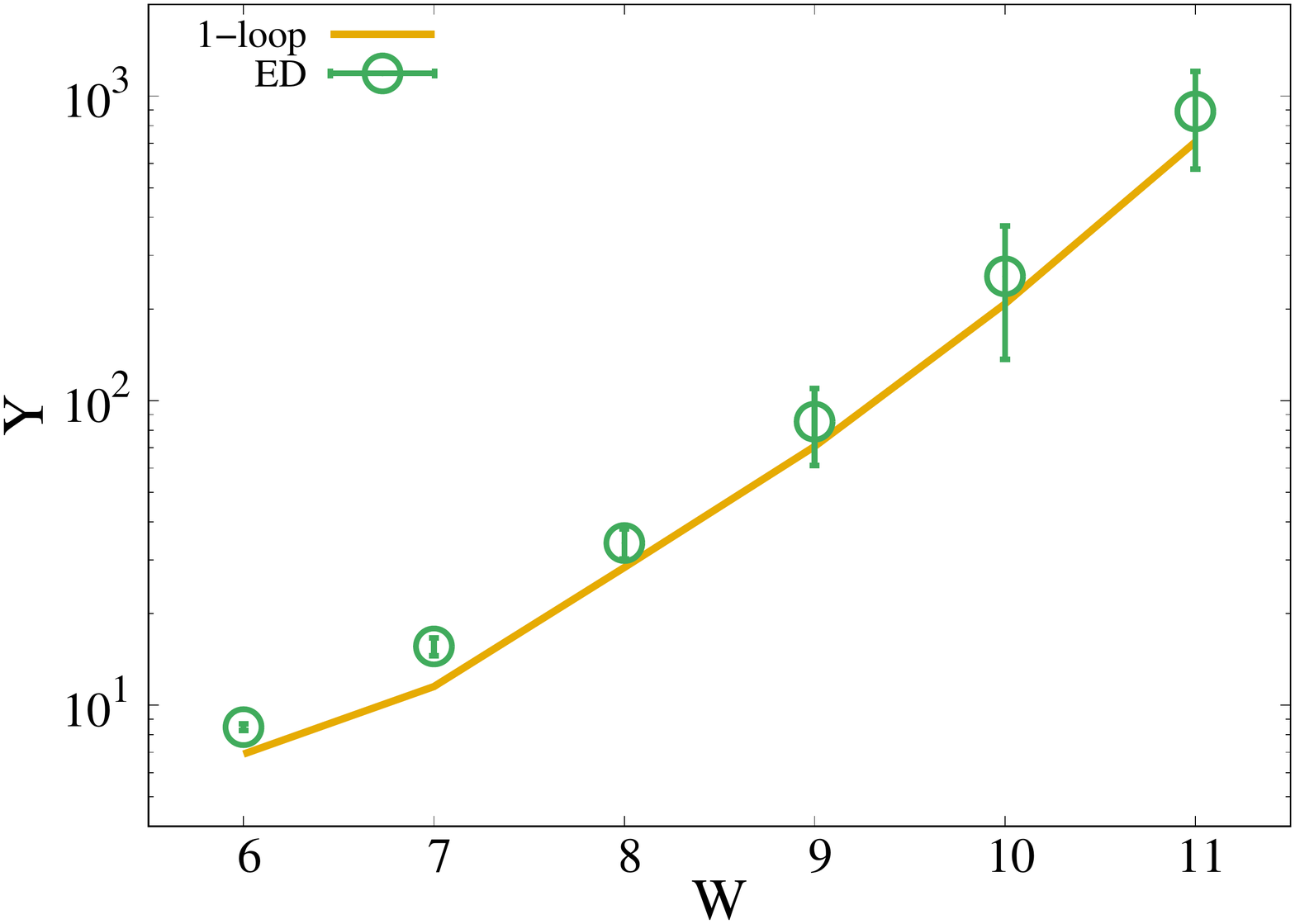} 
\vspace{-0.4cm}
\caption{1-loop value of the IPR at order $1/N$ on RRGs of $N$ nodes (Eq.~\eqref{eq:I2}, continuous line), compared with the IPR obtained from EDs (circles) as a function of the disorder $W$.
\label{fig:IPRcorr}}
\end{figure}

Putting all these results into Eq.~\eqref{eq:iprRRG} and summing over the 1-loop diagrams with $\avg{{\cal N}_G (L,L_1)}$ given by Eq.~\eqref{eq:BLpaths} we can finally estimate the $1/N$ corrections on finite RRGs (in the $\eta \to 0$ limit) to the IPR. One finally obtains (see App.~\ref{app:deltaG2} for more details)
\begin{equation} \label{eq:IPR}
I_2 \simeq \frac{Y_{\rm 1loop}}{N} \, , \textrm{~~~~with~~} Y_{\rm 1loop} \simeq \frac{1}{\pi \rho} \, \frac{\Lambda}{1 - \frac{k}{\lambda}} \, ,
\end{equation}
which has the same asymptotic behavior (up to a multiplicative constant) of the exact expression of Ref.~\cite{mirlintikhonov}, $I_2 = (3/N) \avg{\rho^2} / \avg{\rho}^2$. In fact these two results are connected and if one was able to compute analytically the multiplicative constant in Eq.~\eqref{eq:IPR} above, one should be able to recover the exact expression of Ref.~\cite{mirlintikhonov} from the $1$-loop corrections. In our computation the $1$-loop corrections to $\avg{|G_\eta|^2}$ are divergent at $\eta=0$, since the loop contribution decays as $k^{-L}$, that multiplied by the number of loop, which is $O(k^L)$, gives a non decreasing contribution which makes  the sum over $L$ diverge. For this reason we need to add an imaginary regulator to make the sum converge and estimate the $1/\eta$ divergence that one gets in the $\eta \to 0$ limit. The exact $k^{-L}$ decay at $\eta=0$ is {\it likely} associated with the existence of the Goldstone mode related to spontaneous symmetry breaking in the delocalized phase, see Eq.~(48) or Ref.~\cite{mirlin,mirlin1}. On the other hand in Ref.~\cite{mirlintikhonov} the  leading $1/N$ term of the IPR is computed directly integrating explicitly on the manifold of symmetry breaking saddle points.

The $1$-loop result can be compared with the numerical estimation of the average IPR of the eigenvectors of the Anderson model on RRGs of large but finite sizes obtained from EDs (see App.~\ref{app:deltaG2}). For $N$ large enough (i.e. $N \gg \Lambda$) the IPR behaves as $I_2 \approx Y_{\rm ED}/N$ (see Fig.~\ref{fig:ipr}). The values of $Y_{\rm ED}$ are plotted in the bottom pannel of Fig.~\ref{fig:BLcorr} showing an excellent quantitative agreement with the predictions of Eq.~\eqref{eq:IPR}.

\section{Conclusions} \label{sec:conclusions}

In this paper we have studied the corrections to the BL solution of  AL in the metallic regime due to the presence of one loop of varying length $L$ using a combination of analytical and numerical techniques. The outcomes of this computation have been applied to obtain for the first time corrections to the BL solution on i) RRG of finite size $N$ and ii) euclidean lattices in finite dimension. In both cases we found that corrections are huge and this has deep consequences.

In the first case we show that the $1/N$ corrections to the average values of observables such as the typical DoS and the IPR have prefactors that diverge exponentially approaching the critical point. In fact, as discussed in the introduction, some recent works have claimed that the extrapolations of numerical results obtained from EDs of large but finite RRGs seem to be in contrast with the expected asymptotic behavior on the BL, generating a strong controversy about the existence of a putative extended but non-ergodic phase in a broad disorder range for $W<W_c$~\cite{noi_nonergo,scardicchio1,ioffe1,ioffe3,refael2019,pinorrg}. The analysis of the $1/N$ corrections presented above  provides a clear and transparent explanation for the origin of such controversies, as it indicates that the extrapolations of results obtained from EDs of large but finite samples which are affected by exponentially strong corrections on the relevant observables compared to their infinite BL counterpart and likely fail to capture the correct asymptotic behavior~\cite{mirlinRRG,Bethe,levy,lemarie}. Our results also fully support the notion of volumic scaling discussed in Ref.~\cite{lemarie}. This behavior is in striking contrast with conventional phase transitions, which are instead characterized by a less dramatic algebraic divergence of corrections (see App.~\ref{app:perco} for the case of the percolation transition). The exponential divergence of the $1$-loop corrections also explains the strong differences of the physical properties of the Anderson model on the RRG and on loop-less Cayley trees~\cite{mirlinCT,garel,noiCT}, and  provides an interpretation of the anomalous subdiffusive behavior observed numerically in finite-size samples at finite times~\cite{bera2018,detomasi2020}, since the standard diffusive behavior is expected to be recovered only for $N \gg \Lambda$ and for very large times. 

We have also combined the computation of the 1-loop corrections with the $M$-layer expansion, a novel technique  which has been recently developed to treat problems in which mean-field theory is only available on the BL~\cite{mlayer,mrfim,msg1,msg2,mboot}. This approach allows to study problems in finite dimension using a tunable parameter $M$: when $M$ goes to infinity the BL solution is recovered while the problem in finite dimension, say three, can be studied as an expansion in powers of $1/M$. The corresponding expansion can be written as a sum of topological diagrams whose evaluation requires the analysis of lattices with loops. We have  then been able to show that $1/M$ corrections in finite dimensional lattices also diverge exponentially at the critical point in any finite dimension. This implies that the exotic critical behavior of the BL solution is destroyed by corrections in any finite dimension~\cite{noilarged,garcia,mirlin94,castellani,dobro}, at variance with ordinary second-order phase transitions for which mean-field theory is correct above the upper critical dimension. Remarkably, the hypothesis that the upper critical dimension of AL is infinity is decades-old~\cite{mirlin94,castellani}, and our computation provides the first concrete and quantitative evidence for its validity, providing a rigorous framework to reconcile the scaling hypothesis with the exotic critical behavior observed on the BL~\cite{mirlin94}.

As explained above, from the RG perspective the topological diagrams appearing in the $M$-layer construction play exactly the same role as  Feynman diagrams in the perturbative field-theoretical expansion. Naturally, the ultimate goal would be to calculate the non-mean-field critical exponents by employing techniques similar to field-theoretical perturbative expansion, which involves re-summing the series of diverging diagrams~\cite{SFT,lebellac,ZJ,cardy}. In this endeavor, we encounter essential singularities at the critical point, prompting the pursuit of a possible critical series expressed in terms of powers of the correlation volume $\Lambda$. This is certainly a very promising direction for future investigations. 

Finally, the loop expansion discussed in this paper could be also applied in the context of the MBL transition to study the spectral properties of the problem when the many-body quantum dynamics is recasted as a single-particle diffusion in the Hilbert space, which, for a system of $n$ interacting degrees of freedom, is typically (non-random) a $n$-dimensional graph with many loops of all sizes. 

\appendix

\section{M-layer expansion for random percolation} \label{app:perco}

This appendix is devoted to the reader which is not familiar with the $M$-layer construction. In order to illustrate how the method works, we apply it to the simplest statistical mechanics model with a second order phase transition, i.e. random percolation. In particular we will show that the $M$-layer expansion allows one to recover the critical series of the diagrammatic  loop expansion of the corresponding cubic field theory. A very detailed and general explanation of the $M$-layer approach can be found in Ref.~\cite{mlayer}. The reader who is already familiar with the $M$-layer approach can skip this whole section and jump directly to Sec.~\ref{app:Anderson}.

\subsection{Exact solution on the Bethe lattice}

Random percolation is defined as follows: a given node of the lattice is occupied with probability $p$ and empty with probability $1-p$, independently of the occupation of all other nodes. The order parameter of the transition is $P$, defined as the probability that a given node belongs to the percolating cluster which spans the whole lattice.

The problem can be solved exactly in the  $M \to \infty$ limit, i.e. on the infinite BL. In order to do so, one first writes a self-consistent equation for the probability $P_{\rm cav}$ that a cavity site, in absence of one of its neighbors, belongs to a semi-infinite percolating branch, as a function of the same cavity probability on the neighboring nodes:
\begin{equation} \label{eq:Pcav}
P_{\rm cav} = p - p(1-P_{\rm cav})^k \, ,
\end{equation}
$k+1$ being the total local connectivity of the lattice. Once the fixed point of this equation is found, one obtains an equation for the probability that the node belongs to the percolating cluster in presence of all its neighbors:
\begin{equation}
P = p - p(1-P_{\rm cav})^{k+1} \, .
\end{equation}
From the solution of these equations one finds a critical value of the occupation probability, $p_c = 1/k$, below which $P$ is identically equal to $0$ and above which $P$ is strictly positive. $P$ vanishes linearly  approaching the critical point from above as
\begin{equation} \label{eq:P}
P \simeq \frac{2(k+1)}{k-1} \, (p-p_c) \, .
\end{equation}

\subsection{Leading order behavior of the $2$-points correlation function and the Ginzburg criterion} \label{app:GP}

We now study the behavior of the $2$-points correlation function, defined as the probability that two nodes at distance $L$ on the infinite BL belong to the same cluster. In the $M \to \infty$ limit this is simply given by:
\begin{equation}
C_{\rm BL}(L) = p^L = k^{-L} e^{-L/\xi_{\rm BL}} \, ,
\end{equation} 
where the correlation length $\xi_{\rm BL}$ can be formally defined as:
\begin{equation}
    \xi_{\rm BL} =- \frac{1}{\ln(pk)} \approx \frac{p_c}{|p_c - p|} \, .
\end{equation}
Although random percolation is a simple problem, the behavior of the $2$-points correlation function found on the BL is quite general to many second order phase transitions, and reminds for instance the behavior of the density-density correlator of the Anderson model on the BL in the localized phase~\cite{verba,zirn,mirlin,mirlin1,mirlintikhonov} (see Eq.~\eqref{eq:corr}).

As explained in the main text, at leading order of the $M$-layer expansion, where no loops are present, any correlation function on the $d$-dimensional $M$-layered lattice is strictly related to the one on the BL~\cite{mlayer}: In particular one can formally show that a generic correlation between two lattice sites at distance $r$ on the $d$-dimensional euclidean $M$-layered lattice is given by Eq.~\eqref{eq:Cd} where ${\cal N}(r,L)$ is given by the number of non-backtracking paths of length $L$ on the $M$-layered lattice between the two sites at distance $r$ in the euclidean space, Eq.~\eqref{eq:NrL}. 

Going to the momentum space and performing the sum over $L$ one finds the standard Gaussian propagator:
\begin{equation} \label{eq:Cdq}
\begin{aligned}
\hat{C}_d (q) & = \frac{1}{M} \sum_{L=1}^{\infty} e^{-L (q^2 + \xi_{\rm BL}^{-1})}\simeq \frac{1}{M} \, \frac{1}{q^2 + \xi_{\rm BL}^{-1}} \, , \\ 
C_d (r) & = \frac{1}{M} \, \frac{e^{-r/\xi_d}}{r^{d-2}} \, .
\end{aligned}
\end{equation}
We immediately notice that the correlation length on the euclidean lattice is given by the square root of the correlation length on the BL, $\xi_d = \sqrt{\xi_{\rm BL}} \propto |p-p_c|^{-1/2}$. This is due to the fact that a particle that diffuses freely on the BL is found at distance $L$  from the origin after $L$ steps, while it is found at distance $\sqrt{L}$ from the origin on the $d$-dimensional lattice (see also Sec.~\ref{app:ema}).

From the knowledge of the 2-point correlation function at the leading order one can apply the Ginzburg criterion which consists in requiring that finite-dimensional fluctuations do not modify the mean-field critical behavior. To this aim, as in Sec.~\ref{sec:ginzburg}, we define a local order parameter $P_i = (1/M) \sum_{\alpha=1}^M P_{i,\alpha}$, defined as the fraction of layers of the $M$-layer replicated lattice such that the site $i$ belongs to the percolating cluster. In the large-$M$ limit $P_i$ is essentially equal to $P$ on the BL, Eq.~\eqref{eq:P}, on all the sites of the original lattice with small $O(1/M)$ fluctuations around the mean, and the mean-field theory becomes exact. Nonetheless according to the Ginzburg criterion we have to check that the prefactor of $1/M$ of these fluctuations does not diverge at the transition in order for the mean-field predictions to be valid. Setting $r=0$ in Eq.~\eqref{eq:Cd} we obtain the fluctuations of $P_i$ at the order $1/M$ at a given position $i$ of the $M$-layered euclidean lattice among different layers:
\[
\langle P_{i,\alpha} P_{i,\beta} \rangle - P^2 = C_d(r=0) \simeq \frac{1}{M} \, \xi_{\rm BL}^{1 - d/2} \int_1^\infty {\rm d} x \, \frac{e^{-x}}{x^{3/2}} \, .
\]
Applying the requirement that these fluctuations are smaller than the square of the local order parameter, one finds an upper critical dimension equal to $d_U = 6$:
\begin{equation}
\frac{1}{M} \, \xi_{BL}^{1-d/2} \propto \frac{1}{M} \, (p-p_c)^{d/2-1} \ll (p-p_c)^{2} \, .
\end{equation}
Therefore for $d<d_U=6$, no matter how large $M$ is, there will always be a region close enough to $p_c$ where the fluctuations of the local order parameter are much larger than its mean. Therefore the BL solution should not be trusted in dimension smaller than $6$.

\subsection{$1$-loop corrections to the $2$-points correlation function}

\begin{figure}
\vspace{0.3cm} 
\includegraphics[width=0.482\textwidth]{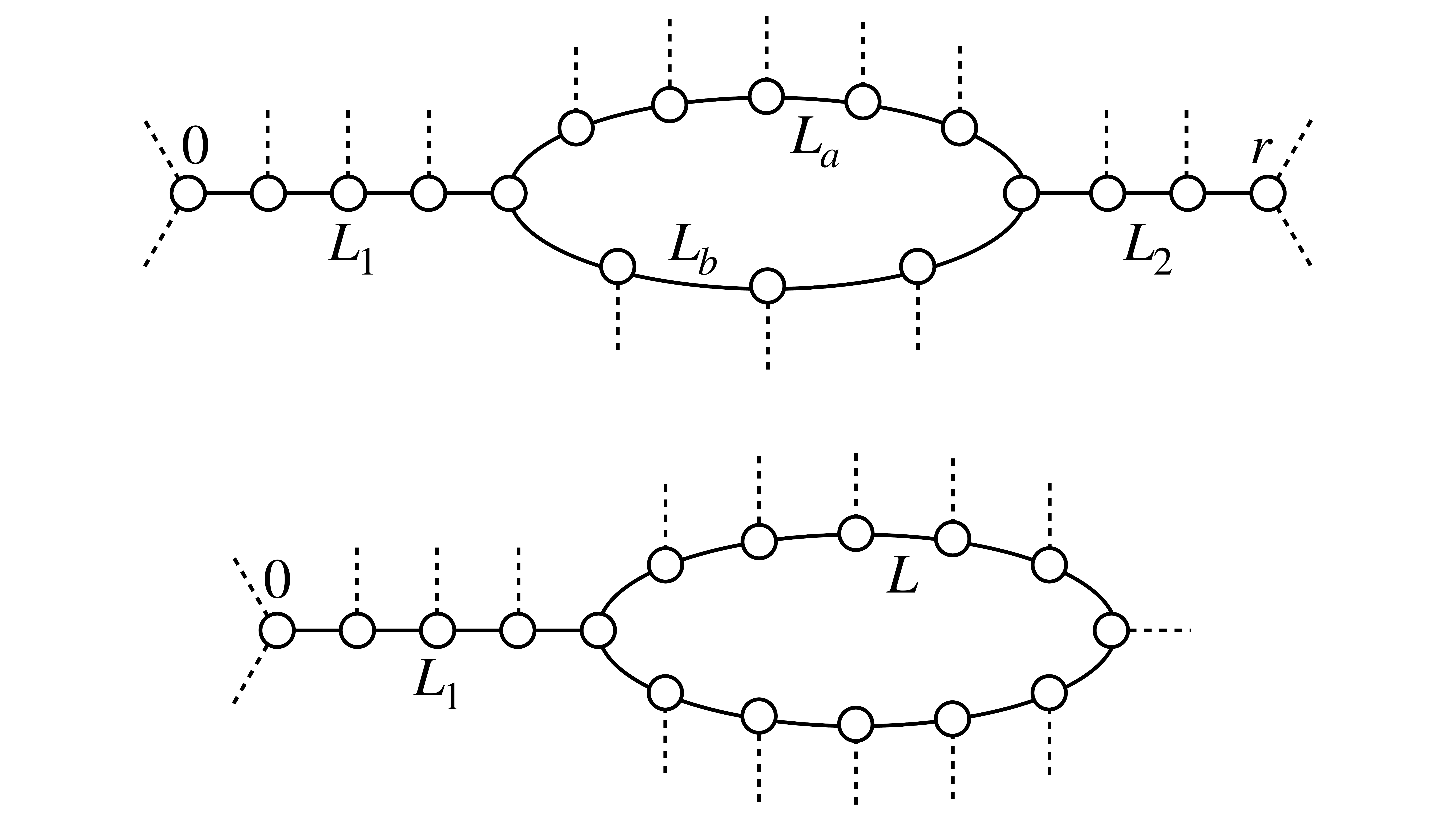}  
 \caption{\label{fig:diagrams} Topological diagrams relevant for the $1$-loop  corrections of two-points (top) and one-point (bottom) observables.}
\end{figure}

We can now go a step forward and compute the 1-loop corrections to the correlation function. The 1-loop contributions come from the top diagram given in Fig.~\ref{fig:diagrams}, constructed considering a portion of the original lattice, replacing the internal lines with appropriate one-dimensional chains, and attaching to the internal points the appropriate number of infinite Bethe trees to restore the correct local connectivity of the original model, which is $k+1=3$ in the example of Fig.~\ref{fig:diagrams}. These diagrams are called ``fat-diagrams'': They  are  topological structures analogous to Feynman diagrams, but preserving the local finite-connectivity nature of the original lattice.
Subtracting the contribution of disconnected diagrams~\cite{mlayer}, one finds:
\begin{widetext}
\begin{equation}
\begin{aligned}
C_{\rm BL}^{1{\rm loop}} (L_1,L_2,L_a,L_b) & = p^{L_1+1} p^{L_2+1} \left( p^{L_a-1} +  p^{L_b-1} - p^{L_a + L_b - 2} \right) -  p^{L_1 + L_2 + L_a + 1} - p^{L_1 + L_2 + L_b + 1} \\
& =  - p^{L_1 + L_2 + L_a + L_b} \, .
\end{aligned}
\end{equation}
The corresponding contribution in $d$ dimensions is obtained by summing over all possible values of the legs of the internal lines, $L_1$, $L_2$, $L_a$, $L_b$, with the corresponding geometric factor that counts the number of such topological diagrams on the $M$-layer lattice between two points at distance $r$ in the euclidean space. Going again to the Fourier space and performing the sum over the lengths of the legs one finds:
\begin{equation}
\begin{aligned}
C_d^{1{\rm loop}} (q) & = \frac{1}{M^2} \! \sum_{L_1,L_2,L_a,L_b} \!\!\!  \frac{k^{L_1+L_2+L_a+L_b} e^{-q^2[L_1 + L_2 + (L_a^{-1} + L_b^{-1})^{-1}]}}{(L_a + L_b)^{d/2}} \, C_{\rm BL}^{\rm 1loop} (\vec{L}) \\
& = - \frac{1}{M^2} \left( \frac{1}{q^2 + \xi_{\rm BL}^{-1}} \right)^2 \int {\rm d}^d \vec{q}_1 \, \frac{1}{q_1^2 + \xi_{\rm BL}^{-1}} \, \frac{1}{|\vec{q} - \vec{q}_1|^2 + \xi_{\rm BL}^{-1}}  \, ,
\end{aligned}
\end{equation}
where $(q^2 + \xi_{\rm BL}^{-1})^{-1}$ the Gaussian propagators found at the leading order, Eq.~\eqref{eq:Cdq}.  Note that this is exactly the same contribution, with exactly the same numerical prefactor, coming from the perturbative 1-loop Feynman diagrams of the corresponding cubic field-theory obtained from the Fortuin-Kasteleyn mapping to the  $q$-state Potts model in the limit $q \to 1$~\cite{phi3}. In particular, the integral on the right hand side of the equation above diverges in dimensions smaller than $6$. One can rigorously show that this is in fact a general result: For any generic observable, at any order, the $M$-layer expansions reproduces the critical series of the  diagrammatic loop expansion of the corresponding field-theory~\cite{mlayer}. For the specific case of percolation, such diagrammatic loop expansion can be obtained without resorting to the Fortuin-Kasteleyn mapping.
\end{widetext}

Hence, complementing this result with higher order terms and the 3-point function, one can compute the non mean-field critical exponents in $d$ dimensions resumming the critical series of the diagrammatic loop expansion through the same recipes of the field theoretical perturbative expansion inspired by RG arguments, as done in Ref.~\cite{phi3}, and as explained in standard textbooks~\cite{lebellac,cardy,ZJ}.

\subsection{$1$-loop correction to the order parameter in the percolating phase} \label{app:DeltaP}

One can also apply a similar approach to compute the finite-dimensional corrections to the order parameter $P$. At 1-loop level these corrections are given by the contribution of the bottom diagrams of Fig.~\ref{fig:diagrams}:
\begin{equation}
\Delta \! P_{\rm 1loop} = {1 \over M} \,
\sum_{L_1=0}^\infty \sum_{L=3}^\infty {\cal N} (L,L_1) \, \delta P(L,L_1) \, ,
\end{equation}
where $\delta P(L,L_1) = P_{\rm 1loop}(L,L_1) - P_{\rm BL} $ is the difference between the probability that the node $0$ belongs to the percolating cluster in presence of the loop and in absence of it. The geometric factor ${\cal N} (L,L_1)/M$ corresponds once again to the number of these diagrams that one finds embedded in the $M$-layer lattice, which at asymptotically large $L$ and $L_1$ is given by the number of diagrams in the original $d$-dimensional euclidean lattice, ${\cal N} (L,L_1) \simeq k^{L + L_1}/L^{d/2}$, Eq.~\eqref{eq:loops}, divided by the number of layers.

After a tedious but easy computation one finds that $\delta P$ for fixed $L$ and $L_1$ is given by the following expression:
\begin{equation}
\begin{aligned}
\delta P(L,L_1)  = & - P_{\rm cav} \left(p \left (1-P_{\rm cav}\right)^{k-1} \right)^{L_1 + 1} \\
& \,\,\,\, \times \bigg[ (1 - p) L \left (p \left(1-P_{\rm cav}\right)^{k-1}\right)^L \\
& \qquad \,\, + (2 - P_{\rm cav}) \left (p \left (1-P_{\rm cav}\right)^{k-1}\right)^{L+1} \bigg] \, .
\end{aligned}
\end{equation}
It is important to notice that the dependence on the length of the loop $L$ and the length of the external leg $L_1$ completely factorizes. The same kind of factorization will also be found for the $1$-loop corrections to $1$-point observables in the Anderson model (see Sec.~\ref{app:deltalnrho} and \ref{app:deltaG2}). The physical origin of this factorization can be understood in the following way: the presence of the loop induces a modification of the value of the probability of belonging to the percolating cluster on the node at the base of the loop. Using the cavity recursion relation~\eqref{eq:Pcav}, a small modification of this probability propagates linearly along the chain of length $L_1$. As a consequence, $\delta P$ is the product of a response function, which depends only on $L_1$, times the amplitude of the perturbation induced by the loop, which depends only on $L$.

Performing the sum over $L$ and $L_1$ with the appropriate geometric factor, one finally finds the 1-loop corrections to $P$. The asymptotic behavior of the corrections close to $p_c$ behaves as:
\begin{equation}
\Delta P_{\rm 1 loop} \simeq \frac{1}{M} \, \left \{
\begin{array}{ll}
a_d + b_d (p - p_c) & \textrm{for~} d>6 \, ,\\
a_d + b_d^\prime (p - p_c)^{d/2-2} & \textrm{for~} d<6 \, , 
\end{array}
\right .
\end{equation}
which turns out to be different below and above the upper critical dimension. Note again that the correction is small at large values of $M$ but below $d_U=6$ there is always a region of values of $p$ close to $p_c$ where the prefactor of the $1/M$ correction to the mean-field result for the order parameter is much larger than its value and therefore the mean-field critical exponents should not be trusted.

\section{1-loop corrections to the Bethe Lattice solution of Anderson localization} \label{app:Anderson}

In this appendix we provide more details and numerical results related to several points discussed in the main text concerning 1-loop corrections to the BL solution of AL.

\subsection{Exact self-consistent equations on the infinite Bethe lattice} \label{app:cavity}

As explained in the main text, the central object of our analysis is the probability distribution of the Local Density of States (LDoS), $\rho_i (E) = \sum_\alpha |\psi_\alpha (i)|^2 \delta(E - E_\alpha )$, which plays the role of the order parameter function for AL~\cite{mirlin94}. The statistics of the LDoS is encoded in the statistics of the elements of the resolvent, $G_{ij} = ({\rm i} \eta {\cal I} - {\cal H})^{-1}_{ij}$, where ${\cal I}$ is the identity matrix, ${\cal H}$ is the Anderson Hamiltonian, (Eq.~(1) of the main text), and $\eta$ is an infinitesimal imaginary regulator that softens the pole singularities in the denominator.
On the infinite BL the diagonal elements of $G$ verify an exact self-consistent recursion relation~\cite{Abou-Chacra}: Taking a site $i$ of BL and removing one of its neighbors, say site $j$,  a recurrence equation (which becomes asymptotically exact in the thermodynamic limit) can be obtained for  the diagonal elements on site $i$  of  the (cavity) Green's function of the modified model which describes the system where $j$ has been removed, $G_{i \to j}$, in terms of the diagonal elements of the cavity Green's functions on the neighbors of $i$ in absence of node $i$ itself, $G_{m \to i}$:
\begin{equation} \label{eq:Gcav}
G_{i \to j}^{-1} = -\epsilon_i - {\rm i} \eta - t^2 \! \! \! \sum_{m \in \partial i / j} G_{m \to i} \, ,
\end{equation}
where, without loss of generality, we have set $E=0$. The diagonal elements of the resolvent of the original Anderson tight-binding model can then be expressed in terms of these cavity Green's function as:
\begin{equation} \label{eq:G}
G_{ii}^{-1} = -\epsilon_i - {\rm i} \eta - t^2 \! \sum_{m \in \partial i} G_{m \to i} \, .
\end{equation}
We set $t=1$ throughout.

The statistics of the LDoS, as well as the IPR are encoded in the statistics of the diagonal elements of the resolvent, according to Eqs.~\eqref{eq:ldos} and~\eqref{eq:I2}.

\begin{figure*}
\includegraphics[width=0.482\textwidth]{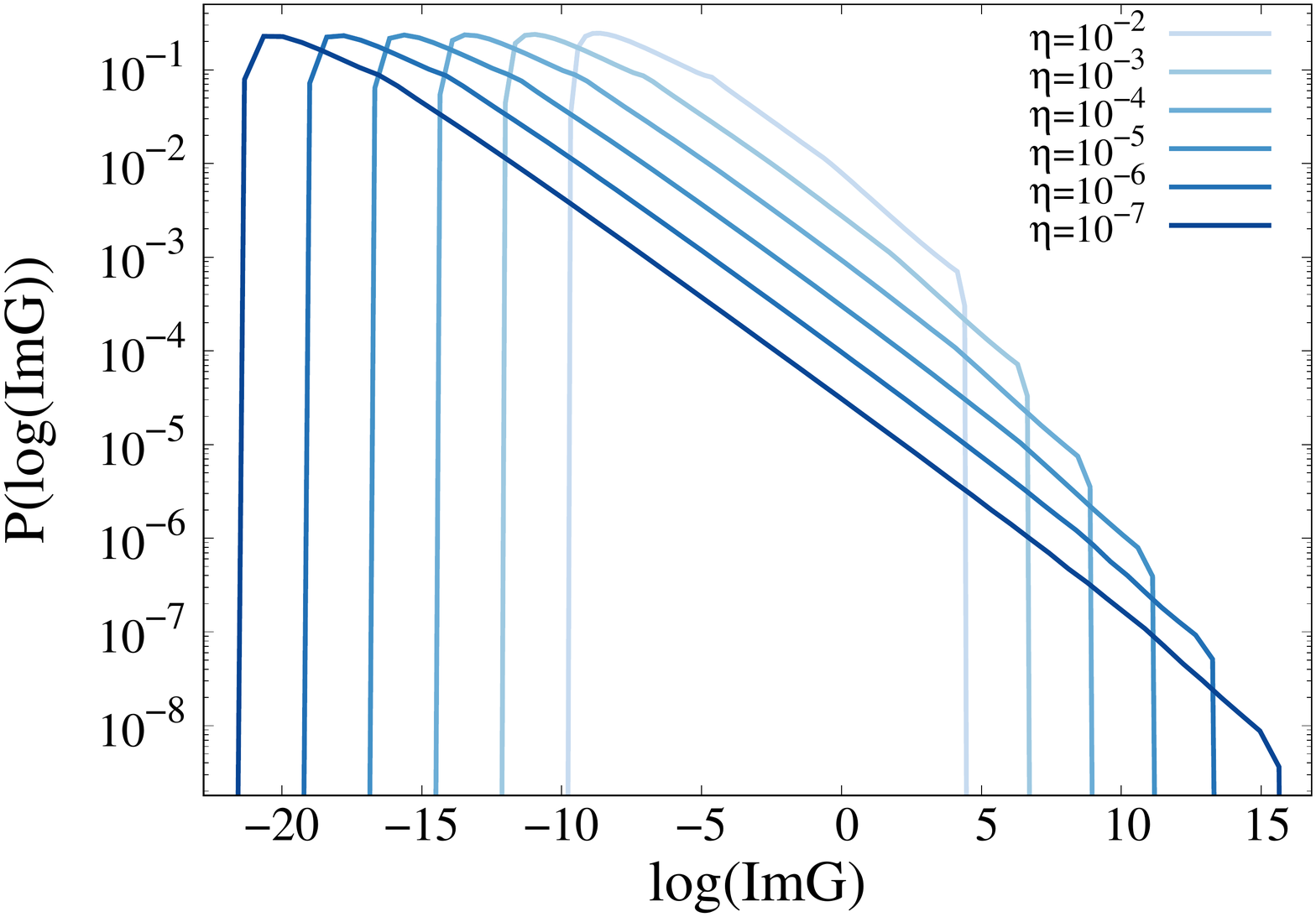}  \hspace{-0.1cm} 
\includegraphics[width=0.482\textwidth]{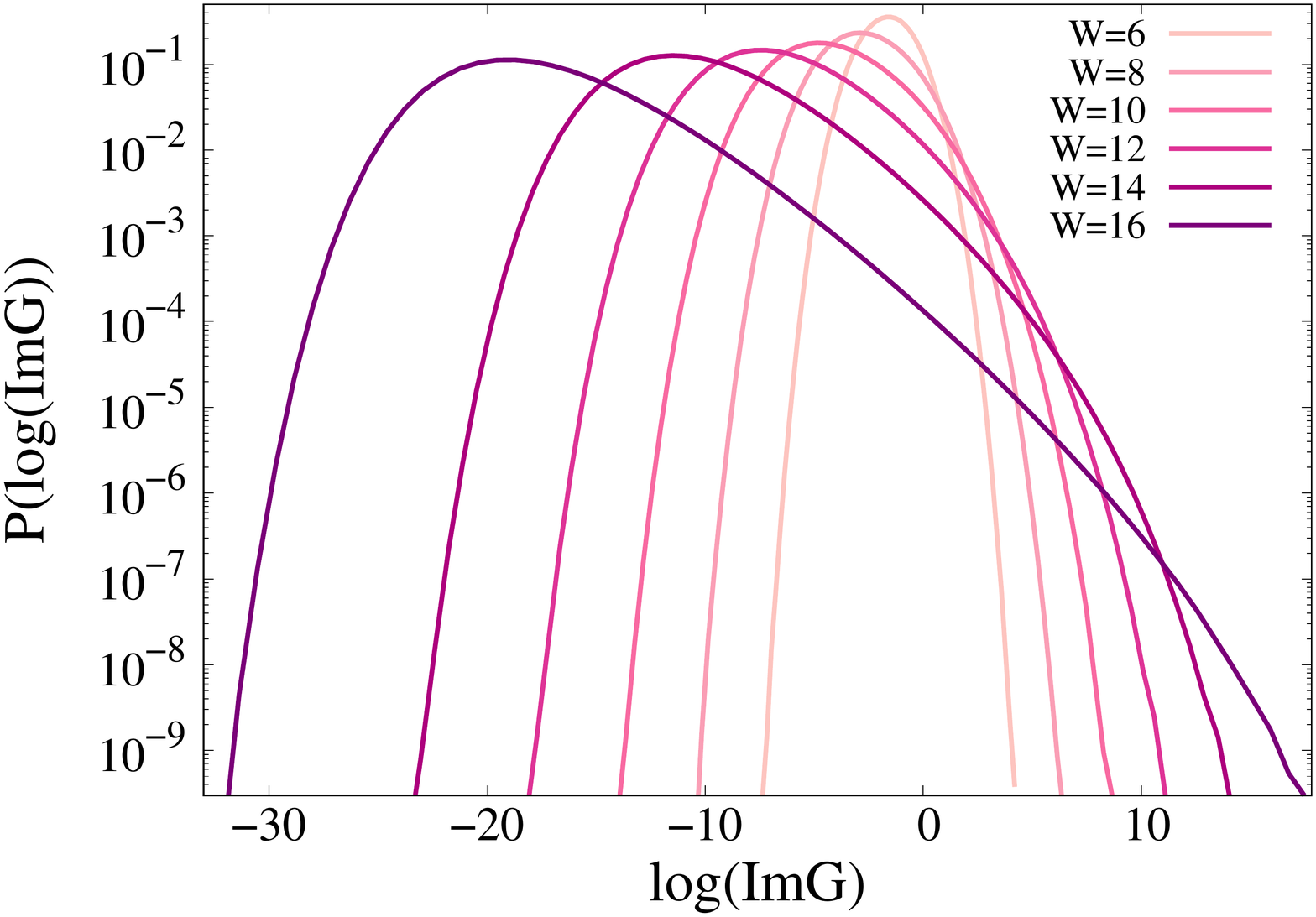}  
 \caption{\label{fig:Prho} Left: Probability distribution of the logarithm of the imaginary part of the Green's function for $k=2$ and $W=24$, deep into the localized phase (the transition point is at $W_c \approx 18.17$~\cite{tikhonov_critical}), and for several values of $\eta$ from $10^{-2}$ to $10^{-7}$. Right: Probability distribution of the logarithm of the imaginary part of the Green's function for several values of the disorder across the delocalized phase in the $\eta \to 0^+$ limit. The plots are obtained using the population dynamics algorithm with a pool of $\Omega=2^{26}$  elements.}
\end{figure*}

 Eq.~(\ref{eq:Gcav}) should be in fact interpreted as a self-consistent integral equation for the probability distribution of $P(G_{i \to j})$:
\[
P (G) = \int {\rm d} p(\epsilon) \prod_{m=1}^k P (G_m) \, \delta \left ( G^{-1}  + \epsilon + {\rm i} \eta + \! \sum_{m =1}^k G_m \right) \, .
\] 
This equation can be solved numerically using population dynamics algorithms with arbitrary numerical precision~\cite{mezard,Bethe,large_deviations,tikhonov_critical}: The probability distribution is approximated by the empirical distribution of a large pool of $\Omega$ elements $G_\alpha$, $P(G) \simeq \sum_{\alpha=1}^\Omega \delta (G - G_\alpha)$; At each iteration step $k$ instances ($k$ being the branching ratio) of $G$ are extracted from the sample and a value of $\epsilon$  is taken from the uniform distribution; A new instance of $G$ is generated using Eq.~(\ref{eq:Gcav}) and inserted in a random position of the pool until the process converges to a stationary distribution. Once the distribution of the cavity Green's function is obtained, one can compute the probability distribution of the Green's function of the original problem from Eq.~(\ref{eq:G}):
\[
\tilde{P} (G) = \int {\rm d} p(\epsilon) \prod_{m=1}^{k+1} P (G_m) \, \delta \left ( G^{-1}  + \epsilon + {\rm i} \eta + \! \sum_{m =1}^{k+1} G_m \right) \, .
\] 
The numerical data shown in this paper are obtained with pools of size ranging from $\Omega=2^{26}$ to $\Omega=2^{28}$, and for $k=2$ and $k=5$. The effect of the finiteness of the size pool has been discussed in detail in Ref.~\cite{tikhonov_critical}, where it has been shown that finite pool size effects become strong close to the transition point. However in the numerical analysis described below we will consider values of the disorder far enough from $W_c$, such that our pool sizes are sufficiently large to avoid any significant finite-$\Omega$ corrections.

As shown in the left panel of Fig.~\ref{fig:Prho}, in the insulating phase $P({\rm Im} G)$ is singular in the $\eta \to 0$ limit: It has a maximum in the region ${\rm Im}G \sim \eta$ and power-law tails $P({\rm Im} G) \sim \sqrt{\eta}/({\rm Im} G)^{3/2}$ with a cutoff at $\eta^{-1}$. Hence the main contribution to the moments $\langle ({\rm Im} G)^q \rangle \propto \eta^{1-q}$ comes from the cutoff at ${\rm Im} G \sim \eta^{-1}$ for $q \ge 1/2$ (and hence the IPR is of order $1$), while the normalization integral is dominated by the region ${\rm Im} G \sim \eta$, and the typical value of ${\rm Im} G$ (i.e. $I_{q \to 0}$) is of order $\eta$. This behavior reflects the fact that in the localized phase wave-functions are exponentially localized on few $O(1)$ sites where $\rho_i$ takes very large values, while the typical value of the LDoS is exponentially small and vanishes in the thermodynamic limit for $\eta \to 0^+$. In the metallic phase, instead, $P({\rm Im} G)$ is unstable to introduction of an arbitrary small but finite imaginary part, i.e., $P({\rm Im} G)$ converges to a non-singular $\eta$-independent distribution for $\eta \to 0^+$ (right panel of Fig.~\ref{fig:Prho}). However, upon approaching the critical disorder from below $P({\rm Im} G)$ becomes very broad and asymmetric, and a (large) characteristic scale $\Lambda$, playing a role analogous to that of $\eta^{-1}$ in the localized phase, spontaneously emerges: The probability distribution has a sharp maximum for ${\rm Im} G \sim \Lambda^{-1}$ followed by a power law decay $P({\rm Im} G) \sim ({\rm Im} G)^{-3/2}$ with a cutoff at ${\rm Im} G$ of order $\Lambda$~\cite{mirlin94,large_deviations}. Such $\Lambda$ is found to diverge exponentially at the critical point according to Eq.~\eqref{eq:rhotypBL}~\cite{efetov_bethe,zirn,mirlin,mirlin1,verba,tikhonov_critical,mirlintikhonov,large_deviations} with $\nu_{\rm del} = 1/2$, and can be interpreted as the {\it correlation volume} of typical eigenstates: On finite RRGs of $N$ nodes and for $W \lesssim W_c$ the wave-functions have $O(N/\Lambda)$ bumps localized in a small region of the graph where the amplitude is of order $\Lambda/N$ (to ensure normalization), separated by regions of size $\ln \Lambda$ where the amplitude is very small.

\begin{figure*}
\includegraphics[width=0.482\textwidth]{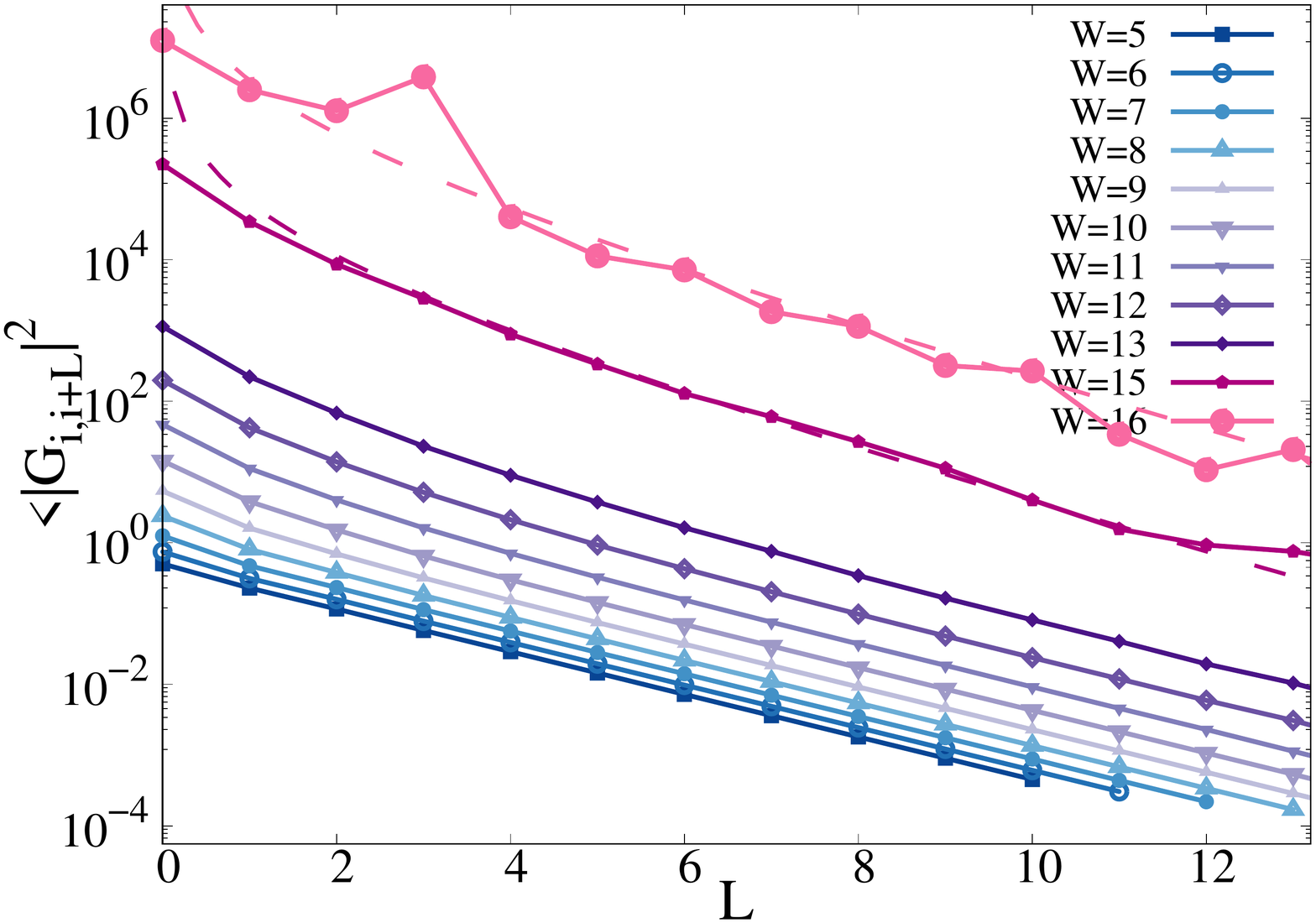} \hspace{-0.1cm} \includegraphics[width=0.482\textwidth]{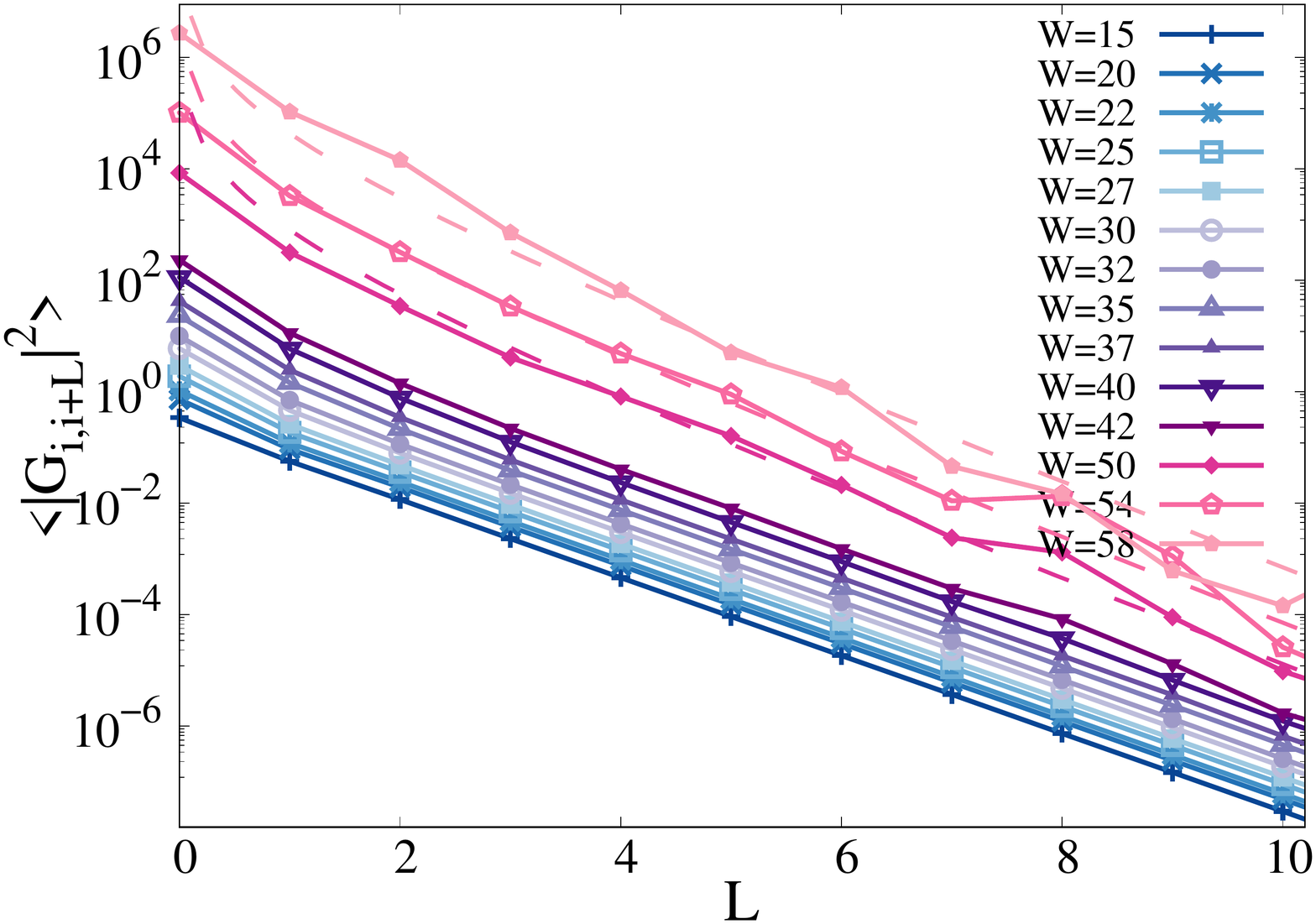}  

\includegraphics[width=0.482\textwidth]{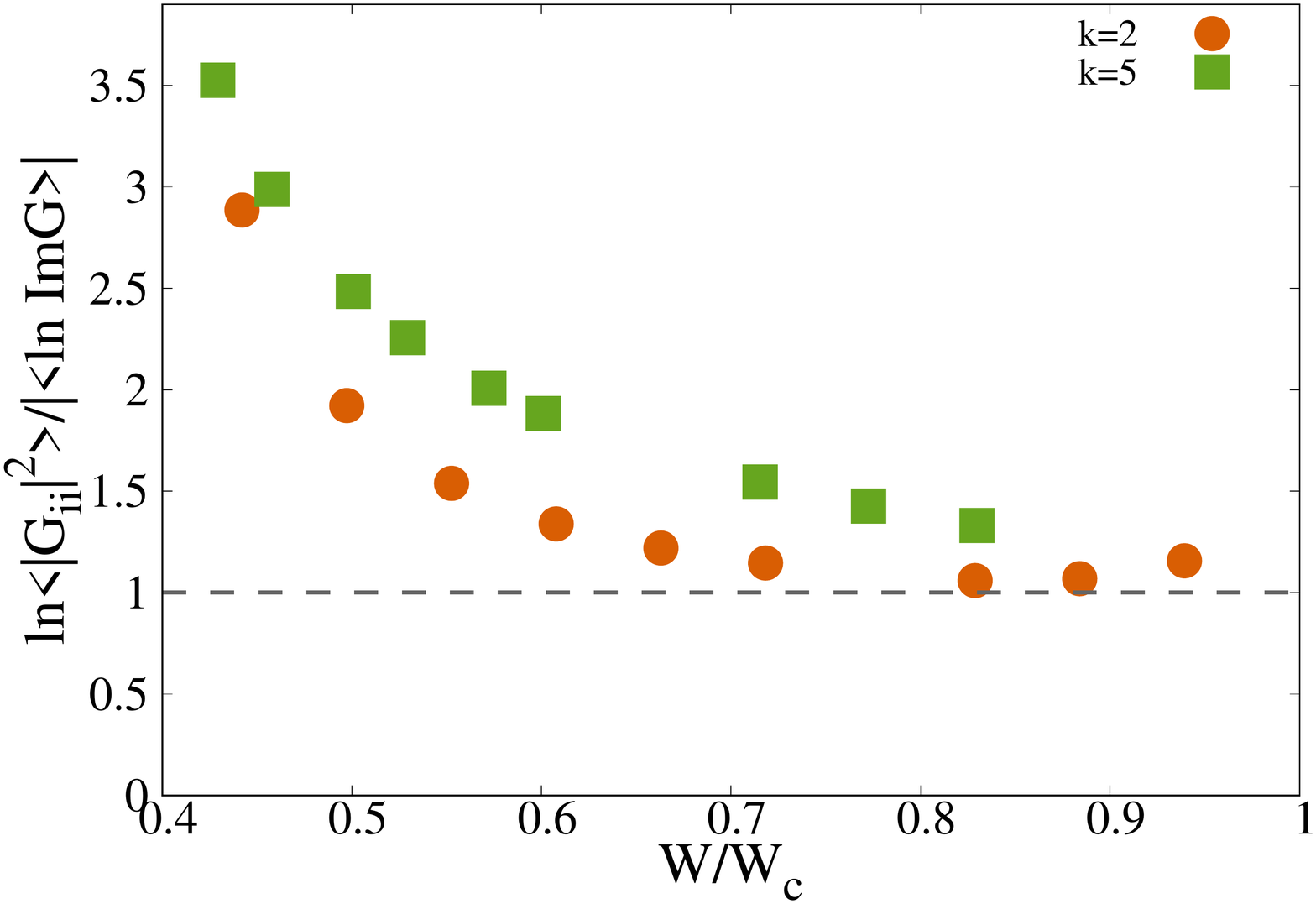} \hspace{0.1cm} \includegraphics[width=0.469\textwidth]{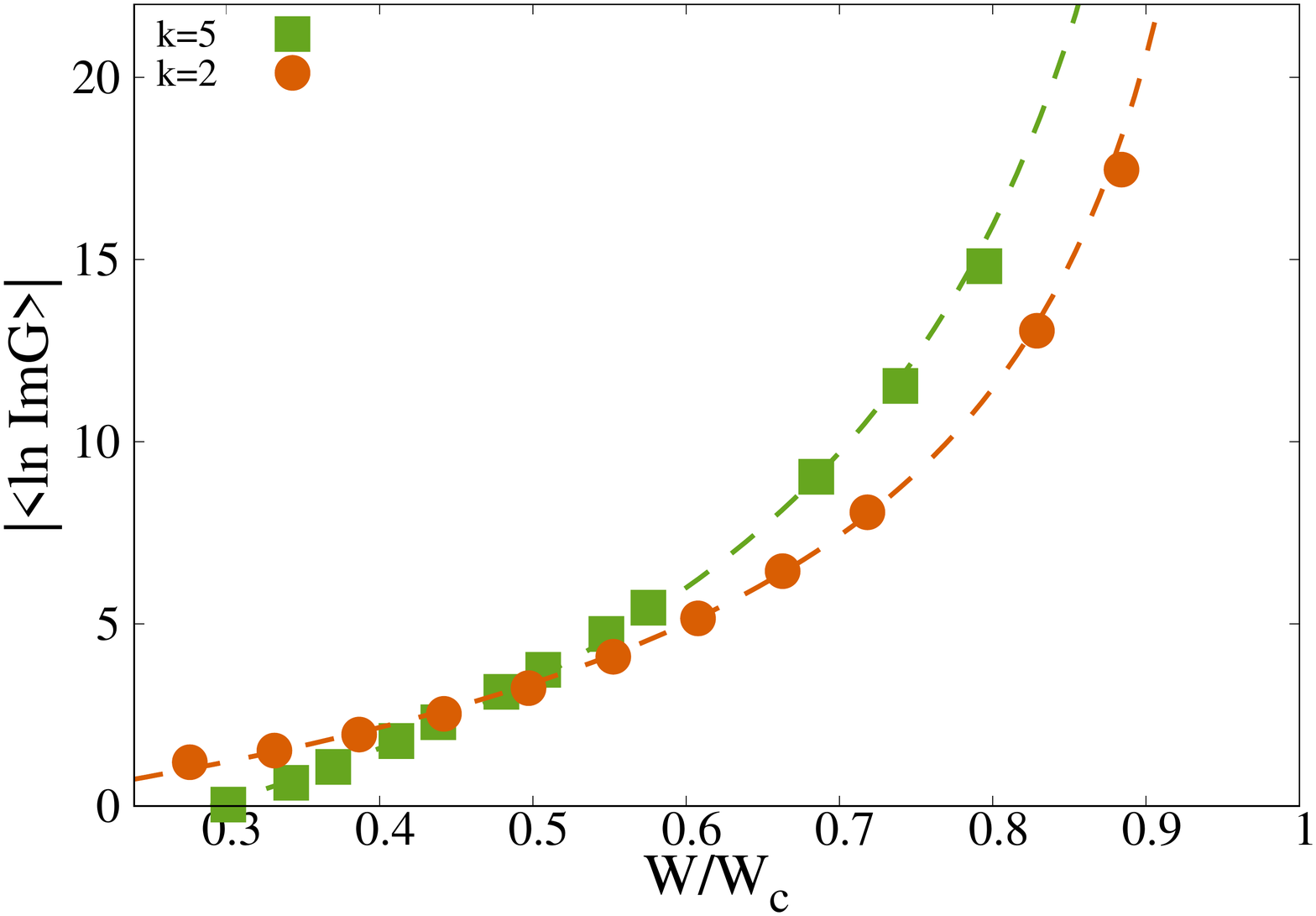}  

\caption{Top panels: $\avg{|G_{i,i+L}|^2}$ as a function of $L$ for several values of $W$ across the metallic phase for $k=2$ (left) and $k=5$ (right). Sufficiently close to the transition point the data are well fitted by Eq.~\eqref{eq:G0L}. The transition point are $W_c \approx 18.17$ for $k=2$~\cite{tikhonov_critical} and $W_c \approx 73$ for $k=5$. Bottom-left panel: $\ln \avg{|G|^2}/|\avg{\ln {\rm Im} G}|$ as a function of $W/W_c$ for $k=2$ and $k=5$, showing that the ratio approaches $1$ for $W$ close enough to $W_c$. Bottom-right panel: Plot of $|\avg{\ln {\rm Im} G}|$ as a function of $W/W_c$ for $k=2$ and $k=5$. The dashed lines are fits of the data according to Eq.~\eqref{eq:rhotypBL}: $\ln \Lambda \simeq A/\sqrt{W_c-W} + B$, with $A\approx 39$ 
and $B\approx -10.5$ for $k=2$ (to be compared to the best numerical estimation for $A$ for $k=2$ reported in Ref.~\cite{tikhonov_critical}, yielding $A \simeq 32$), and $A\approx 130$ and $B\approx-18$ for $k=5$.
\label{fig:G0L2}}
\end{figure*}

\subsection{Numerical results for the correlation function $\langle |G_{i,i+L}|^2 \rangle$}  \label{app:2points}

In this section we provide a few numerical data on the correlation function $\langle |G_{i,i+L}|^2 \rangle$ in the delocalized phase (and in the $\eta \to 0^+$ limit), which is related to the spectral representation of the probability that a particle starting in $i$ at time $0$ reaches the node $i+L$ (at distance $L$ from $i$) after infinite time. Its critical behavior close to $W_c$ when the transition is approached from the metallic side can be determined exactly from the supersymmetric treatment~\cite{efetov_bethe,zirn,mirlin,mirlin1,verba,mirlintikhonov}: 
\begin{equation} \label{eq:G0L}
\avg{|G_{i,i+L}|^2} \simeq \Lambda \frac{k^{-L}}{L^{3/2}} \, ,
\end{equation}
where $\Lambda$ is the correlation volume which diverges exponentially at $W_c$, $\ln \Lambda \propto (W_c-W)^{-1/2}$. Note that the value in $L=0$ of is related to the inverse participation ratio.

$\langle |G_{i,i+L}|^2 \rangle$ is plotted in the top panels of Fig.~\ref{fig:G0L2} for $k=2$ (left) and $k=5$ (right), showing that close enough to the critical point the numerical data are well fitted by Eq.~\eqref{eq:G0L}. (The density-density correlation $\avg{\rho_i \rho_{i+L}}$ exhibits the same decay as $\langle |G_{i,i+L}|^2 \rangle$ for not too large distances, $L < (\ln \Lambda)^3$, whereas it features an extra exponential decay for $L > (\ln \Lambda)^3$ and saturates at the value given by its disconnected part~\cite{mirlintikhonov,zirn,efetov_viehweger}.)

In the bottom-left panel of Fig.~\ref{fig:G0L2} we plot the ratio between $\ln \langle |G_{i,i}|^2 \rangle$ and (minus) the logarithm of the typical DoS, $|\langle \ln {\rm Im} G \rangle |$ as a function of the disorder (divided by $W_c$) for $k=2$ and $k=5$. This plot shows that sufficiently close to the critical point this ratio tends to one, in agreement with the intuitive argument given in the main text suggesting that the prefactor of the IPR and the inverse of the typical DoS are both proportional to the correlation volume $\Lambda$:
\begin{equation}
    \label{eq:order_parameter}
    \avg{|G_{i,i+L}|^2} \simeq \left( e^{\avg{\ln {\rm Im} G_{ii}}} \right)^{-1} \simeq \Lambda \, .
\end{equation}
Since $\langle \ln {\rm Im} G \rangle$ is much easier to compute numerically than $\langle |G_{i,i}|^2 \rangle$ (the latter is related to the second moment of the probability distribution of the LDoS, which becomes extremely broad as the localization transition is approached, while the former is proportional to the $q \to 0$ moment of the distribution and fluctuates much less than the latter), we will use $e^{-\langle \ln {\rm Im} G \rangle}$ as a proxy of the correlation volume $\Lambda$ throughout.

In the bottom-right panel of Fig.~\ref{fig:G0L2} we show the increase of $-\langle \ln {\rm Im} G \rangle$ as a function of the disorder for $k=2$ and $k=5$. The dashed lines are fits of the logarithm of the correlation volume as $\ln \Lambda \simeq A/\sqrt{W_c-W} + B$, with coefficients $A$ and $B$ given in the legend.

\subsection{Exact decimation equations} \label{app:decimation}

\begin{figure}
    \includegraphics[width=0.46\textwidth]{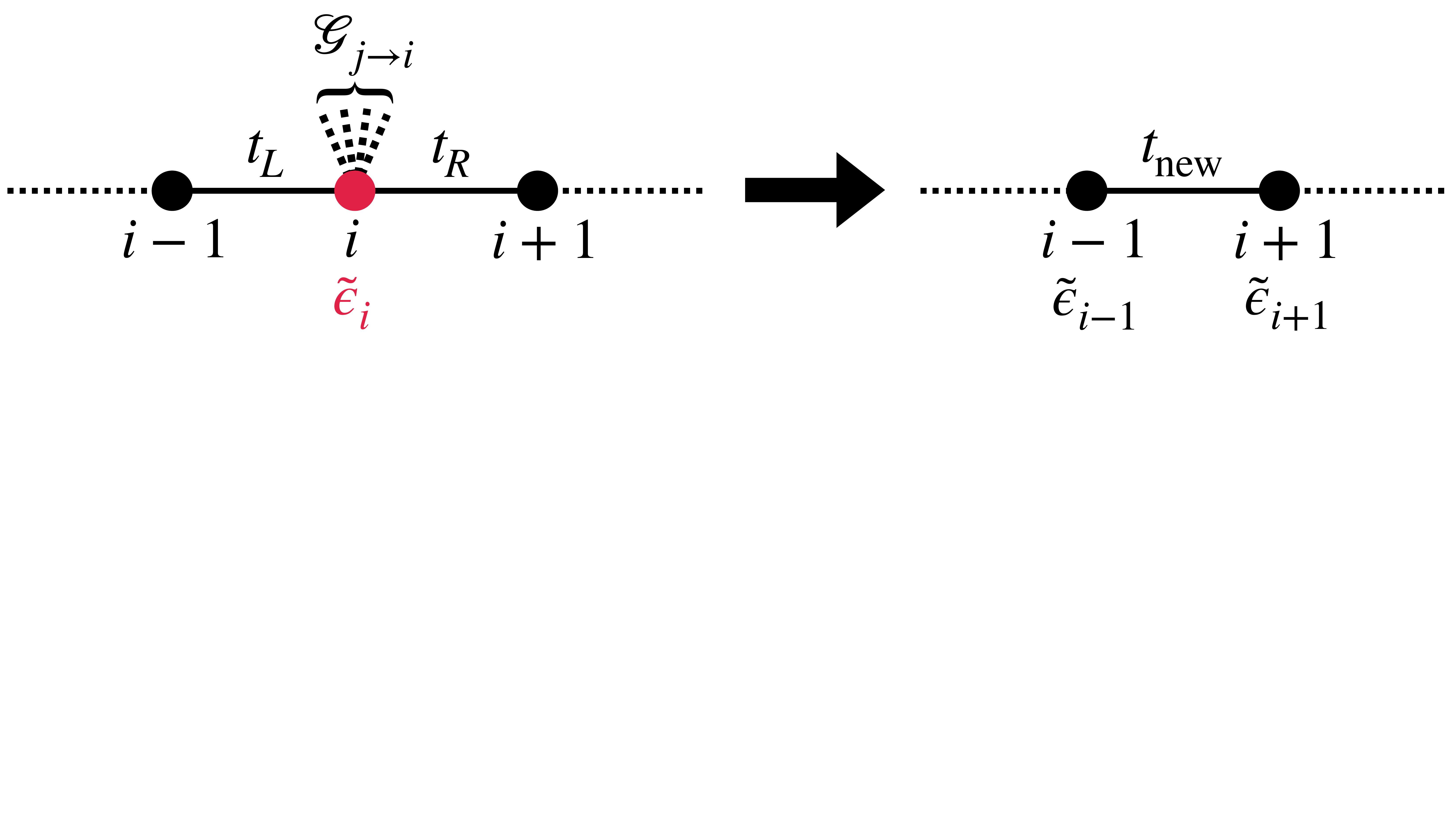}
	\vspace{-2.6cm}
	\caption{\label{fig:diag2} Pictorial sketch of the exact decimation procedure which is used to integrate out a site $i$, see Eqs.~(\ref{eq:decimation}). }
\end{figure}

In presence of closed loops, such as the ones appearing in Fig.~\ref{fig:1loopdiagram} of the main text and in Fig.~\ref{fig:diagrams}, Eqs.~\eqref{eq:Gcav} and~\eqref{eq:G} are no longer correct. A convenient way to compute the observalbes of interest is to perform an exact decimation procedure which  allows  one  to  integrate  out progressively all the intermediate nodes on the on the lines of the diagrams. The decimation procedure is discussed in details in Refs.~\cite{aoki,dobro,noilarged} and is schematically depicted in Fig.~\ref{fig:diag2}. 

\begin{widetext}
Let us consider a node $i$ of the lattice connected with the node $i-1$ by the hopping $t_L$, with the node $i+1$ by the hopping $t_R$ and with $k-1$ nodes $j$ which are the roots of semi-infinite branches of the loop-less BL. The effective on-site energy on site $i$ is defined as $\tilde{\epsilon}_i$. The nodes $j$ are connected to $i$ by the hopping $t$. The semi-infinite branches originating from those sites can be integrated out exactly using the recursion relations~\eqref{eq:Gcav}, yielding the cavity Green's functions $G_{j \to i}$ on each one of these nodes. As illustrated in the figure, when the node $i$ is integrated out, it yields a modification of the random potential on its left and right neighbors, and generates a new hopping amplitude between them:
\[
\begin{aligned}
& e^{-\frac{i}{2} [ - \tilde{\epsilon}_{i+1} \phi_{i+1}^\star \phi_{i+1} - \tilde{\epsilon}_{i-1} \phi_{i-1}^\star \phi_{i-1}]} \int {\rm d} \phi_i {\rm d} \phi_i^\star \, e^{ \frac{{\rm i}}{2} \tilde{\epsilon}_i \phi_i^\star \phi_i - \frac{{\rm i}}{2} \sum_{j=1}^{k-1}  \frac{\phi_j^\star \phi_j }{G_{j \to i}} + \frac{i}{2} t_L (\phi^\star_i \phi_{i-1} + \phi^\star_{i-1} \phi_i ) + \frac{i}{2} t_R (\phi^\star_i \phi_{i+1} + \phi^\star_{i+1} \phi_i ) } \\
& \qquad \qquad \qquad \propto \, e^{-\frac{i}{2} [ - \tilde{\epsilon}_{i+1} \phi_{i+1}^\star \phi_{i+1}  - \tilde{\epsilon}_{i-1} \phi_{i-1}^\star \phi_{i-1}  - t_{\rm new} (\phi^\star_{i+1} \phi_{i-1} + \phi^\star_{i-1} \phi_{i+1})]} \, , 
\end{aligned}
\]
with:
\begin{equation} \label{eq:decimation}
\begin{aligned}
	\tilde{\epsilon}_{i-1} & \longrightarrow \tilde{\epsilon}_{i-1} + \frac{t_L^2}{\tilde{\epsilon}_i  - t^2 \sum_{j \in \partial i / \{i-1,i+1\}}  G_{j \to i} } \, , \\
		\tilde{\epsilon}_{i+1} & \longrightarrow \tilde{\epsilon}_{i+1}  + \frac{t_R^2}{\tilde{\epsilon}_i  - t^2 \sum_{j \in \partial i / \{i-1,i+1\}}  G_{j \to i} } \, , \\
		t_{\rm new} & =  - \frac{t_L t_R}{\tilde{\epsilon}_i  - t^2 \sum_{j \in \partial i / \{i-1,i+1\}} G_{j \to i} }  \, ,
\end{aligned}
\end{equation}
with the initial conditions $\tilde{\epsilon}_i = \epsilon_i + {\rm i} \eta$ and $t_L=t_R=t$. The sum in the denominators runs over the $k-1$ neighbors  $j$ of $i$ which are {\it not} along the loop (i.e. all the neighbors of $i$ except sites $i+1$ and $i-1$) which are connected to semi-infinite loop-less branches of the BL and which carry the cavity Green's functions $G_{j \to i}$.
\end{widetext}

\subsection{Numerical analysis of the correlation function $C_{\rm BL} (L)$ and the Ginzburg criterion} \label{app:ginzburg}

\begin{figure*}
\includegraphics[width=0.338\textwidth]{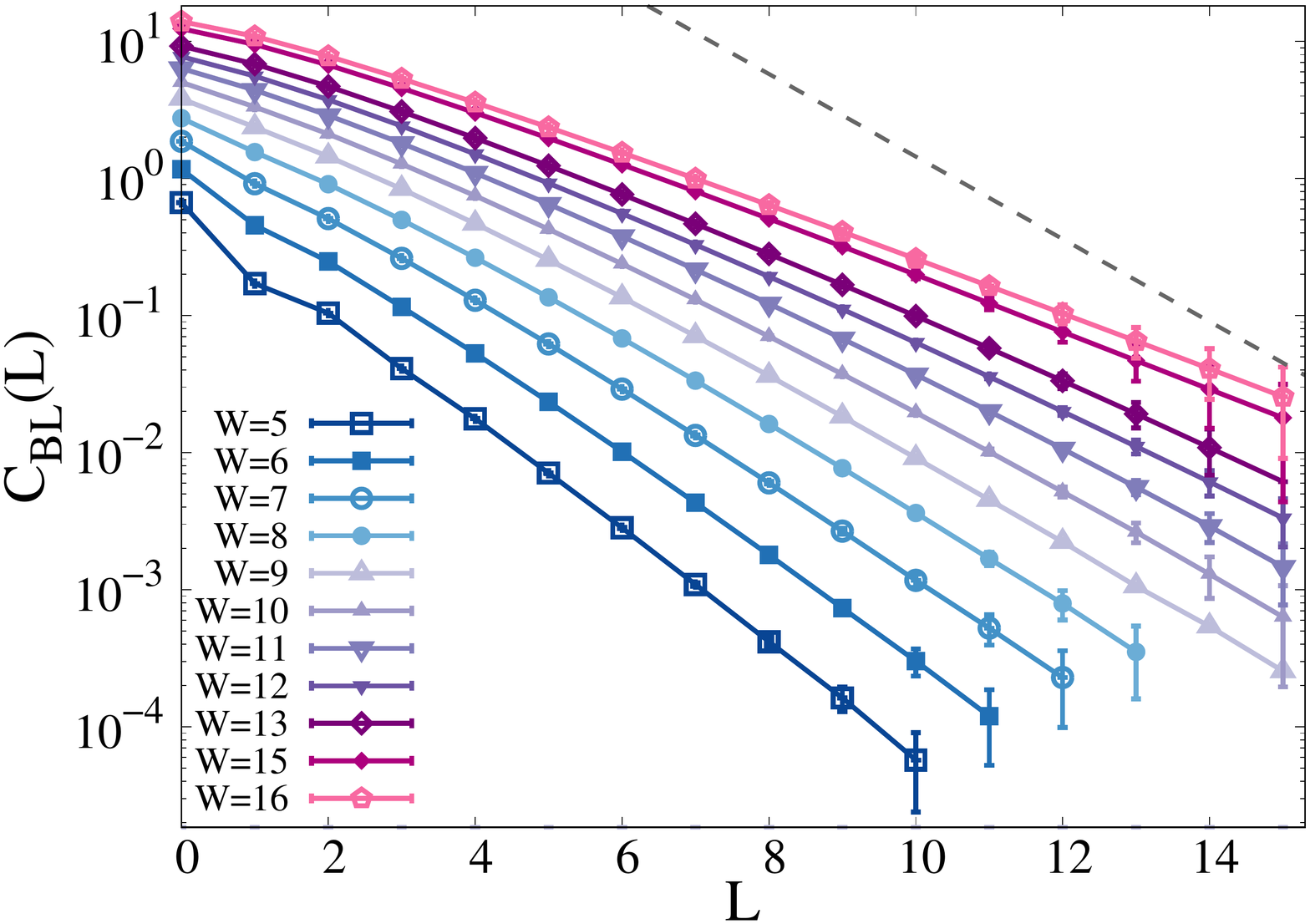} \hspace{-0.33cm} \includegraphics[width=0.338\textwidth]{ginzburg}
\hspace{-0.334cm} \includegraphics[width=0.325\textwidth]{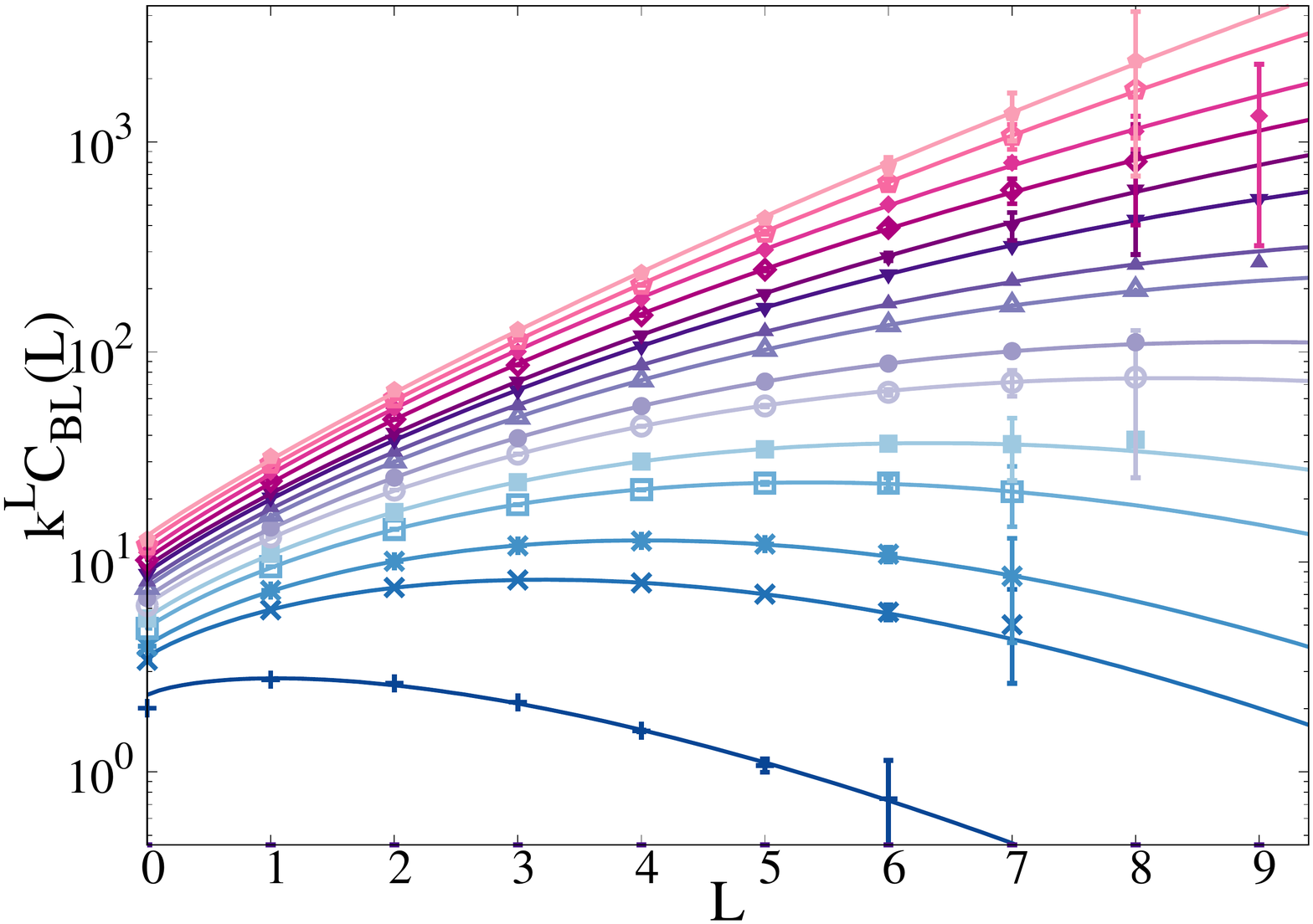} 
\caption{$C_{\rm BL} (L)$ as a function of $L$ for different values of the disorder across the metallic phase and for $k=2$ (left) and $k=5$ (middle). The gray dashed line corresponds to $k^{-L}$. %
Right panel: $k^L C_{\rm BL} (L)$ as a function of $L$ for $k=5$ and for the same values of $W$ as in the middle panel. The continuous lines correspond to the fits of the data according to Eq.~\eqref{eq:fit} with $f(x) = A e^{B x -C x^\beta}$. The best fits are obtained for $\beta \approx 1.32$ for $k=2$ (see Fig.~\ref{fig:ginzburg}) and  $\beta \approx 1.42$ for $k=5$. 
\label{fig:ginzburgSI}}
\end{figure*}

\begin{figure*}
\includegraphics[width=0.338\textwidth]{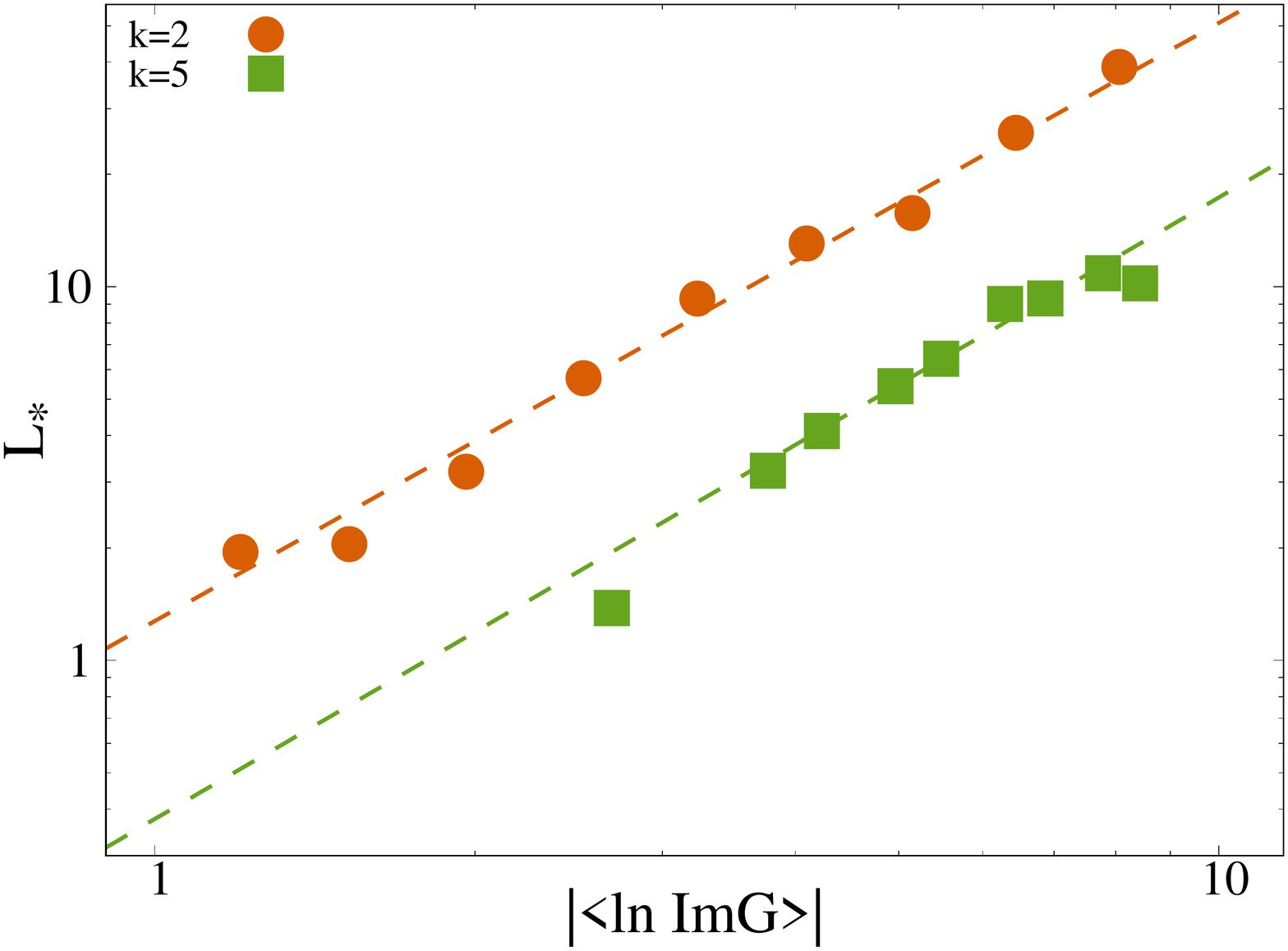} \hspace{-0.33cm} \includegraphics[width=0.338\textwidth]{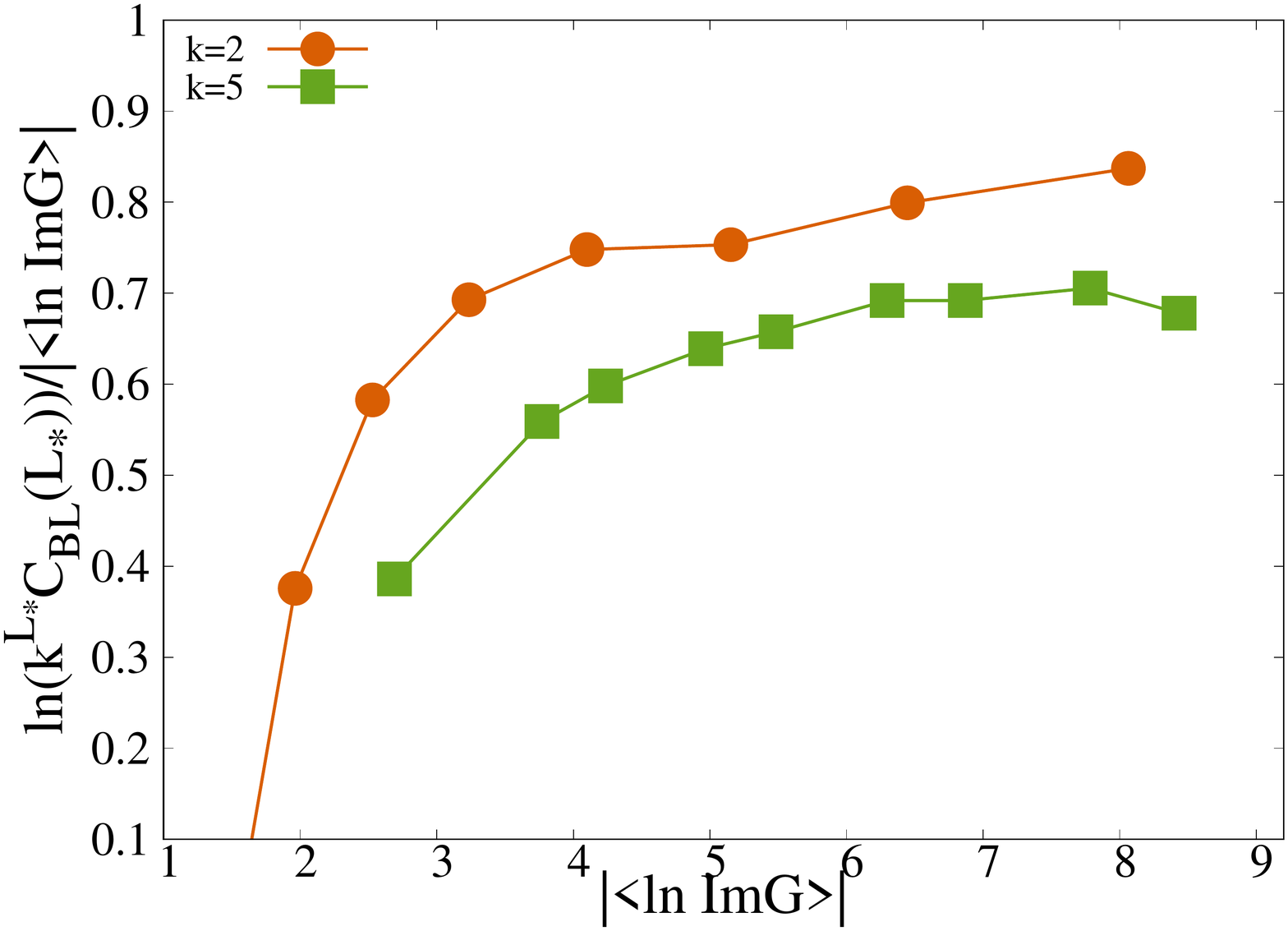} \hspace{-0.33cm} \includegraphics[width=0.338\textwidth]{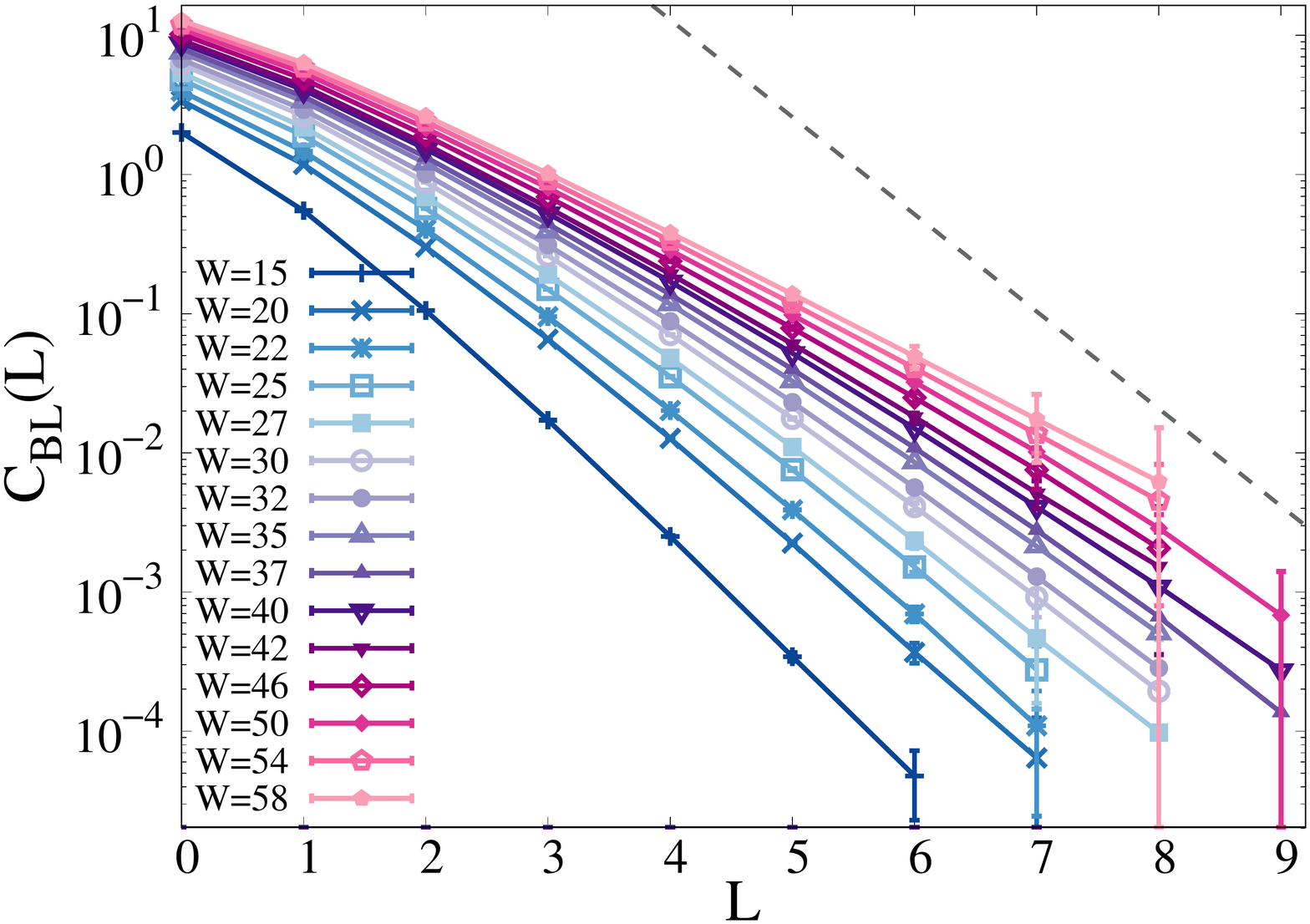}
\caption{Left: Parametric plot of the position of the maximum $L_{\star}$ as a function of $\ln \Lambda = | \avg{\log {\rm Im} G}|$ for $k=2$ (circles) and $k=5$ (squares) in a log-log scale. The dashed lines are power-law fits of the data of the form $L_\star \propto (\ln \Lambda)^{\delta}$, with $\delta \approx 1.7$ for $k=2$, and $\delta \approx 1.42$ for $k=5$. Middle: Parametric plot of the ratio $\log[k^{L_\star} C_{\rm BL} (L_\star)]/| \avg{\log {\rm Im} G}|$ as a function of $| \avg{\log {\rm Im} G}|$ for $k=2$ (circles) and $k=5$ (squares), showing that at large enough disorder $\log[k^{L_\star} C_{\rm BL} (L_\star)] = b | \avg{\log {\rm Im} G}|$, with $b\approx 0.8$ for $k=2$ and $b \approx 0.7$ for $k=5$. Right panel: Plot of the ratio $\ln \chi_\xi/| \avg{\log {\rm Im} G}|$ (where $\chi = \sum_L k^L C_{\rm BL} (L)$) {\it vs} $| \avg{\log {\rm Im} G}|$ for $k=2$ and $k=5$, showing that at large enough disorder one has that $\ln \chi_\xi \propto c \ln \Lambda$, with $c \approx 1.4$ for $k=2$ and $c \approx 1$ for $k=5$. 
\label{fig:max}}
\end{figure*}

In this section we perform a thorough numerical analysis of the asymptotic behavior of the connected correlation function $C_{\rm BL} (L) = \langle \ln \rho_i \ln \rho_{i+L} \rangle_c$  between two nodes at distance $L$ on the infinite BL (Eq.~\eqref{eq:CBL} of the main text). As explained above, in fact, $\avg{\ln \rho_i}$ can be considered as a proxy for the order parameter distribution function. In the top panels of Fig.~\ref{fig:ginzburgSI} $C_{\rm BL}$ is plotted as a function of $L$ for several values of the disorder $W$ across the metallic phase  for $k=2$ (left) and for $k=5$ (middle) and for $\eta \to 0^+$. Within the range of distances $L$ in which our results are statistically meaningful (i.e. such that the average of $C_{\rm BL}$ is significantly larger than the statistical error) $C_{\rm BL}(L)$ decreases much more slowly than  $\avg{|G_{i,i+L}|^2}$. In particular for not too large values of $L$ the decrease of $C_{\rm BL}(L)$ is much slower than $k^{-L}$ (gray dashed lines) both for $k=2$ and $k=5$. This is clearly highlighted when $C_{\rm BL}(L)$ is multiplied by  $k^L$ (i.e. the volume of the sphere of radius $L$ on the BL which enters in the number of paths between two points at distance $r$ on the original lattice, Eq.~\eqref{eq:NrL}): Both for $k=2$ (Fig.~\ref{fig:ginzburg} of the main text) and for $k=5$ (right panel of Fig.~\ref{fig:ginzburgSI}) $k^L C_{\rm BL} (L)$ first grows for $L$ small enough, and then decreases at larger $L$, after going through a maximum in $L_\star$. The position of the maximum moves to larger and larger values of $L$ as $W$ is increased and its height grows very rapidly with $W$. We estimate quantitatively the  position and the height of the maximum by performing a parabolic fit of the numerical data in its vicinity. The results of this analysis are shown in Fig.~\ref{fig:max}. The left panel indicates that, at least in the disorder range within which we are able to identify the position of the maximum, $L_\star$ grows as a power of the logarithm of the correlation volume, $L^\star \propto |\avg{\ln {\rm Im} G}|^\delta$. In the middle panel we plot the logarithm of the height of the maximum of $k^L C_{\rm BL} (L)$ divided by the logarithm of our estimator of the correlation volume, showing that the ratio  $\ln[k^{L_\star} C_{\rm BL} (L_\star)]/|\avg{\ln {\rm Im} G}|$ tends to a constant of order $1$ when the disorder is increased both for $k=2$ and $k=5$. This implies that the position of the maximum $L_\star$, and its height  $k^{L_\star} C_{\rm BL} (L_\star)$ behaves accordingly to Eq.~\eqref{eq:Lstar} of the main text, where $a$ and $b$ are constant of order $1$.

We have thus shown that the correlation function $C_{\rm BL} (L)$ has a cutoff on a length scale $L_\star$ which is of the order of some power of the logarithm of the correlation volume $\ln \Lambda$ (note that, as discussed above and in Refs.~\cite{mirlintikhonov,efetov_bethe,zirn,verba}, a similar cutoff is present in the density-density correlation $\langle \rho_i \rho_{i+L}\rangle_c$), suggesting that: 
\begin{equation} \label{eq:fit}
C_{\rm BL} (L) \simeq \kappa(W) \, f \! \left( \frac{L}{L_\star} \right) \, ,
\end{equation}
where $\kappa (W)$ is a disorder dependent prefactor that, according to~\eqref{eq:Lstar}, scales as $\kappa(W) \approx e^{b \ln \Lambda}/k^{L_\star} \approx e^{b \ln \Lambda - (\ln \Lambda)^\delta \ln k}$, and $f(x)$ is a cutoff function which goes to zero faster than $k^{-x}$ for $x \gg 1$. For concreteness and for practical purposes in the following we use a specific functional form for the cutoff function which fits reasonably well the data for all values of $W$ both for $k=2$ and $k=5$, $f(x) = A e^{B x -C x^\beta}$, with $\beta>1$. In order to reduce the number of fitting parameters, we let the coefficients $A$, $B$, and $C$ to adjust freely for every value of $W$, while keeping  the value of $\beta$ fixed for all values of $W$. Eq.~\eqref{eq:fit} fits remarkably well the numerical data both for $k=2$ (Fig.~\ref{fig:ginzburg}) and $k=5$ (right panel of Fig.\ref{fig:ginzburgSI}) in the whole range of values of $L$ and $W$ that we can explore. The best fits are obtained for $\beta \approx 1.32$  for $k=2$ and  $\beta \approx 1.42$ for $k=5$.  Note that $C_{\rm BL} (0) = \avg{(\ln \rho)^2}-\avg{\ln \rho}^2$ increases smoothly as $W$ is increased but stays finite at the critical point.

As explained in the main text, we consider as a local observable $(1/M) \sum_{\alpha=1}^M \ln \rho_{i,\alpha}$, which plays the role of a proxy for the order parameter function. Its average at large values of $M$ converges to the BL result $\ln \rho_{\rm typ}$. The fluctuations of its value at two lattice sites at distance $r$ on the original $d$-dimensional euclidean $M$-layered lattice is given by Eq.~\eqref{eq:Cd}~\cite{mlayer}, where ${\cal N} (r,L)$ is the number of non-backtracking paths of length $L$ connecting the two points at distance $r$ on the original lattice (with $M=1$), given in Eq.~\eqref{eq:NrL}. Hence, setting $r=0$ we obtain the fluctuations of $\ln \rho_{i,\alpha}$ on a given site $i$ among different layers $\alpha$ at the order $1/M$: 
\begin{equation} \label{eq:cd}
C_d(0) = \frac{1}{M} \sum_L \frac{k^L C_{\rm BL} (L)}{L^{d/2}} \, .
\end{equation}
According to the Ginzburg criterion (see App.~\ref{app:GP} for the case of the percolation transition) we have to check that the prefactor of the fluctuations of order $1/M$ does not diverge at the transition in order for the mean-field predictions to be valid. The sum over $L$ in the expression above is dominated by the maximum of $k^L C_{\rm BL} (L)$ at $L=L_\star$, which grows very fast as the disorder is increased. In order to obtain a quantitative estimation of $C_d(0)$ we have computed $\chi = \sum_L k^L C_{\rm BL} (L)$ numerically by plugging directly the ansatz~\eqref{eq:fit} into the sum and using the parameters of the cutoff function $f(x)$ that one obtains from the fit of the numerical data reported in the bottom panels of Fig~\ref{fig:ginzburgSI}. The results of this procedure are illustrated in the right panel of Fig.~\ref{fig:max} where we plot the ratio $\ln \chi/|\avg{\ln {\rm Im} G}|$ as a function of $|\avg{\ln {\rm Im} G}|$, showing that at at large enough disorder both for $k=2$ and $k=5$ one has $\ln \chi \propto c \ln \Lambda$, where $c$ is a constant of order $1$. We thus obtain that the fluctuations of $\ln \rho$ at a given position in the $d$-dimensional space between different layers of the lattice  behave as:
\begin{equation}
C_d(0) \approx \frac{1}{M} \frac{e^{c \ln \Lambda}}{L_\star^{d \delta/2}} \propto \frac{1}{M} (W_c - W)^{d \delta/4} \, e^{\frac{{\rm cst}}{(W_c-W)^{1/2}}} \, ,
\end{equation}
(see Eq.~\eqref{eq:Cd0} of the main text) which grows exponentially fast close to the localization transition in any dimension. Conversely $\avg{\ln \rho}^2$ only grows algebraically close to $W_c$, as  $\avg{\ln \rho}^2 \propto (W_c-W)^{-1}$, indicating that the Ginzburg criterion is never satisfied in any finite $d$.

\subsection{Numerical analysis of the $1$-loop corrections to the typical DoS} \label{app:deltalnrho}

\begin{figure}
	\includegraphics[width=0.46\textwidth]{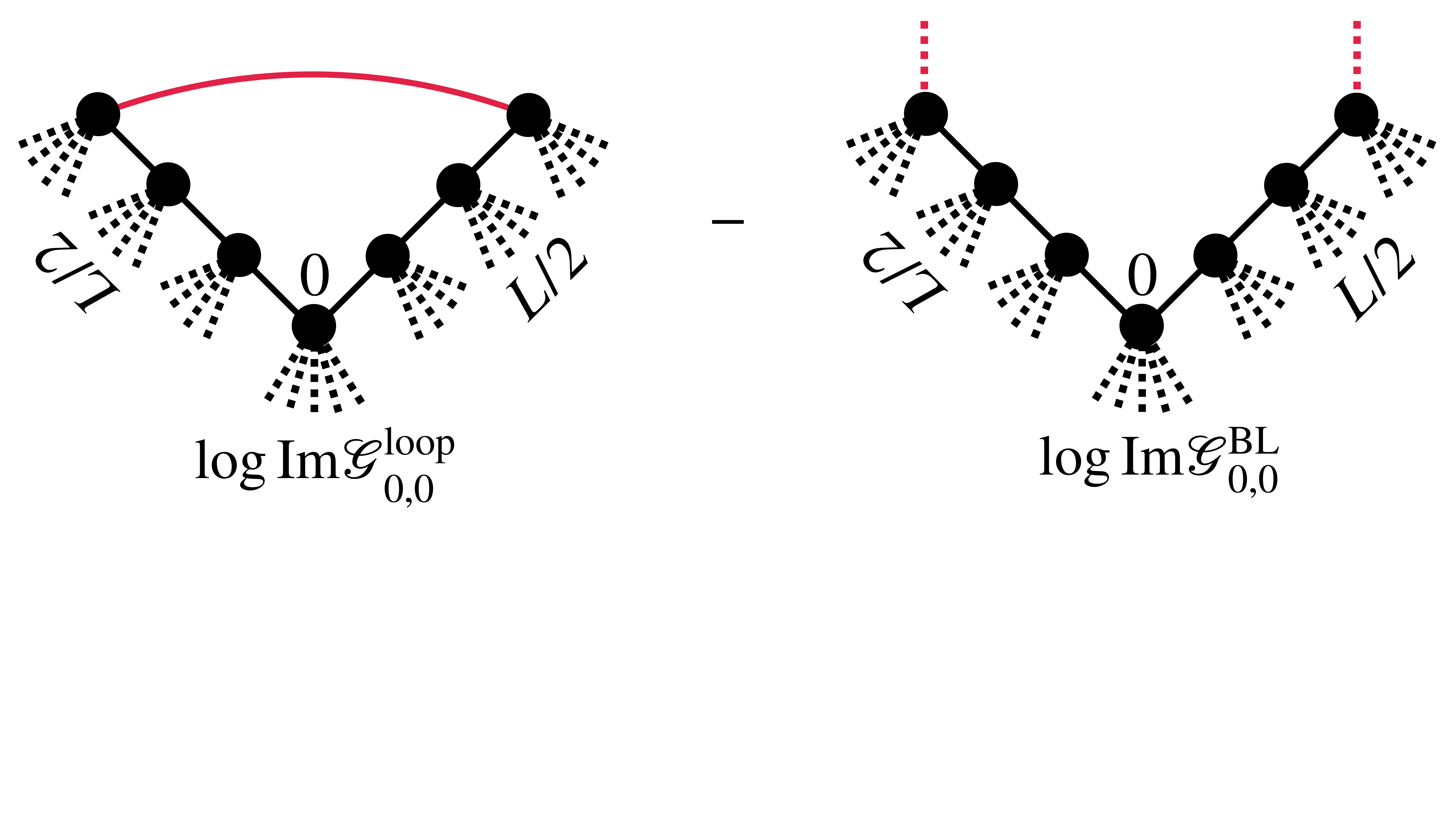}
	\vspace{-1.8cm}
	\caption{\label{fig:loop_sketch} Pictorial illustration of the procedure used to compute the line connected value of the correction at the one loop level to the logarithm of the typical DoS. }
\end{figure}

In this section we present an accurate numerical analysis of the the 1-loop line connected value of the average of the logarithm of the LDoS, $\Delta \! \avg{\ln \rho}_{\rm 1loop}$, given in Eq.~\eqref{eq:rhotyp1loop}, were $\delta [\ln \rho (L,L_1)]$ is the line connected value of $\langle \ln \rho \rangle$ on site $0$ at the 1-loop level,  defined as the difference between the average of $\ln \rho_0$ computed in presence and in absence of the loop, Eq.~\eqref{eq:deltalnrho}. ${\cal N} (L,L_1)$ is the number of 1-loop diagrams for one-point observables in the $d$-dimensional euclidean lattice, Fig.~\ref{fig:1loopdiagram}, whose asymptotic expressions is given in Eq.~\eqref{eq:loops}.

We start by explaining in detail how to compute it (we also use the same strategy to compute the $1$-loop corrections to $\avg{|G_\eta|^2}$, see Sec.~\ref{app:deltaG2} below). Having clean numerical results for $\delta [\ln \rho (L,L_1)]$ is of paramount importance to characterize its dependence on $L$ and $L_1$ thoroughly. In order to avoid the effect of fluctuations induced by rare values of the Green's functions in the far tails of the distribution, it is crucial that the computation of $\im G_0$ with and without the loop is performed using the same realization of the disorder in both cases. The concrete implementation of this procedure is schematically illustrated in Fig.~\ref{fig:loop_sketch}. More specifically, we consider a site $0$ with $k-1$ branches of the infinite BL attached to it (carrying the cavity Green's functions $G_{j \to 0}$), and with two outgoing chains of length $L/2$; Each site of these chains is also attached to $k-1$ branches of the loop-less infinite BL, which carry the corresponding cavity Green's function. A loop of length $L$ is obtained by connecting the two nodes at the end of the two chains with an edge (full red line of the left drawing).  In order to compute ${\rm Im} G_0^{\rm loop}$ in presence of the loop we thus need to extract $(k-1)(L+1)$ cavity Green's function from the pool and $L+1$ random energies from the box distribution, and integrate out progressively all the $L$ nodes of the loop using the exact decimation procedure illustrated above, Eqs.~(\ref{eq:decimation}). The line connected value of the logarithm of the LDoS on site $0$ at the 1-loop level is obtained by subtracting the logarithm of the LDoS on site $0$ when the loop is removed. This is done by attaching to each of the two nodes at the end of the two chains of length $L/2$ one more branch of the infinite tree (red dashed lines of the right drawing). The computation of ${\rm Im} G_0^{\rm BL}$ is thus performed using the same $L+1$ random energies and the same $(k-1)(L+1)$ cavity Green's function as ${\rm Im} G_0^{\rm loop}$, plus only two extra cavity Green's functions extracted from the pool. 

\begin{figure*}

\includegraphics[width=0.337\textwidth]{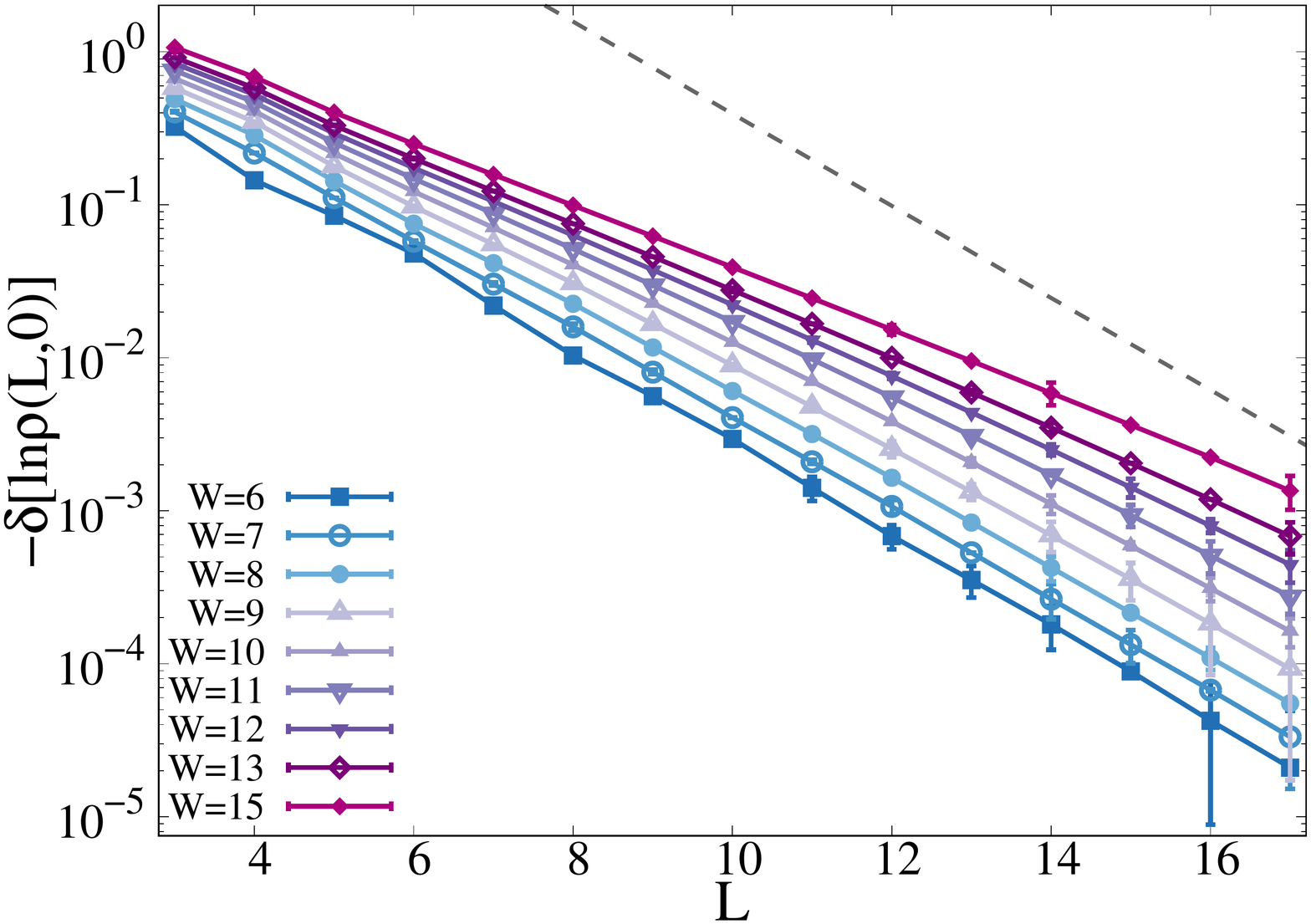} \hspace{-0.33cm} \includegraphics[width=0.337\textwidth]{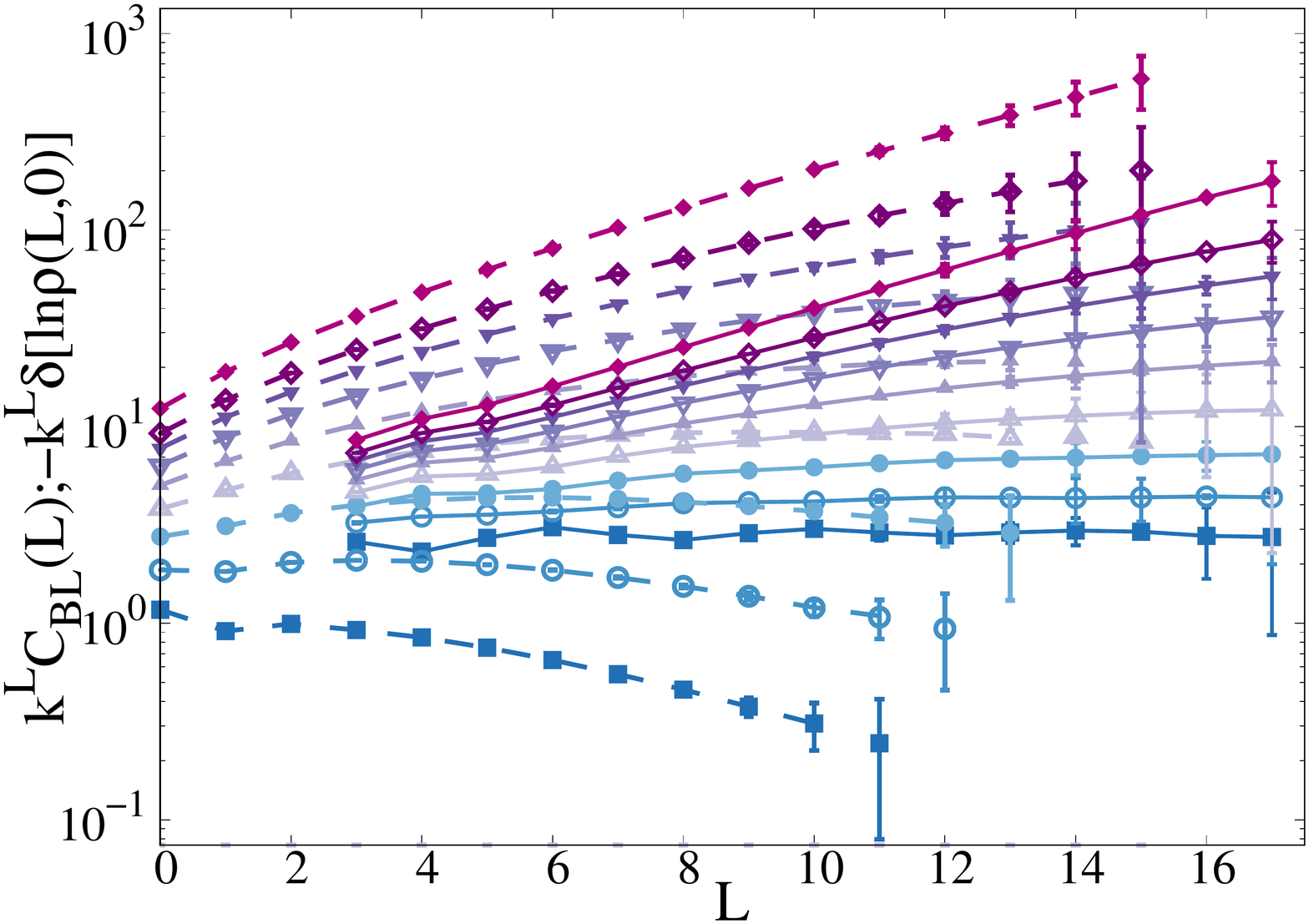} \hspace{-0.33cm} \includegraphics[width=0.34\textwidth]{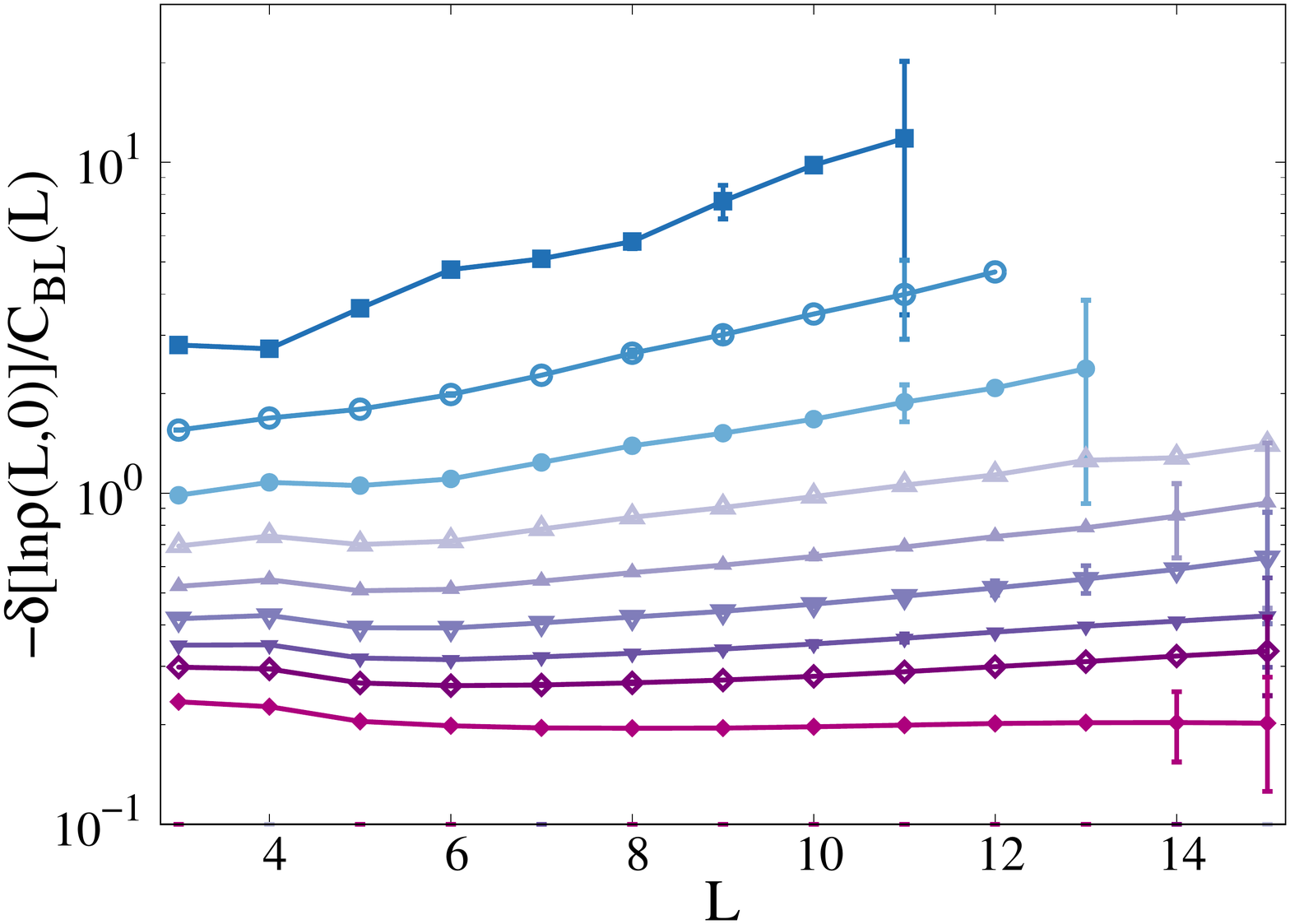} 

\hspace{-0.25cm}\includegraphics[width=0.341\textwidth]{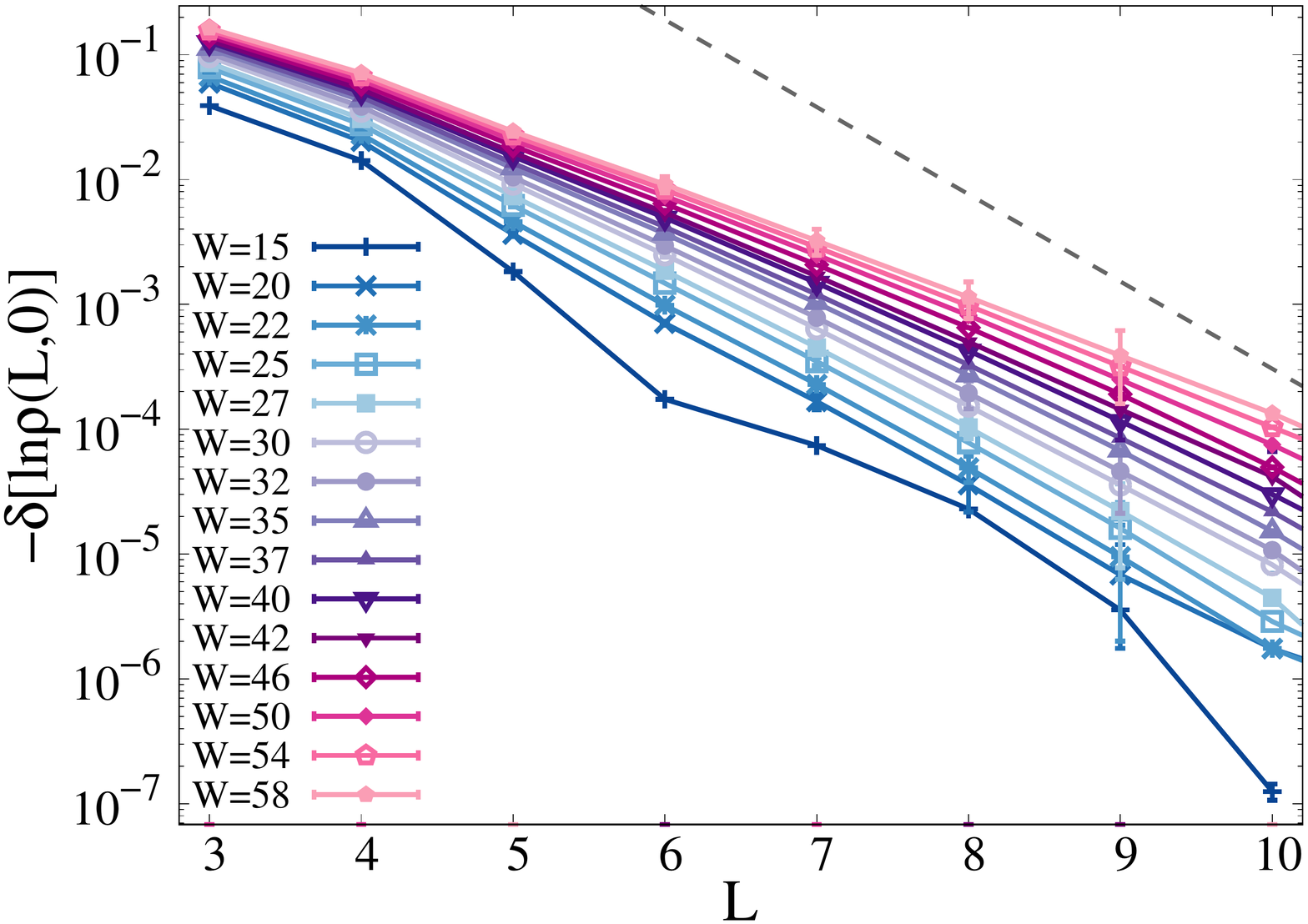} \hspace{-0.315cm} \includegraphics[width=0.328\textwidth]{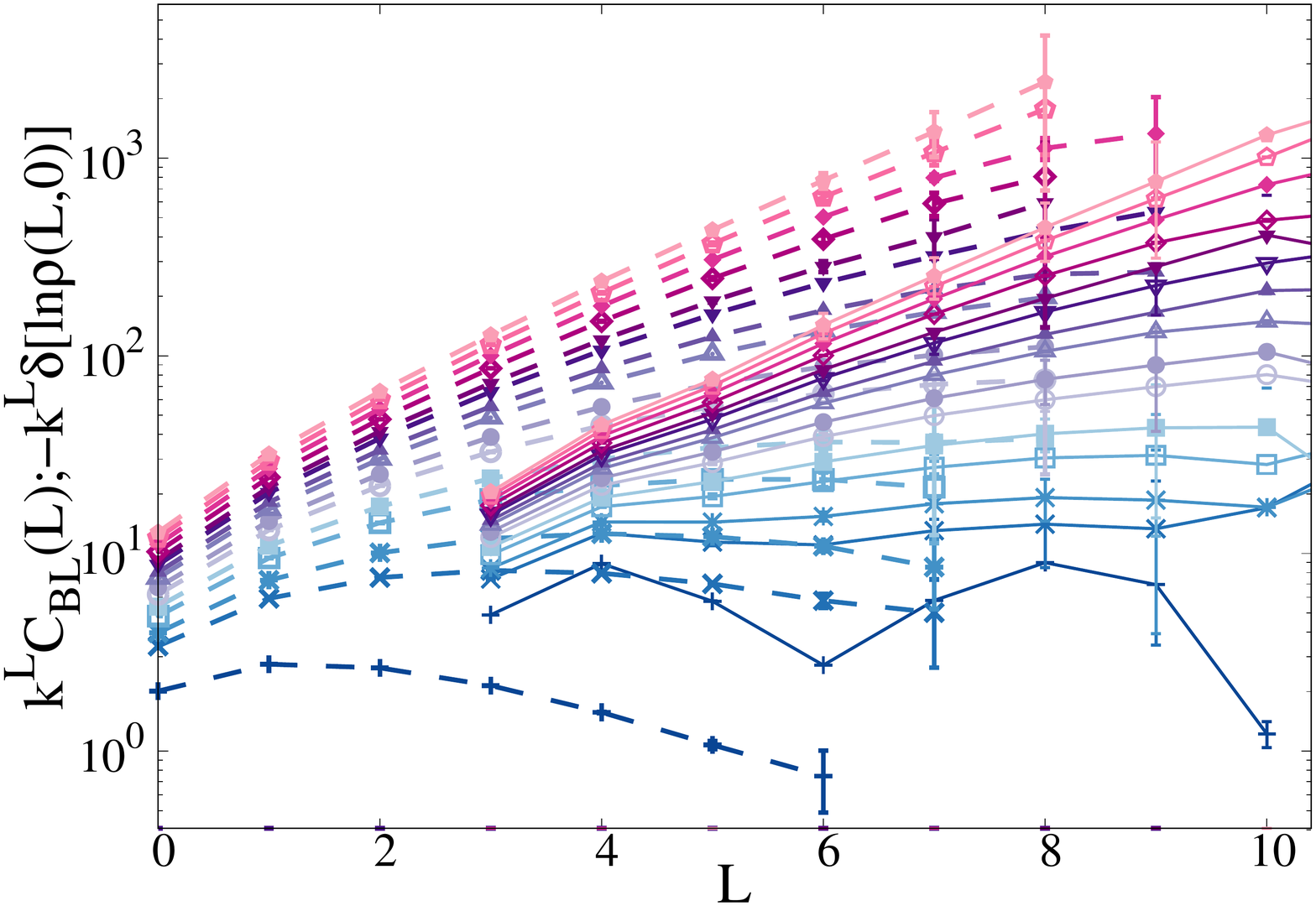} \hspace{-0.158cm} \includegraphics[width=0.34\textwidth]{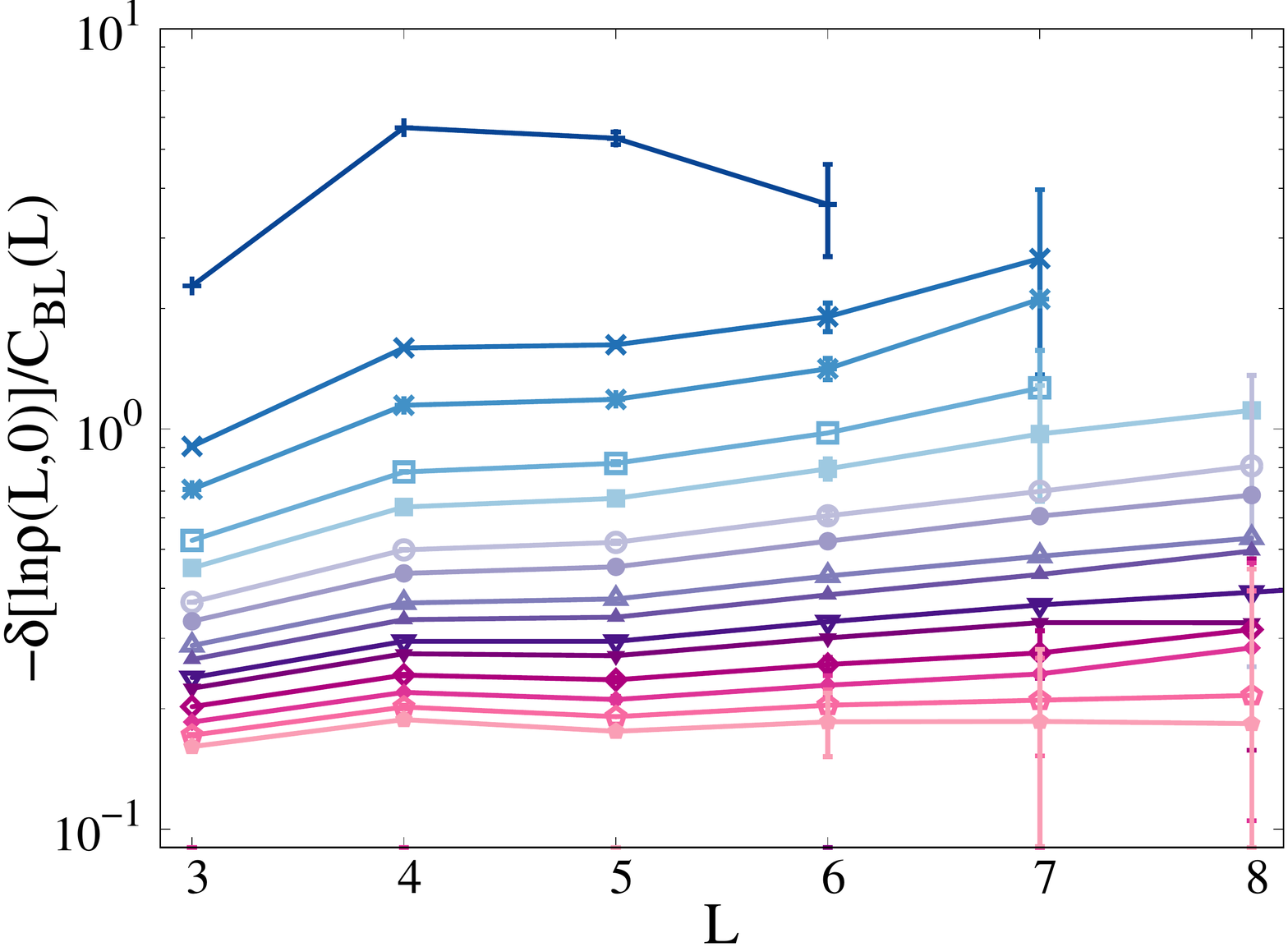} 
\caption{Left panels: Line connected average $\ln |\delta[ \ln \rho (L,0)]|$ as a function of $L$ for different values of the disorder across the metallic phase for $k=2$ (top) and $k=5$ (bottom). The black dashed-dotted lines correspond to $k^{-L}$ which is the slowest exponential decay which would ensure the convergence of the 1-loop corrections. Middle panels: $k^L C_{\rm BL} (L)$ (dashed curves) and $- k^L \delta[ \ln \rho (L,0)]$ (continuous curves) as a function of $L$ for different values of the disorder for $k=2$ (top) and $k=5$ (bottom). Right panels: Logarithm of the ratio $-\delta[ \ln \rho (L,0)/C_{\rm BL} (L)$ for several values of the disorder across the metallic phase for $k=2$ (top) and $k=5$ (bottom). The ratio becomes independent of $L$ at large enough disorder and decreases smoothly as $W$ is increased. 
\label{fig:loop}}
\end{figure*}

We start by analyzing the dependence of $\delta [\ln \rho (L,L_1)]$ on $L$ and $L_1$ in the limit $\eta = 0^+$. As explained in the main text, we find that, similarly to the case of percolation discussed in Sec.~\ref{app:DeltaP}, the dependence on $L$ and $L_1$ completely factorizes as $\delta [\ln \rho (L,L_1)] = f_a(L) f_r(L_1)$. Such factorization property can be understood by realizing that $\delta [\ln \rho (L,L_1)]$ is essentially a response function: It measures the variation of $\ln \rho$ on site $0$ due to the variation of the  LDoS on a site at distance $L_1$ from $0$ due to the presence of a loop of length $L$ originating from it. The length of the loop sets the amplitude of the perturbation. Since a small perturbation of the value of one of the cavity propagators in the right hand side of the BL iteration relations~\eqref{eq:Gcav} propagates linearly along a chain of the tree, it is natural to expect that the dependence on $L$ and $L_1$ of $\delta [\ln \rho (L,L_1)]$ factorizes in terms of the product of the amplitude of the perturbation (which depends only on $L$) times the response function (which depends only on $L_1$).

In order to check these ideas  we start by fixing the length of the external chain $L_1$ to $0$. We find $|\delta [\ln \rho (L,0)]|$ is negative, as expected, since finite-dimensional fluctuations are supposed to reduce the value of the critical disorder. In left panels of Fig.~\ref{fig:loop} $-\delta [\ln \rho (L,0)] \equiv -f_a(L)$ is plotted as a function of the length of the loop $L$  for $k=2$ (top) and $k= 5$ (bottom) and for different values of the disorder across the metallic phase. (Note that by defining $\delta [\ln \rho (L,0)]$ as $f_a(L)$ we have implicitly set $f_r(0)=1$.)  Comparing these plots with the ones of $C_{\rm BL} (L)$ (left and middle panels of Fig.~\ref{fig:ginzburgSI}, one notices that in fact $|\delta [\ln \rho (L,0)]|$ resembles, at least qualitatively, to the connected correlation function of $\ln \rho_i$. In particular, the exponential decay of $|\delta [\ln \rho (L,0)]|$ appears to be slower than $k^{-L}$ (gray dashed line) already at very small disorder, so that multiplying by the number of diagrams and summing over $L$ in~Eq.~\eqref{eq:rhotyp1loop},  would produce a divergence of $\Delta \! \avg{\ln \rho}_{\rm 1loop}$ even very far from the transition point. Hence, a cutoff similar to the one introduced in Eq.~\eqref{eq:fit} is necessary to ensure convergence. 

To make the comparison between $|\delta [\ln \rho (L,0)]|$ and $C_{\rm BL} (L)$ more quantitative, in the middle panels of Fig.~\ref{fig:loop} we show on the same plots  $k^L C_{\rm BL} (L)$ (dashed curves) and $k^L \delta [\ln \rho (L,0)]$ (continuous curves) both for $k=2$ (top) and for $k=5$ (bottom). Although at small disorder $k^L \delta [\ln \rho (L,0)]$ seems to decrease even more slowly than $k^L C_{\rm BL} (L)$ at large $L$, with a maximum shifted to larger values of $L$ and possibly a less sharp cutoff, at large enough disorder the two functions are  essentially proportional. This is confirmed by the bottom panes of Fig.~\ref{fig:loop}, where we plot the ratio $-\delta [\ln \rho (L,0)]/ C_{\rm BL} (L)$ as a function of $L$ for several values of $W$ across the delocalized phase, for $k=2$ (left) and $k=5$ (right). These plots show that upon increasing the disorder the ratio becomes roughly constant within our numerical accuracy, and close enough to the transition it is essentially independent of $L$ both for $k=2$ and $k=5$. Moreover, the ratio decreases smoothly as the disorder is increased and approaches a finite value for $W \to W_c$.  These plots thus suggest that sufficiently close to the transition point one has: 
\begin{equation} \label{eq:psi}
\delta [\ln \rho (L,0)] \approx - \varphi \, C_{\rm BL} (L) \, .
\end{equation}
The coefficient $\varphi$ has a smooth dependence on $W$: It decreases with $W$ and tends to a finite value $\varphi \sim [0.1,0.2]$ close to the critical point both for $k=2$ and $k=5$.

This behavior can be rationalized as follows: If the the logarithm of the imaginary part of the Green's function on two nodes at distance $L$ placed at the two ends of a chain embedded in the BL are correlated, then if one connects the two nodes to the same site $0$ producing a loop, the average value of $\ln {\rm Im} G_0$ will be modified compared to its BL counterpart (i.e. the case in which the loop is absent). Conversely, if the the logarithm of the imaginary part of the Green's function on the two nodes at the two ends of the chain are uncorrelated, then if one connects the two nodes to the same site $0$, $\ln {\rm Im} G_0$ will be on average the same as on the infinite BL in which the loop  is absent and all the neighbors of $0$ are uncorrelated. 

\begin{figure*}
\includegraphics[width=0.482\textwidth]{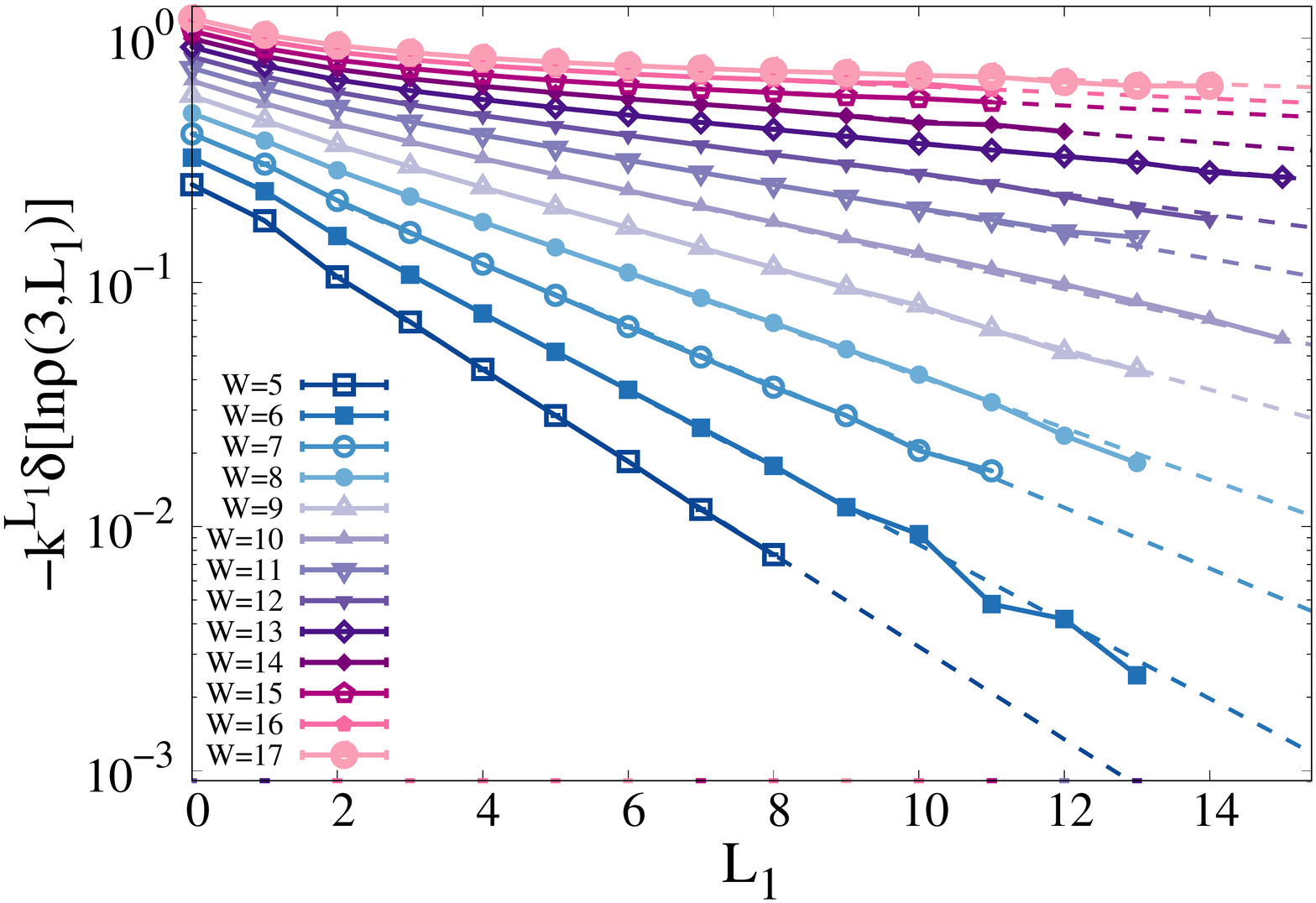}  \hspace{-0.1cm} 
\includegraphics[width=0.482\textwidth]{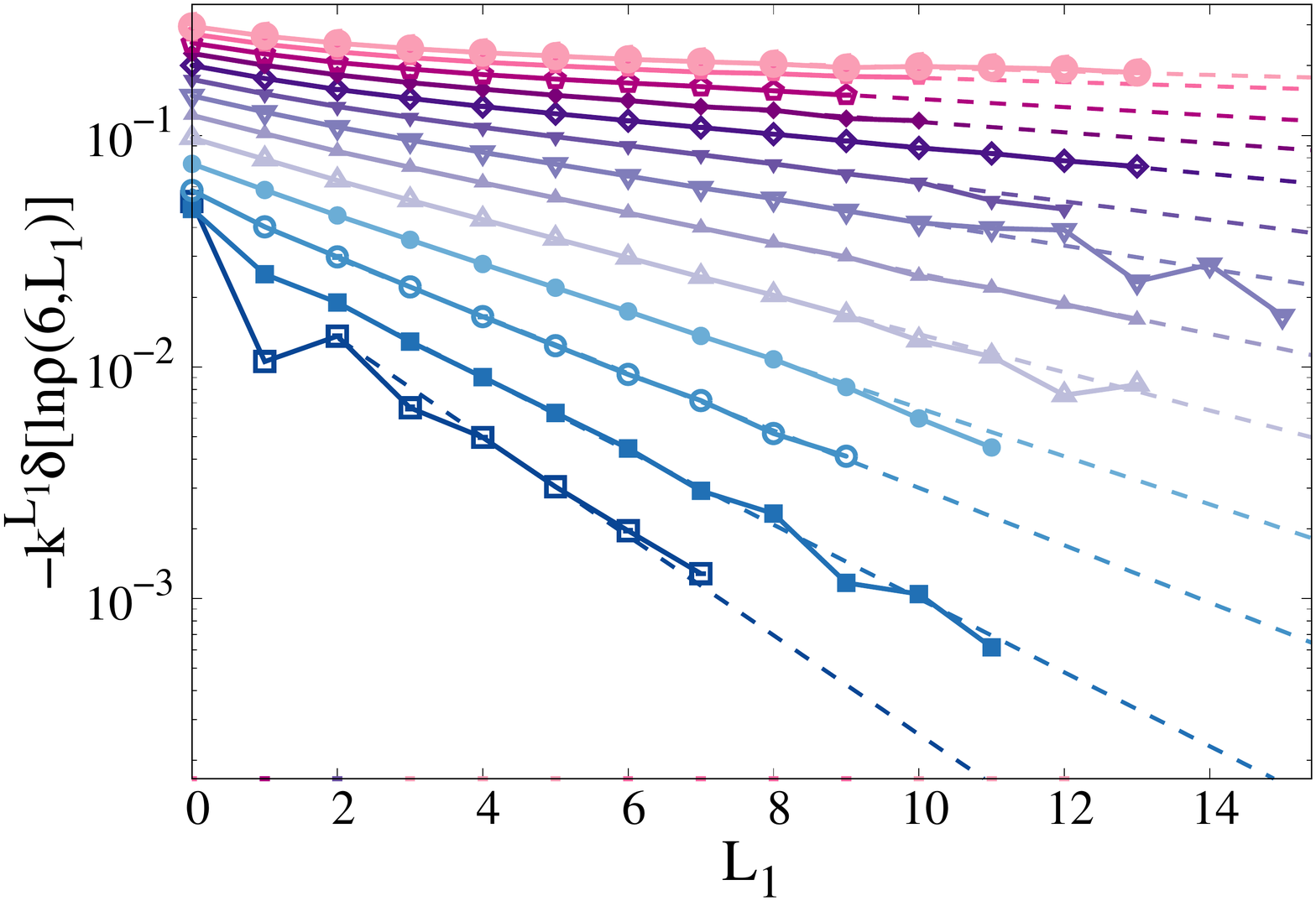}  

\includegraphics[width=0.482\textwidth]{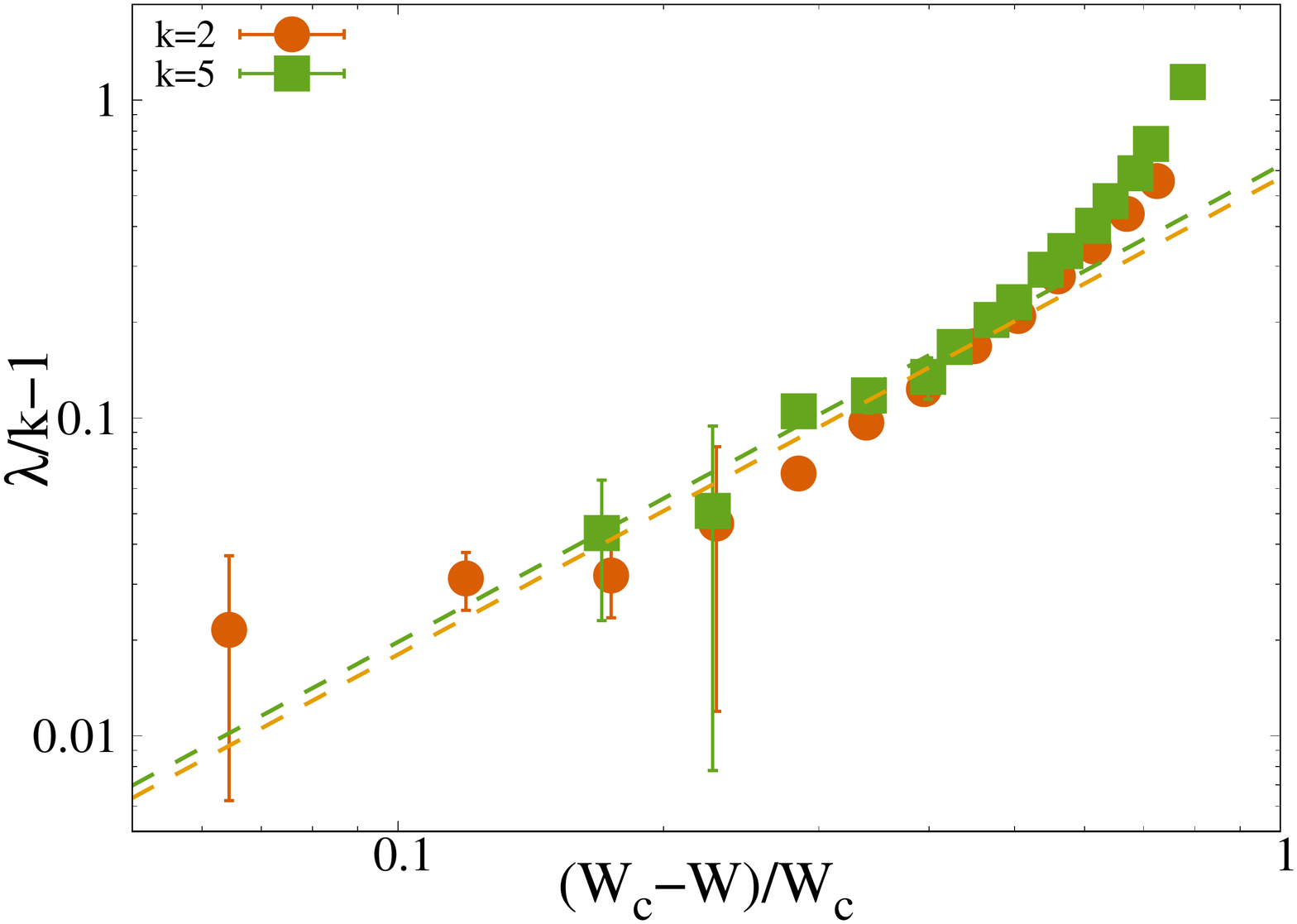} \hspace{-0.1cm} 
\includegraphics[width=0.482\textwidth]{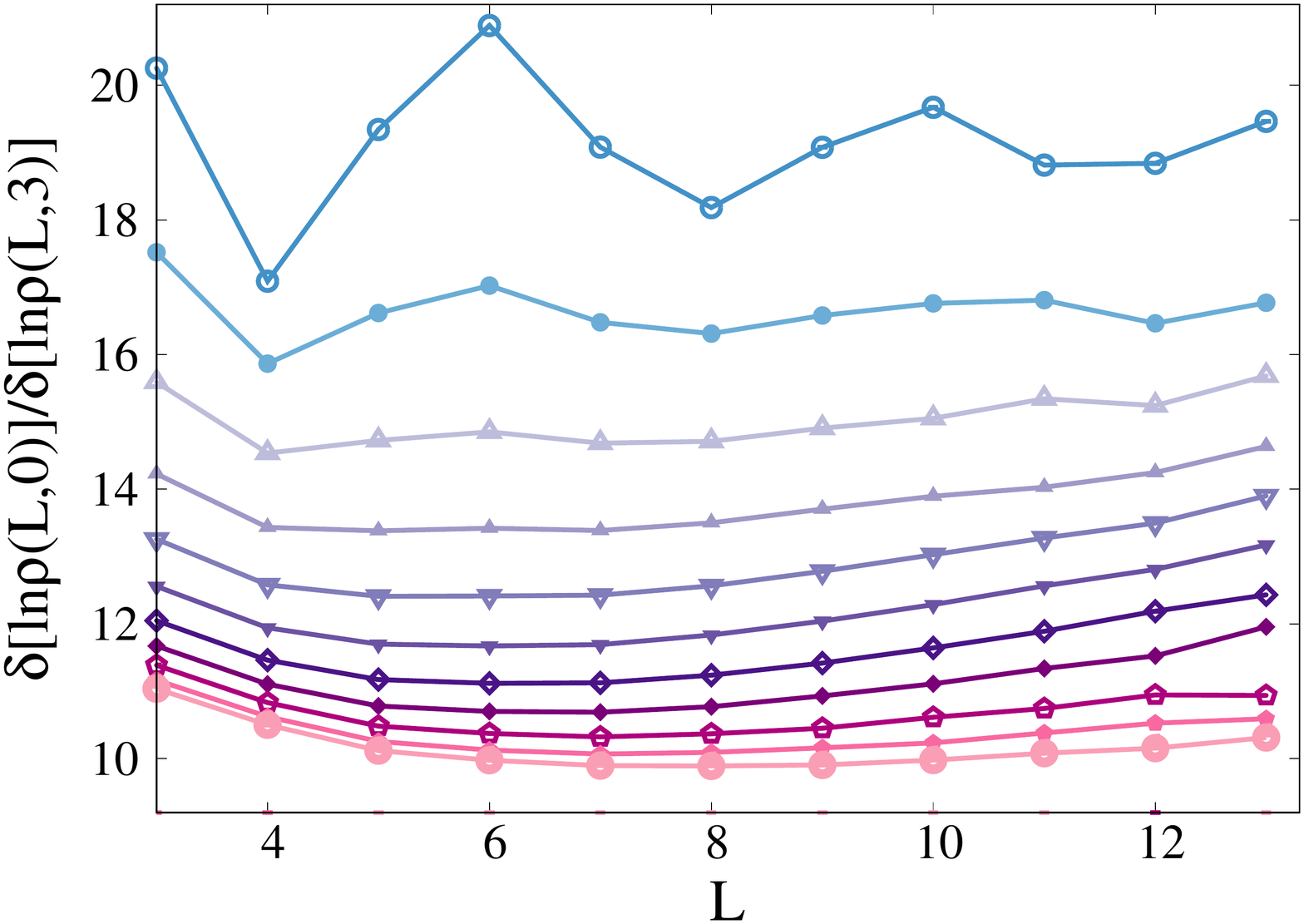}  
 
 \caption{Top panels: $-\delta [\ln \rho (L,L_1)]$  multiplied by $k^{L_1}$ as a function of $L_1$ for several value of the disorder across the delocalized phase for $k=2$ and for $L=3$ (left) and $L=6$ (right). Bottom-left panel: Log-log plot of $\lambda/k - 1$ as a function of the relative distance from the critical disorder, $(W_c-W)/W_c$, for $k=2$ (circles) and $k=5$ (squares). The decay rate $\lambda$ is extracted from a fit of the data shown in the top panels. The dashed lines are fits of the form $\lambda(W) - k \propto (1-W/W_c)^{3/2}$. Bottom-right panel: Ratio $\delta [\ln \rho (L,L_1=0)]/\delta [\ln \rho (L,L_1=3)]$ as a function of $L$ for the several values of $W$ in the metallic phase.
\label{fig:loop1LA}}
\end{figure*}

We now examine the dependence of $\delta [\ln \rho (L,L_1)]$ on the length of the external chain $L_1$. Below, for succinctness we will show the data for $k=2$ only, as for $k=5$ we find essentially the same behavior. We start by setting the length of the loop $L$ to a fixed value $L_0$ and study the behavior of $k^{L_1} \delta [\ln \rho (L=L_0,L_1)]$ varying $L_1$. The numerical results for several values of the disorder across the metallic phase are plotted in the top panels of Fig.~\ref{fig:loop1LA} for $L_0=3$ (i.e. the shortest possible loop) and $L_0=6$. The left ($L_0=3$) and right ($L_0=6$) panels are qualitatively identical, which supports the idea that the dependence of $\delta [\ln \rho (L,L_1)]$ on $L$ and $L_1$ factorizes as $\delta [\ln \rho (L,L_1)] = f_a(L) f_r(L_1)$. The curves in the right panel are globally shifted to smaller values compared to those in the left panel due to the fact that $f_a(3)$ is larger than $f_a(6)$.

At large enough $L_1$ the 1-loop correction $\delta [\ln \rho (L,L_1)]$ decreases exponentially with $L_1$, as $\lambda^{-L_1}$. The decay rate $\lambda$ can be extracted from a fit of the data and is plotted in the bottom-left panel of Fig.~\ref{fig:loop1LA} both for $k=2$ and for $k=5$: It is larger than $k$ for $W < W_c$ (thereby ensuring the convergence of the sum over $L_1$ in Eq.~\eqref{eq:rhotyp1loop}), and decreases smoothly as $W$ is increased, approaching $\lambda \to k$ for $W \to W_c$. The numerical values of $\lambda$ for $k=2$ and $k=5$ are strikingly close to each other when plotted as a function of the relative distance from the critical point $(W_c-W)/W_c$. Close enough to $W_c$ the data are well fitted by a power-law behavior of the form $\lambda - k \propto (1 - W/W_c)^{\omega}$ with $\omega \approx 3/2$.

The factorization property of $\delta [\ln \rho (L,L_1)]$ is further supported by the bottom-right plot of Fig.~\ref{fig:loop1LA}, in which we show the behavior of the ratio $\delta [\ln \rho (L,L_1=3)]/\delta [\ln \rho (L,L_1=0)]$ as a function of $L$ for several values of the disorder across the extended phase. In agreement with our intuition, the ratio is found to be essentially independent of $L$, and decreases as $W$ is increased as $f_r(0)/f_r(3) \approx (\lambda(W))^3$.

All in all, the numerical results reported in this section indicate that for $L$ and $L_1$ large enough, sufficiently close to the localization threshold, and in the $\eta = 0^+$ limit, the line connected value of $\langle \ln \rho \rangle$ at the 1-loop level behaves as:
\begin{equation}
\delta [\ln \rho (L,L_1)] \approx - \varphi \, C_{\rm BL} (L) \lambda^{-L_1} \, ,
\end{equation}
where $\varphi$ is a positive coefficient of order $1$ which decreases smoothly as $W$ is increased and approaches a finite value in the interval $[0.1,0.2]$ close to $W_c$, $\lambda$ is larger than $k$ at small disorder and approaches $k$ for $W \to W_c$ as  $\lambda - k \propto (1 - W/W_c)^{\omega}$ (with $\omega \approx 3/2$), and the properties of $C_{\rm BL} (L)$ have been discussed in the previous section. Plugging this expression into Eq.~\eqref{eq:rhotyp1loop}, one obtains that close to the critical points the 1-loop corrections to the logarithm of the typical DoS are given by:
\begin{equation}
\Delta \! \avg{\ln \rho}_{\rm 1loop} \approx - \frac{\varphi}{M} \sum_L \frac{k^L C_{\rm BL}(L)}{L^{d/2}} \sum_{L_1} \left( \frac{k}{\lambda} \right)^{L_1} \, .
\end{equation}
As discussed above, $k^L C_{\rm BL} (L)$ has a very pronounced maximum at $L_\star \propto \left( \ln \Lambda \right)^{\delta}$ where $ k^L_\star C_{\rm BL} (L_\star) \simeq e^{b  \ln \Lambda}$ (see Eq.~\eqref{eq:Lstar} and Fig.~\ref{fig:max}). The first sum can be thus evaluated within the saddle point approximation yielding:
\[
\begin{aligned}
\sum_L \frac{k^L C_{\rm BL}(L)}{L^{d/2}} & \propto \frac{e^{b \ln \Lambda }}{\left( \ln \Lambda \right)^{\frac{d \delta}{2}}} + O(\ln \Lambda) \\ 
& \propto (W_c - W)^{\frac{\delta d}{4}} e^{{\rm cst} / (W_c - W)^{\frac{1}{2}}} \, .
\end{aligned}
\]
The sum over $L_1$ simply gives:
\[
\sum_{L_1} \left( \frac{k}{\lambda} \right)^{L_1} \simeq \frac{\lambda}{\lambda - k} \propto (W_c - W)^{-\omega} \, .
\]
One thus finally obtains the result given in Eq.~\eqref{eq:Deltaavglnrho} of the main text.
Hence $\Delta \! \avg{\ln \rho}_{\rm 1loop}$ diverges to minus infinity exponentially fast at the localization transition and are overwhelmingly larger than the order parameter itself in any dimension and for any finite value of $M$, in fully agreement with the outcome of the Ginzburg criterion. The prediction above is plotted as a dashed line in the top panel of Fig.~\ref{fig:BLcorr} of the main text.

\begin{figure*}
\includegraphics[width=0.482\textwidth]{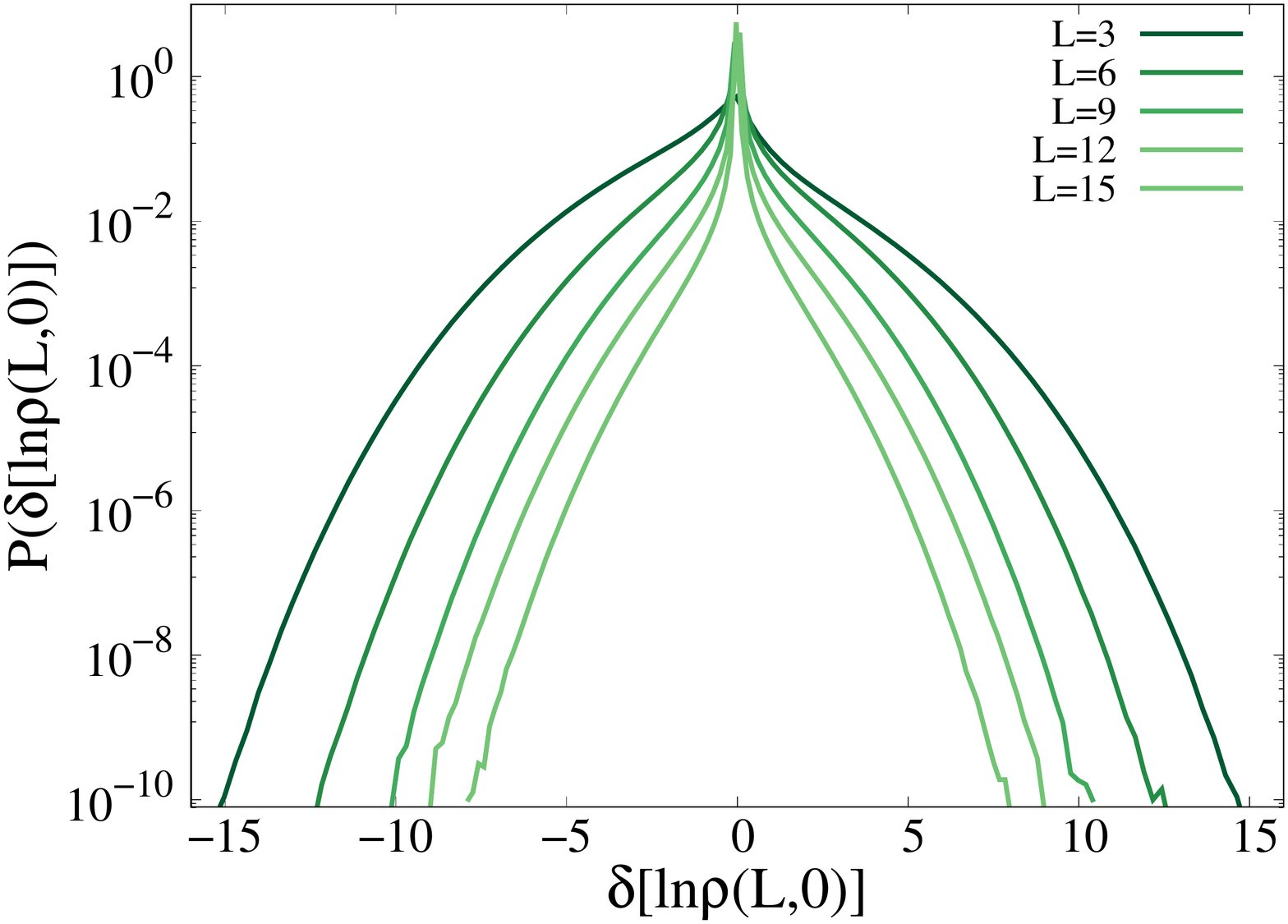}  \hspace{-0.1cm} 
\includegraphics[width=0.482\textwidth]{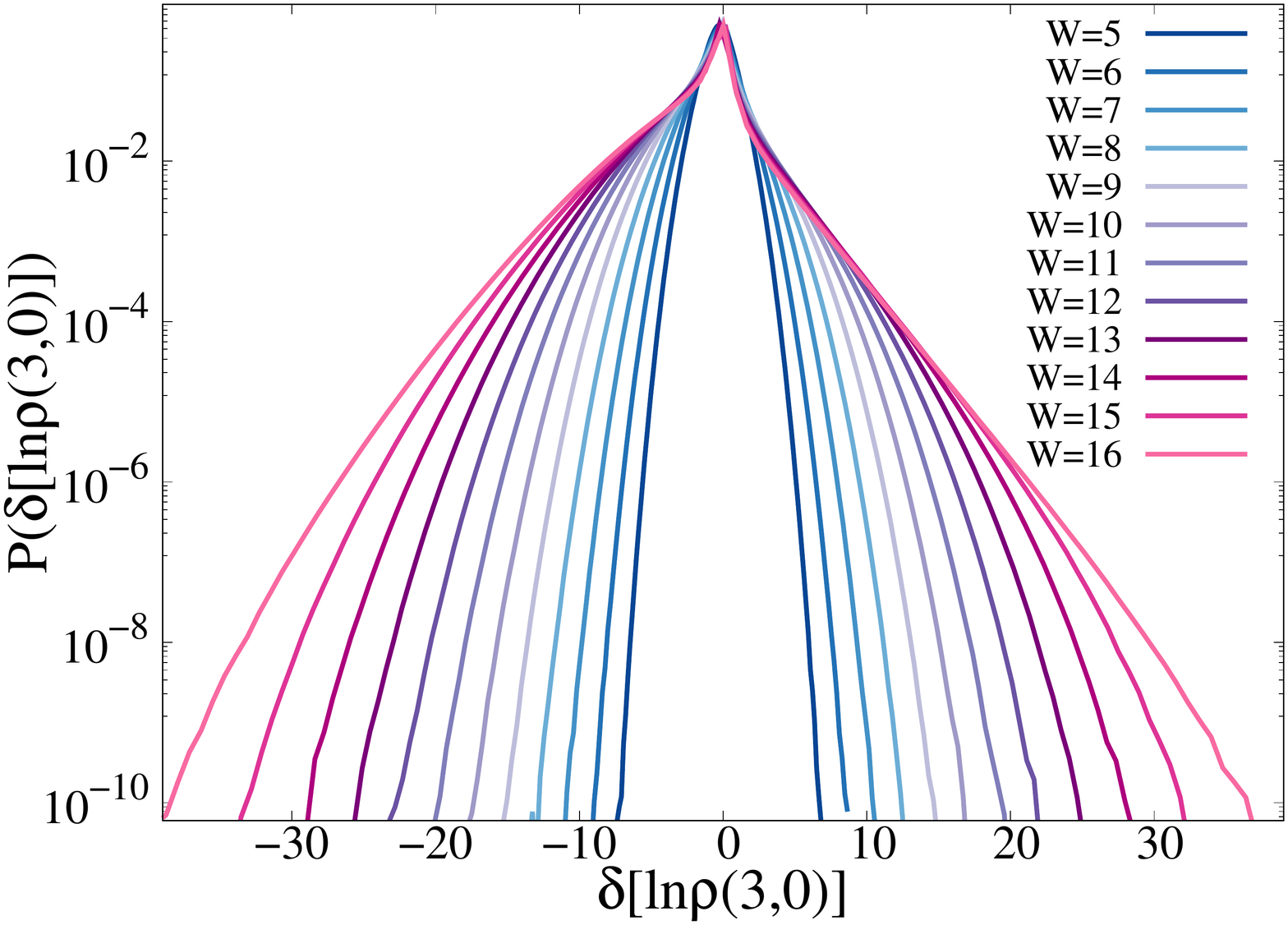}  

 \caption{Left: Probability distribution $P(\delta [\ln \rho (L,0)])$ for $W=9$ (and for $k=2$) varying the length of the loop $L$. Right: Probability distribution $P(\delta [\ln \rho (3,0)])$ for $L=3$ varying the disorder strength (and for $k=2$).
\label{fig:P_delta}}
\end{figure*}

It is also instructive to discuss the behavior of the full probability distribution of $\delta [\ln \rho (L,L_1)]$. For simplicity below we only consider the case $L_1=0$, and vary the disorder $W$ and the length of the loop $L$. In Fig.~\ref{fig:P_delta} we plot $P(\delta [\ln \rho (L,0)])$ for fixed $W=9$ varying $L$ (left panel) and for fixed $L=3$ varying $W$ (right panel). The plots show that the probability distributions are quite broad and extend to values of $\delta [\ln \rho (L,L_1)]$ much larger than the average.  In particular the distribution functions become broader upon increasing the disorder or decreasing $L$. This observation explains why obtaining a statistically reliable signal for $\avg{\delta [\ln \rho (L,L_1)]}$ out of the statistical noise is computationally difficult, especially at large disorder. $P(\delta [\ln \rho (L,0)])$ are characterized by a cusp in $0$, which becomes more pronounced at larger $L$, and exponential tails, whose slope decrease when $W$ is increased. The minus sign of the averages $\avg{\delta [\ln \rho (L,L_1)]}$ comes from the fact that the distributions are slightly asymmetrical, with a slight excess of weight on the negative side.

\subsubsection{$1/N$ Corrections to the typical DoS on Bethe lattices of finite size} \label{app:Ncorrlnrho}

As explained in the main text, the knowledge of the asymptotic behavior of $\delta [\ln \rho (L,L_1)]$ at large $L$ and $L_1$ in the delocalized phase can be exploited to estimate the finite-size corrections of order $1/N$ to the (logarithm of the) typical DoS on RRGs of large but finite sizes. 

A loose intuitive argument that gives the asymptotic expression the average number of 1-loop diagrams having the structure shown in Fig.~\ref{fig:1loopdiagram} of the main text goes as follows: For a $(k+1)$-RRG of $N$ vertices (with $N$ large), the number of nodes at distance $L$ from site $0$ are $(k+1) k^{L-1}$. Each one of these nodes can be potentially the ``root'' of a loop of length $L$. Since there are $\sim k^{L}$ nodes at distance $L$ from the root, the probability that one of those vertices coincide with the root of the loop is, asymptotically in the large $N$ limit, simply given by the ratio $k^{L}/N$. However, this is only a rough estimation. In fact a loop passing through the root can be obtained in several ways, e.g. if one of the $\sim k^{L^\prime}$ sites at distance $L^\prime$ from $0$ coincides with one of the $\sim k^{L - L^\prime}$ sites at distance $L- L^\prime$ from $0$. A precise combinatorial computation of the average number of circuits of length $L$ in a RRG of $N$ nodes (in the large $N$ limit) has been carried out in Refs.~\cite{rrg,guilhem,remi}. Starting from this result, as explained in the main text, one can compute the average number of 1-loop diagrams, Eq.~\eqref{eq:BLpaths}, having the structure shown in Fig.~\ref{fig:1loopdiagram} originating from a given node (labeled as $0$) of the RRG. One can then plug this expression into Eq.~\eqref{eq:expansion} to evaluate the 1-loop corrections of order $1/N$ to the logarithm of the typical DoS on finite RRGs at the order $1/N$, given in Eq.~\eqref{eq:rhotyp1loopRRG}. The connection between $1/N$ corrections and topological loops in the graph has noticed in Refs. \cite{ferrari,lucibello}. In particular in the context of Spin-Glass models it was shown explicitly that $1/N$ corrections to the free energy on Erd\"os-R\'enyi graphs and  RRGs obtained in a conventional way from the integration of Gaussian fluctuations around the saddle-point in a replica computation can indeed be interpreted as a sum of contributions from loops of length $L$ times the number of loops in the RRG, see Eqs.~(1) and~(14) in~\cite{lucibello}. Such an alternative derivation could be also repeated in this context.

As explained in the main text, in order to compare the numerical results obtained for RRGs of finite size one has to replace the $\delta$-function in the definition of the LDoS, Eq.~\eqref{eq:LDoS}, with a smooth Lorentzian function $\delta (x)  \rightarrow  \eta / [\pi (x^2 + \eta^2)]$, depending on a parameter $\eta$. We thus need to determine the effect of introducing a finite (but small) regulator on the asymptotic behavior of the $1$-loop line connected value of the average of the logarithm of the LDoS.

\begin{figure*}
\includegraphics[width=0.482\textwidth]{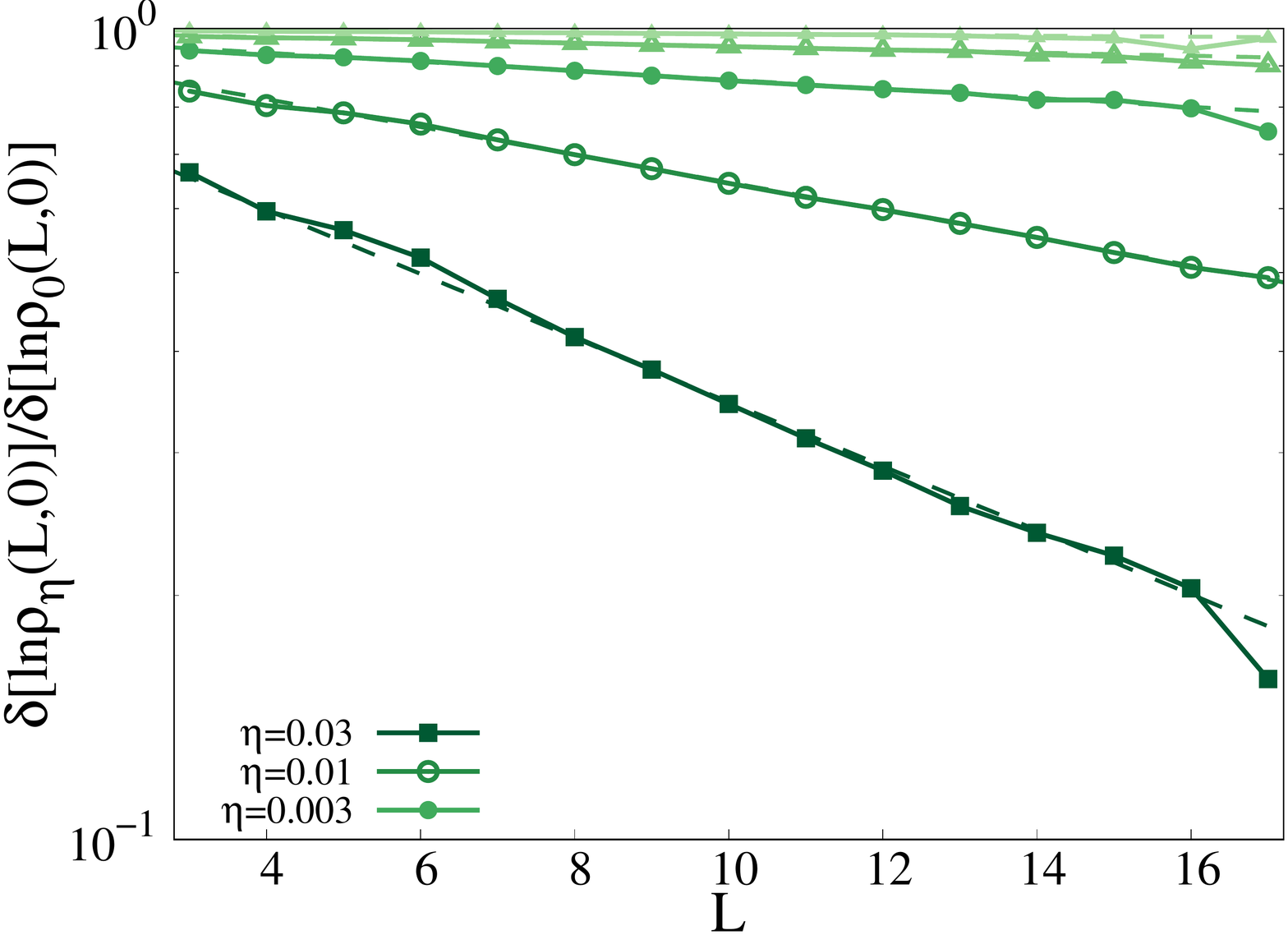} \hspace{-0.1cm} \includegraphics[width=0.482\textwidth]{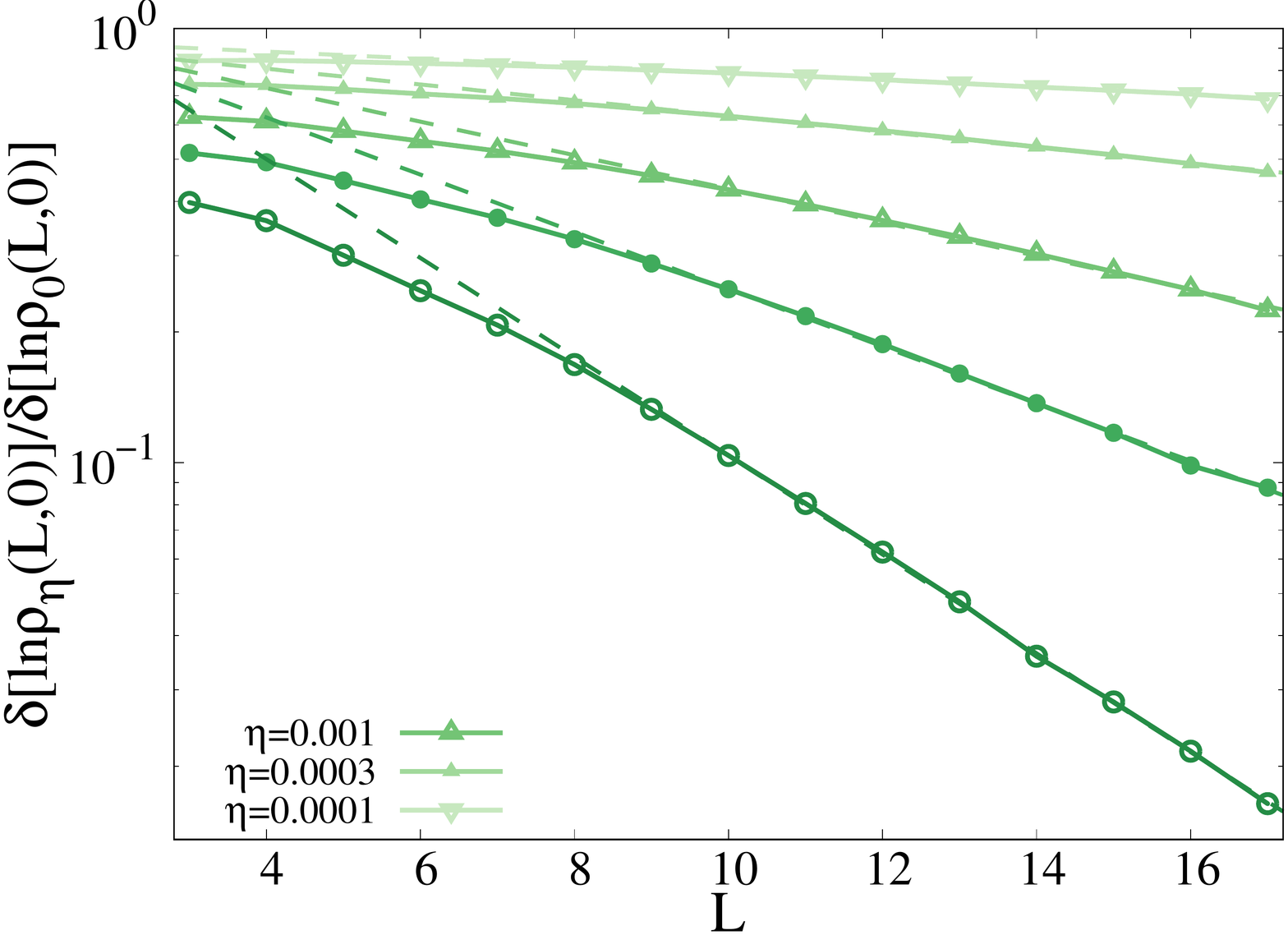} 

\includegraphics[width=0.482\textwidth]{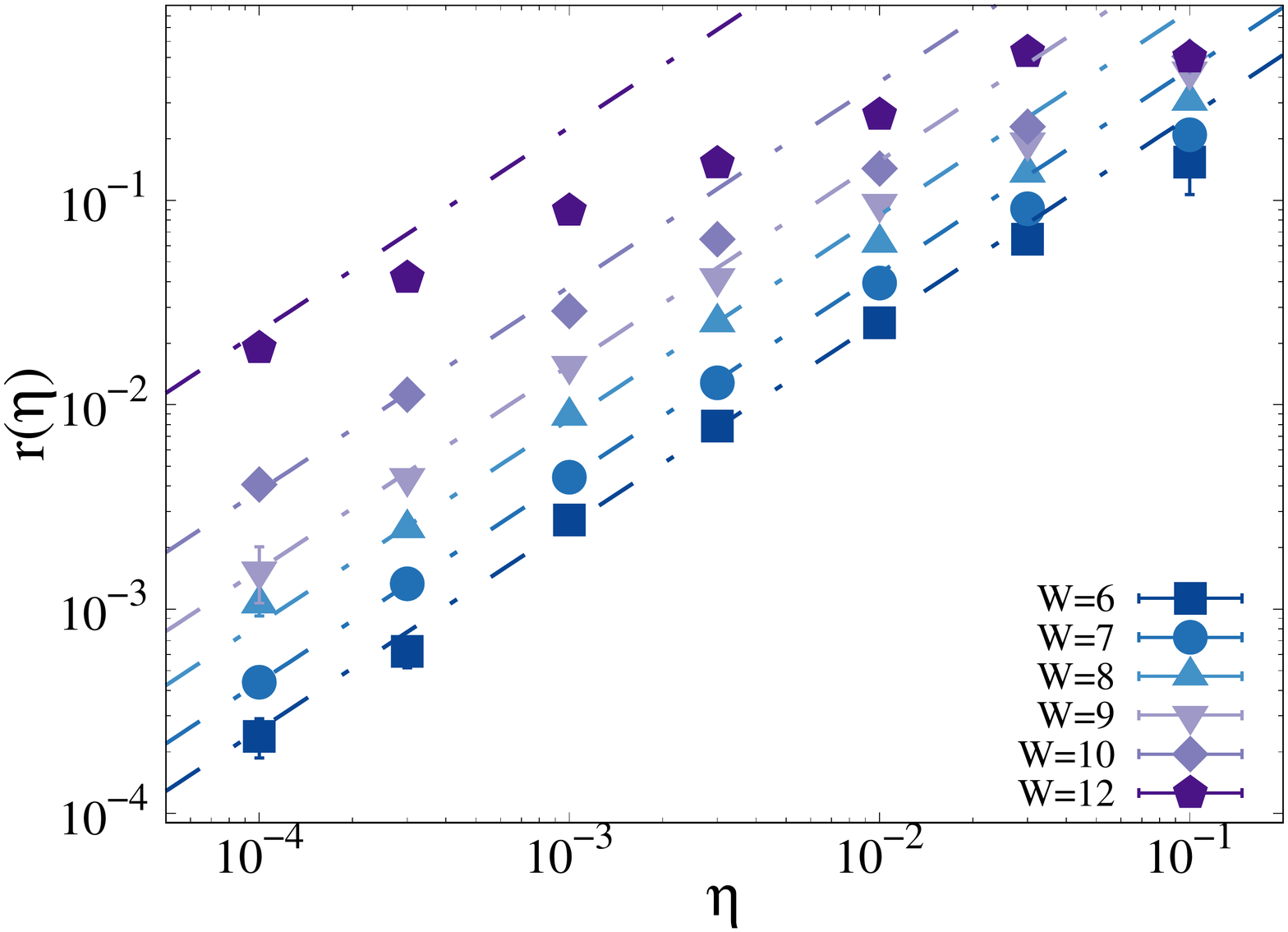} \hspace{-0.1cm} \includegraphics[width=0.482\textwidth]{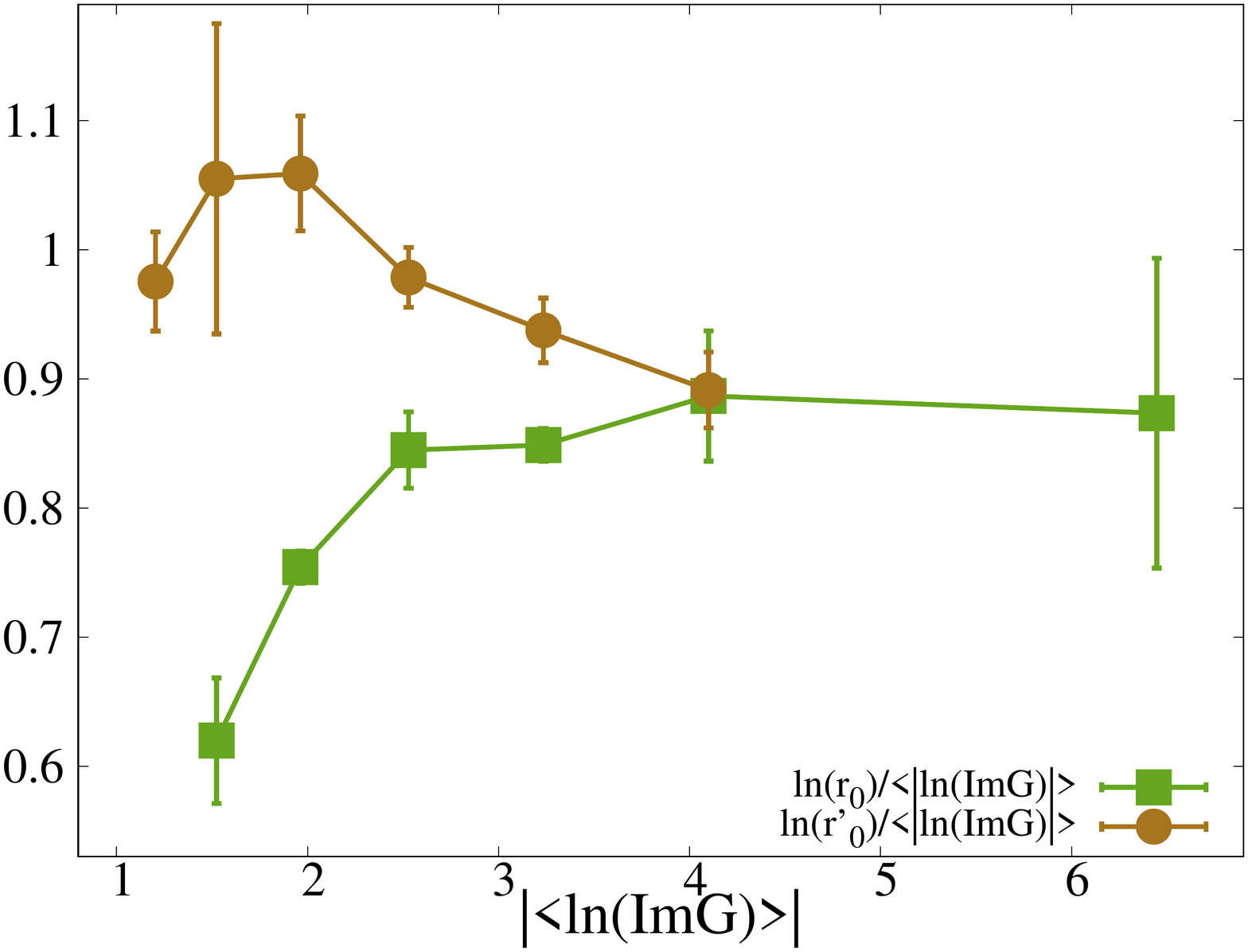} 
\caption{Top panels: $\delta [\ln \rho_\eta (L, 0)]/\delta [\ln \rho_{0^+} (L, 0)]$ as a function of the length of the loop $L$ for $W=7$ (left) and $W=12$ (right) and several values of $\eta$ (and for $k=2$). The dashed lines are exponential fits of the data of the form~\eqref{eq:etaTYP} at large enough $L$. Similar plots are found for other values of $W$. Bottom left: $r(\eta,W)$ as a function of $\eta$ for several values of the disorder across the delocalized phase. The dashed-dotted lines are fits of the data of the form $r(\eta,W) = r_0(W) \eta$. Bottom right: $\ln r_0/|\avg{\ln {\rm Im} G}|$ and $\ln r_0^\prime/|\avg{\ln {\rm Im} G}|${\it vs} $|\avg{\ln {\rm Im} G}|$, where $r_0$ and $r_0^\prime$ are extracted from the fits of $r(\eta,W)$ and $r^\prime(\eta,W)$ at small $\eta$.
\label{fig:k2IratioTYP}}
\end{figure*}

We start here by focusing on the diagrams with $L_1=0$ first, since they produce the most strongly divergent contribution, and we only show the results for $k=2$, since for $k=5$ we obtain essentially the same behavior. In the top panels of Fig.~\ref{fig:k2IratioTYP} we plot the ratio $\delta [\ln \rho_\eta (L, 0)]/\delta [\ln \rho_{0^+} (L, 0)]$ for $W=7$ (left) and $W=12$ (right) and for a few values of the imaginary regulator in the interval $\eta \in [3 \cdot 10^{-2}, 10^{-4}]$ (similar results are found for other values of $W$). We observe that for $L$ large enough (i.e. $L$ larger than a characteristic scale proportional to $\ln \Lambda$) an exponential decay sets in, with a rate that decreases as $\eta$ is decreased. This implies that the imaginary regulator has the effect of introducing an additional exponential cutoff at large $L$ for the 1-loop corrections to $\avg{\ln \rho}$. In order to characterize the dependence on $\eta$ and $W$ of such exponential cutoff we have fitted the numerical data at large $L$ as (dashed lines) as:
\begin{equation} \label{eq:etaTYP}
\frac{\delta [\ln \rho_\eta (L, 0)]}{\delta [\ln \rho_{0^+} (L, 0)]} \approx e^{- r (\eta,W)L - t (\eta,W)} \, .
\end{equation} 
The results of these fits are shown in the bottom left panel of Fig.~\ref{fig:k2IratioTYP}, where $r(\eta,W)$ is plotted as a function of $\eta$ for several values of $W$. This plot indicates that when $\eta$ is small enough (i.e. when $\eta$ becomes smaller than the inverse of the correlation volume) $r(\eta,W)$ vanishes linearly with $\eta$. A similar behavior is found for $t(\eta,W)$:
\[
\begin{aligned}
r(\eta,W) &\approx  r_0 (W) \, \eta \, , \\
t(\eta,W) & \approx t_0 (W) \, \eta \, .
\end{aligned}
\]
The value of the disorder dependent prefactors $r_0 (W)$ and $t_0 (W)$ can be estimated by performing a linear fit of $r(\eta,W)$ and $t(\eta,W)$ at small $\eta$ (dashed-dotted lines): $r_0$ is found to be proportional to our estimator of the correlation volume $\Lambda \simeq e^{-\avg{\ln {\rm Im} G}}$, and thus to increase very rapidly with $W$, while $t_0$ is found to be of order $1$ and (roughly) independent of $W$. The bottom right panel of the figure we plot $\ln r_0/|\avg{\ln {\rm Im} G}|$ as a function of $|\avg{\ln {\rm Im} G}|$, showing that the ratio tends to a constant of order $1$ at large disorder, i.e. $\ln r_0 \propto \ln \Lambda$.

Assuming that close enough to the critical point $\delta [\ln \rho_{0^+} (L, 0)]$ behaves asymptotically as $- \varphi C_{\rm BL} (L)$, and using the approximate functional form of $C_{\rm BL} (L)$ proposed in Eq.~(\ref{eq:fit}), with $f(x) = A e^{B x -C x^\beta}$, with the parameters $A$, $B$, $C$, and $\beta$ extracted from the fits of the numerical data (see Fig.~\ref{fig:ginzburgSI}), one obtains that the extra exponential cutoff introduced by a small imaginary regulator, Eq.~\eqref{eq:etaTYP}, has the effect of shifting the position and the value of the maximum of $\ln[ k^L |\delta [\ln \rho_{\eta} (L, 0)]|]$ to smaller values. In particular, at the linear order in $\eta$ one has:
\begin{equation} \label{eq:Lstar_eta}
\begin{aligned}
\frac{L_\star(\eta)}{L_\star(0^+)} & \approx  1 - c_1 \Lambda \eta \, , \\
 \frac{\ln \left[ k^{L_\star(\eta)} |\delta [\ln \rho_{\eta} (L_\star(\eta), 0)]| \right] }{\ln \left[ k^{L_\star(0^+)} |\delta [\ln \rho_{0^+} (L_\star(0^+), 0)]| \right]}&  \approx 1 - c_2 \Lambda \eta \, ,
\end{aligned}
\end{equation}
with $c_1$ and $c_2$ of order $1$.
Note, however, that although the corrections to the position and the height of the maximum become very small in the $\eta \to 0$ limit, the prefactor $r_0$  is very large even far away from $W_c$, as it grows proportionally to the correlation volume $\Lambda$. Hence, upon increasing the disorder, one has to consider smaller and smaller values of $\eta$, i.e. such that $\eta \ll \Lambda^{-1}$, in order for the linear approximation given above to be justified. In other words, the effect of the imaginary regulator on the position of the maximum and of its height, and thereby on the value of the $1/N$ corrections to the typical DoS, only become small provided that $\eta \Lambda \ll 1$, which confirms the fact that, in order to be representative of the $\eta \to 0^+$ limit the imaginary regulator must be taken much smaller than the inverse of the correlation volume.

For completeness, we have also studied the effect of the imaginary regulator on $\delta [\ln \rho_\eta (L, L_1)]$ when $L$ is kept fixed and $L_1$ is varied. We find an analogous exponential cutoff as the one found on $L$ (for the sake of succinctness we do not show here the results, since the external leg of the diagram only yields a subdominant contribution to the $1/N$ corrections):
\begin{equation}
\begin{aligned}
\delta [\ln \rho_\eta (L, L_1)] & \approx \delta [\ln \rho_{0^+} (L, L_1)] \\
& \qquad \times e^{- r (\eta,W)L - r^\prime (\eta,W) L_1 - t (\eta,W)} \, .
\end{aligned}
\end{equation}
The rate $r^\prime (\eta,W)$ behaves similarly to $r (\eta,W)$: For $\eta$ small enough $r^\prime (\eta,W)$ is proportional to $\eta$, $r^\prime(\eta,W) \approx  r_0^\prime (W) \eta$, through a constant $r_0^\prime$ which grows as very fast as the disorder is increased. The values of $r_0^\prime$ obtained by fitting $r^\prime(\eta,W)$ at small $\eta$ for several values of $W$ are shown in the bottom right panel of Fig.~\ref{fig:k2IratioTYP}, showing that $\ln r_0^\prime \approx c_3 \ln \Lambda$.  

We can now finally estimate the $1/N$ corrections to $\avg{\ln \rho_\eta}$ at finite $\eta$ on the BL by summing over all $1$-loop diagrams of lengths $L$ and $L_1$. Close to the critical point and in the for $\eta$ small and using the fact that $r_0 \approx \Lambda$, one obtain an asymptotic estimation of the corrections as
\[
\begin{aligned}
\Delta \! \avg{\ln \rho_\eta}_{\rm 1loop} & \approx - \frac{\varphi}{2 N} \, e^{-t_0 \eta} \sum_L k^L C_{\rm BL} (L) \, e^{-r_0 \eta L} \\
& \qquad \qquad \qquad  \times \sum_{L_1} \left ( \frac{k}{\lambda} \, e^{-r_0^\prime \eta} \right)^{L_1} \, 
\end{aligned}
\]
which yields the result given in Eq.~\eqref{eq:Dlnrhoeta} of the main text, where $b$, $c_2$, and $c_3$ are constants of order $1$ which can be evaluated by fitting the numerical data, and $\lambda$ depends on $W$ as indicated by the bottom-left panel of Fig.~\ref{fig:loop1LA}, i.e. $\lambda - k \propto (W_c - W)^\omega$ (with $\omega \approx 3/2$). In Fig.~\ref{fig:BLcorr} of the main text we plot the 1-loop corrections to $\avg {\ln \rho_\eta}$ as a function of $W$ in the delocalized phase predicted by Eq.~\eqref{eq:Dlnrhoeta} within the $M$-layer approach for several value of $\eta$, along with the $\eta \to 0^+$ limiting curve (yellow curve).

\begin{figure*}
\includegraphics[width=0.48\textwidth]{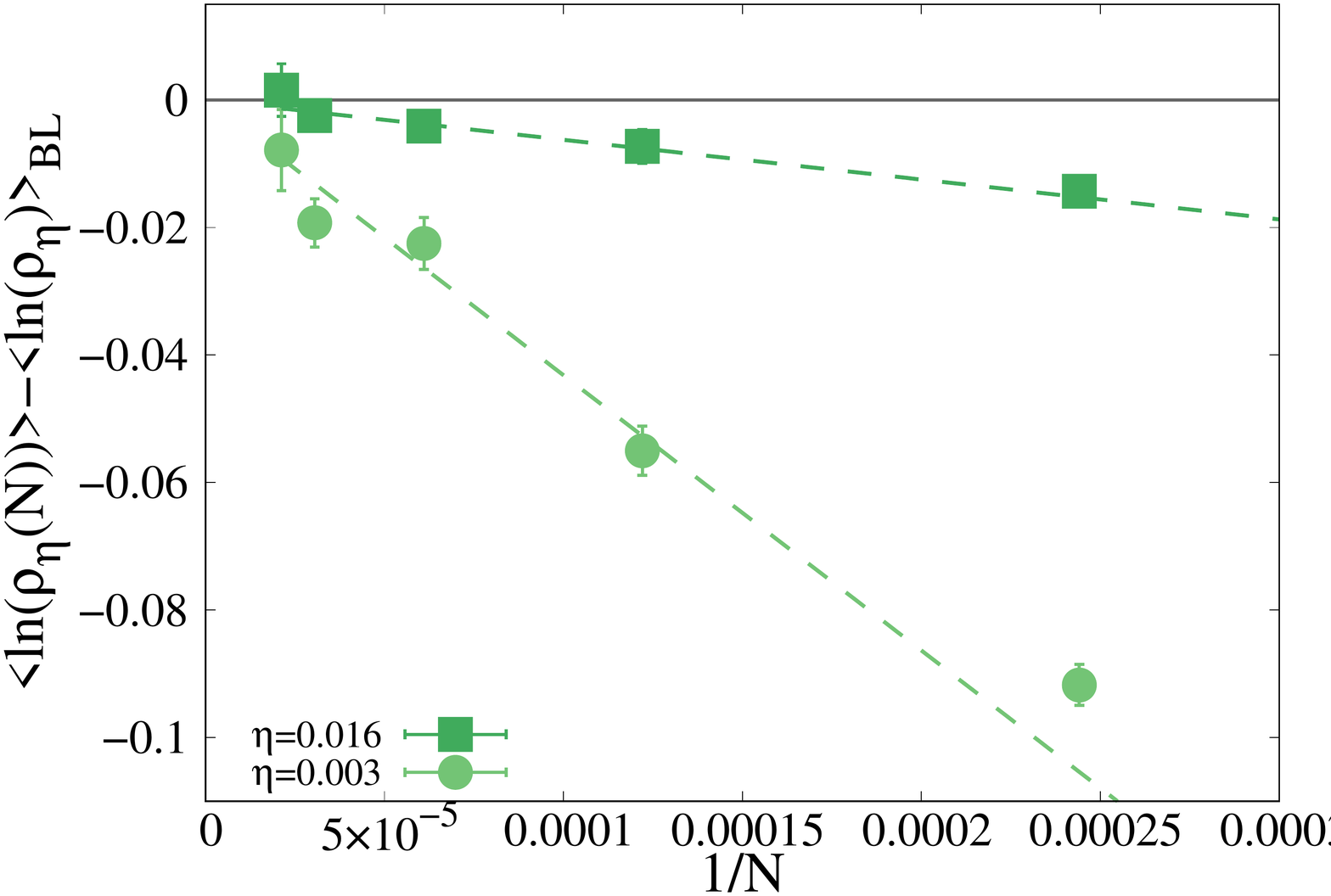} \hspace{-0.1cm} \includegraphics[width=0.48\textwidth]{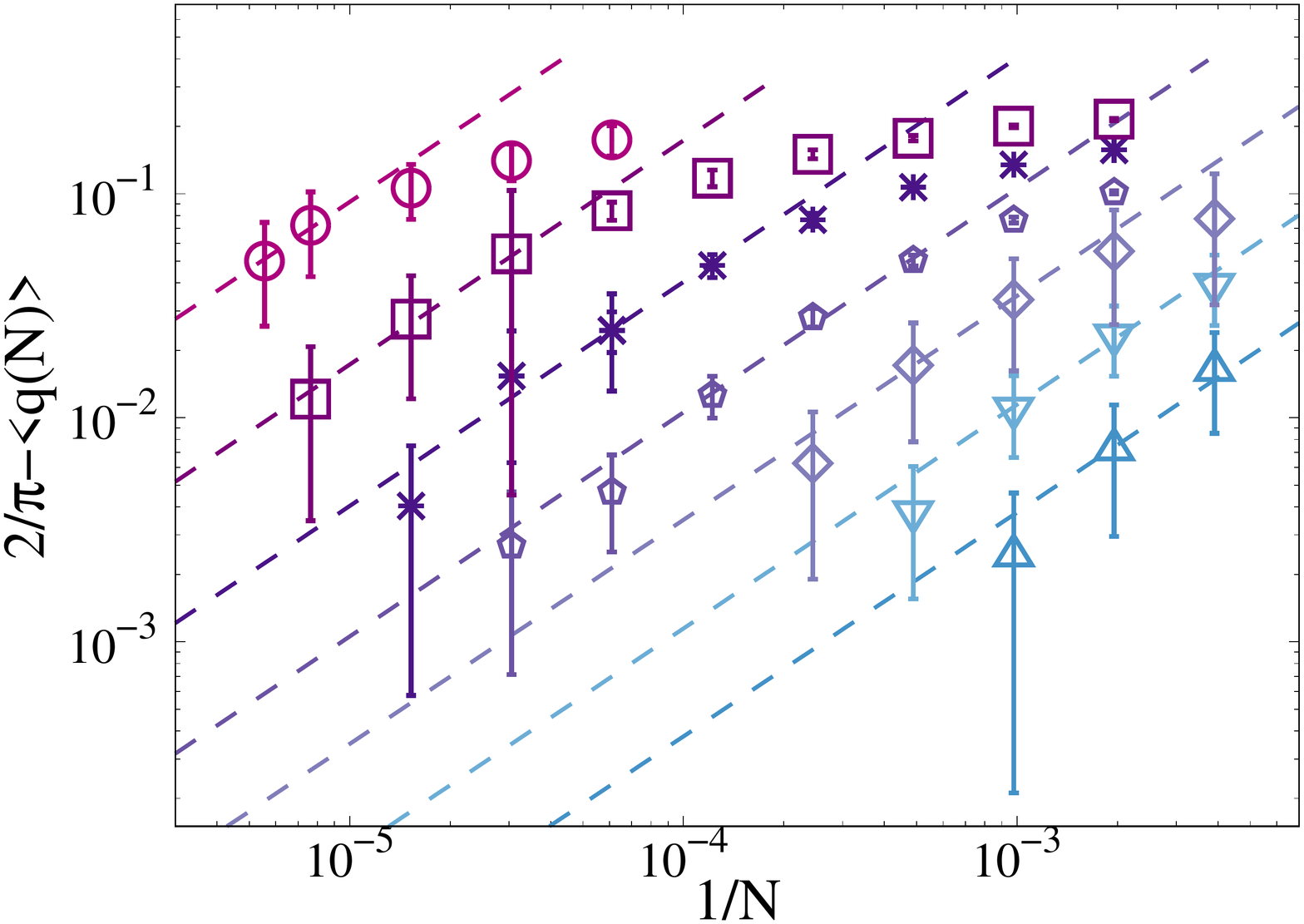} 

\caption{Left: Finite-size corrections of order $1/N$ to the logarithm of the typical DoS at finite $\eta$ for RRGs of $N$ sites and $k=2$: $\avg{\ln \rho_\eta (N)} - \avg{\ln \rho_\eta}_{\rm BL}$ is plotted as a function of $1/N$ for $W=8$ and two values of $\eta$. $\avg{\ln \rho_\eta (N)}$ is computed by averaging $(1/N)\sum_i \ln {\rm Im} G_{ii} (\eta) - \ln [ (1/N) \sum_i {\rm Im} G_{ii} (\eta)]$ over many realizations of RRGs of large but finite sizes $N$. The diagonal elements of the Green's functions are obtained by inverting exactly the matrix ${\cal H} - {\rm i} \eta {\cal I}$. Averages are performed over $2048$ samples for $N=2^{12}$, $512$ samples for $N=2^{13}$, $256$ samples for $N=2^{14}$, $128$ samples for $N=2^{15}$, and $64$ samples for $N=47104$. Similar results are found for others values of $W$ in the delocalized phase. The dashed lines are linear fits of the data at small $1/N$ in the form $\avg{\ln \rho_\eta (N)} - \avg{\ln \rho_\eta}_{\rm BL} \approx \Delta \! \avg{\ln \rho_\eta}/N$. Right: $1/N$ corrections to the average overlap between subsequent eigenstates $\avg{q}$ defined in Eq.~\eqref{eq:Q} on $(k+1)$-RRGs of large but finite sizes and of degree $k+1=3$. The plots show the difference $2/\pi-q(N)$ as a function of $1/N$ for seven values of the disorder from $W=7$ (bottom) to $W=13$ (top) and for $N$ from $512$ to $180224$. The dashed lines correspond to a linear fit at large $N$ of the form $\avg{q(N)}  = 2/\pi + a/N$. A similar behavior is found for the $1/N$ corrections to $\avg{r}$.
\label{fig:BLcorrSI}}
\end{figure*}

These theoretical predictions can be compared to the numerical results of the $1/N$ corrections to the logarithm of the typical DoS (for $\eta$ small but finite) on RRGs of large but finite sizes. In the left panel of Fig.~\ref{fig:BLcorrSI} we show the difference $\avg{\ln \rho_\eta (N)} - \avg{\ln \rho_\eta}_{\rm BL}$ as a function of $1/N$ (with $N$ ranging from $4096$ to $47104$) for $W=8$ and two values of the imaginary regulator, and for $k=2$ (similar results are obtained for others values of $W$). The values of $\eta$ are chosen in such a way that the condition $\eta > \Delta_N = (N \rho)^{-1}$ is always fulfilled. This plot clearly indicates the existence of a regime at large $N$ in which the finite-size corrections to the logarithm of the typical DoS are linear in $1/N$. By performing a linear fit for $1/N$ small of the form $\avg{\ln \rho_\eta (N)} - \avg{\ln \rho_\eta}_{\rm BL} \approx a/N$  (dashed lines), one can evaluate numerically the $1/N$ corrections to the $N \to \infty$ value of $\avg{\ln \rho_\eta}_{\rm BL}$, which can be easily obtained from the solution of the self-consistent equations for the infinite BL, Eqs.~\eqref{eq:Gcav} and~\eqref{eq:G}, via the population dynamics algorithm. The $1/N$ corrections $\Delta \! \avg{\ln \rho_\eta}$ are found to be negative, as expected. The values of $-\Delta \! \avg{\ln \rho_\eta}$ so obtained are reported in the top panel of Fig.~2 of the main text for $\eta=0.016$ and $\eta=0.003$, showing that they agree well with the theoretical prediction of Eq.~\eqref{eq:Dlnrhoeta} for the corresponding values of the regulator. 

Finally, we have also measured numerically the $1/N$ corrections to other observables, namely  the average value of the mutual overlap between two subsequent eigenvectors $\avg{q}$, and the average gap ratio $\avg{r}$, defined in Eq.~\eqref{eq:Q} of the main text. $\avg{q}$ and $\avg{r}$ are both  related to the level statistics, and take different universal values in the delocalized/GOE and in the localized/Poisson phases~\cite{huse,Bethe,large_deviations}. In particular $\langle r \rangle$ is expected to converge to $\langle r \rangle \simeq 0.5306$ in the metallic regime and to $\langle r \rangle \simeq 0.39$ in the insulating regime. Similarly, in the delocalized phase the wave-functions amplitudes are i.i.d. Gaussian random variables of zero mean and variance $1/N$, hence $\avg{q}$ converges to $2/\pi$. Conversely in the localized phase two successive eigenvectors are typically peaked around very distant sites and do not overlap, and therefore $\avg{q} \to 0$ for $N \to \infty$.  $\avg{r}$ and $\avg{q}$ can be directly measured from EDs of RRGs of finite but large sizes and have the advantage of being defined directly in $\eta = 0^+$ limit.  We have computed numerically the $1/N$ corrections to  $\avg{r}$  and $\avg{q}$ by performing linear fits at large $N$ of the form $\avg{q(N)} = 2/\pi + a /N$ and $\avg{r(N)} = 0.5306 + b /N$ for several values of $W$ in the delocalized phase, as in the right panel of Fig.~\ref{fig:BLcorrSI}. The corrections $\Delta \! \avg{q}$ and $\Delta \! \avg{r}$ are both negative. The numerical results for $-N \Delta \! \avg{q}$ and $-N \Delta \! \avg{r}$ are shown as orange and violet dashed lines  in Fig.~\ref{fig:BLcorr} of the main text, and seem to follow the same exponential trend as (minus) the corrections to $\avg{\ln \rho_{0^+}}$.

\subsection{Numerical analysis of the 1-loop corrections to the IPR} \label{app:deltaG2}

\begin{figure}
\includegraphics[width=0.48\textwidth]{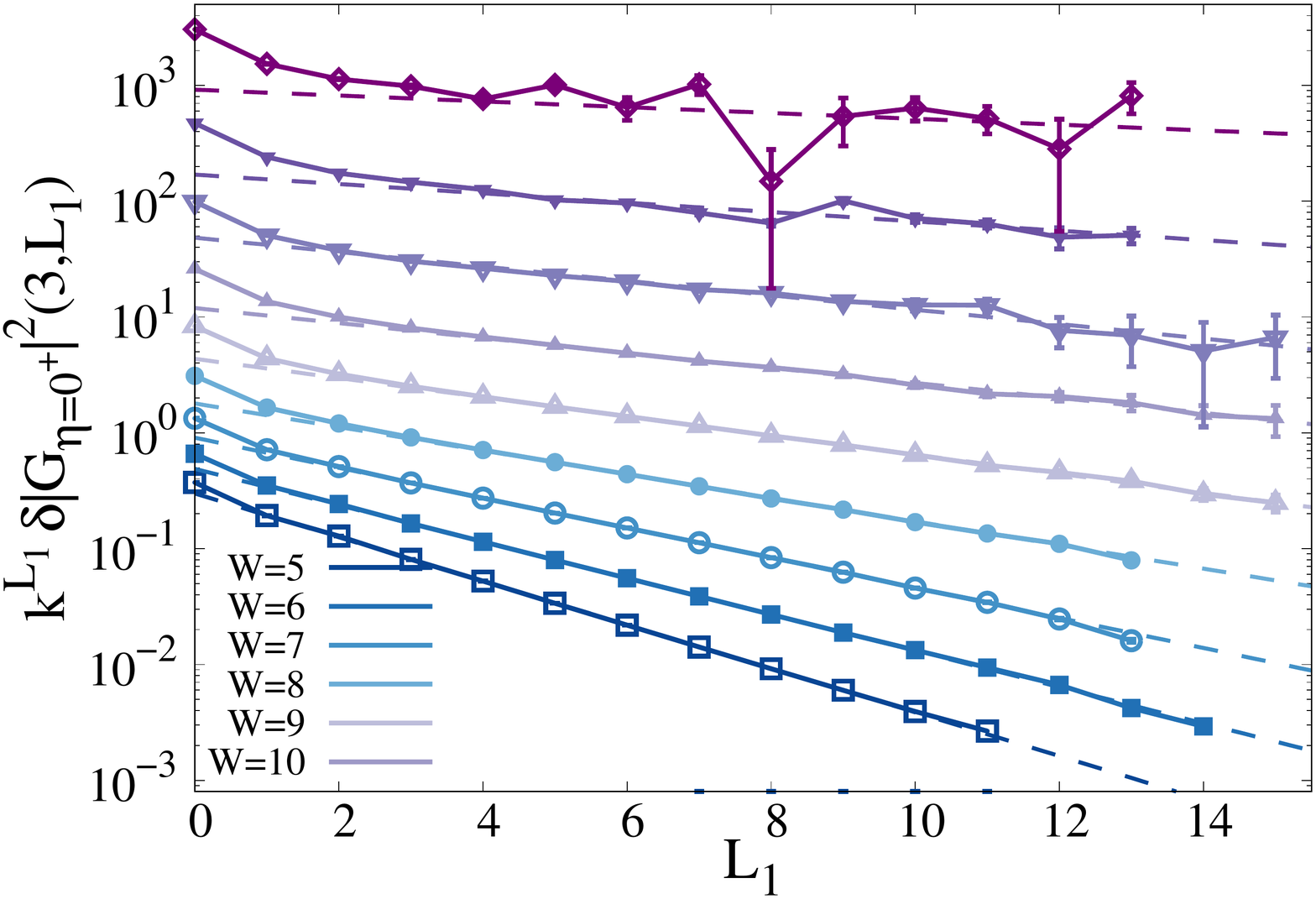}  
\caption{$k^{L_1} \delta |G_{0^+}|^2 (L=3, L_1)$ as a function of $L_1$ (for $L$ fixed and equal to $3$) for several values of $W$ (and for $k=2$). 
\label{fig:k2Iloop}}
\end{figure}

In the thermodynamic limit the IPR is identically equal to zero in the declocalized regime. However for a BL of $N$ nodes (i.e. a RRG) the IPR should scale as $1/N$ times a disorder-dependent constant. Such disorder-dependent constant has been computed within the supersymmetric formalism and is predicted to be proportional to the correlation volume $\Lambda$~\cite{mirlin94,mirlintikhonov}: 
\[ 
I_2 = \frac{3}{N} \, \frac{ \avg{\rho^2} }{ \avg{\rho}^2 } \propto \frac{\Lambda}{N} \, .
\]  
This prediction can be explicitly checked within the $M$-layer approach, as the $1$-loop corrections to the IPR also yield, in turn, the value of the IPR on RRGs of large but finite sizes at the order $1/N$, Eq.~\eqref{eq:iprRRG}. To this aim we need to determine the asymptotic behavior at large $L$ and $L_1$ of $\delta |G_\eta|^2 (L, L_1)$ at small $\eta$, which is the line connected value of $\langle |G_\eta|^2 \rangle$ at the 1-loop level, defined in Eq.~\eqref{eq:deltaG2} of the main text. For conciseness below we only present the numerical analysis for $k=2$, since for $k=5$ we find essentially the same results.

We start by considering the case $\eta=0^+$ first.  The numerical data are shown in Figs.~\ref{fig:k2Iloop_main} and~\ref{fig:k2Iloop}. In Fig.~\ref{fig:k2Iloop_main} we fix $L_1$ to $0$ (no external leg) and plot $k^L \delta |G_{\eta=0^+}|^2 (L, 0)$ as a function of the length of the loop $L$, while in Fig.~\ref{fig:k2Iloop} we fix the length of the loop to $L=3$ and plot $k^{L_1} \delta |G_{\eta=0^+}|^2 (3, L_1)$ as a function of the length of the external leg $L_1$. Analogously to the case of the 1-loop corrections to $\avg{\ln \rho}$ discussed above (see also Sec.~\ref{app:DeltaP} for the case of the percolation transition), we find that the dependence on $L$ and $L_1$ completely factorizes as $\delta |G_{\eta=0^+}|^2 (L, L_1) = g_a(L) g_r(L_1)$. In particular we have explicitly verified that for a given $L_1$ the ratio $\delta |G_{\eta=0^+}|^2 (L, L_1)/\delta |G_{\eta=0^+}|^2 (L, 0) = g_r(L_1)$ is roughly independent on $L$ within our numerical accuracy (note that here we have conventionally set $g_r(0)=1$). Similarly, for a fixed $L$ the ratio $\delta |G_{\eta=0^+}|^2 (L, L_1)/\delta |G_{\eta=0^+}|^2 (3, L_1) = g_a(L)/g_a(3)$ is roughly independent on $L_1$. 

\begin{figure*}
\includegraphics[width=0.338\textwidth]{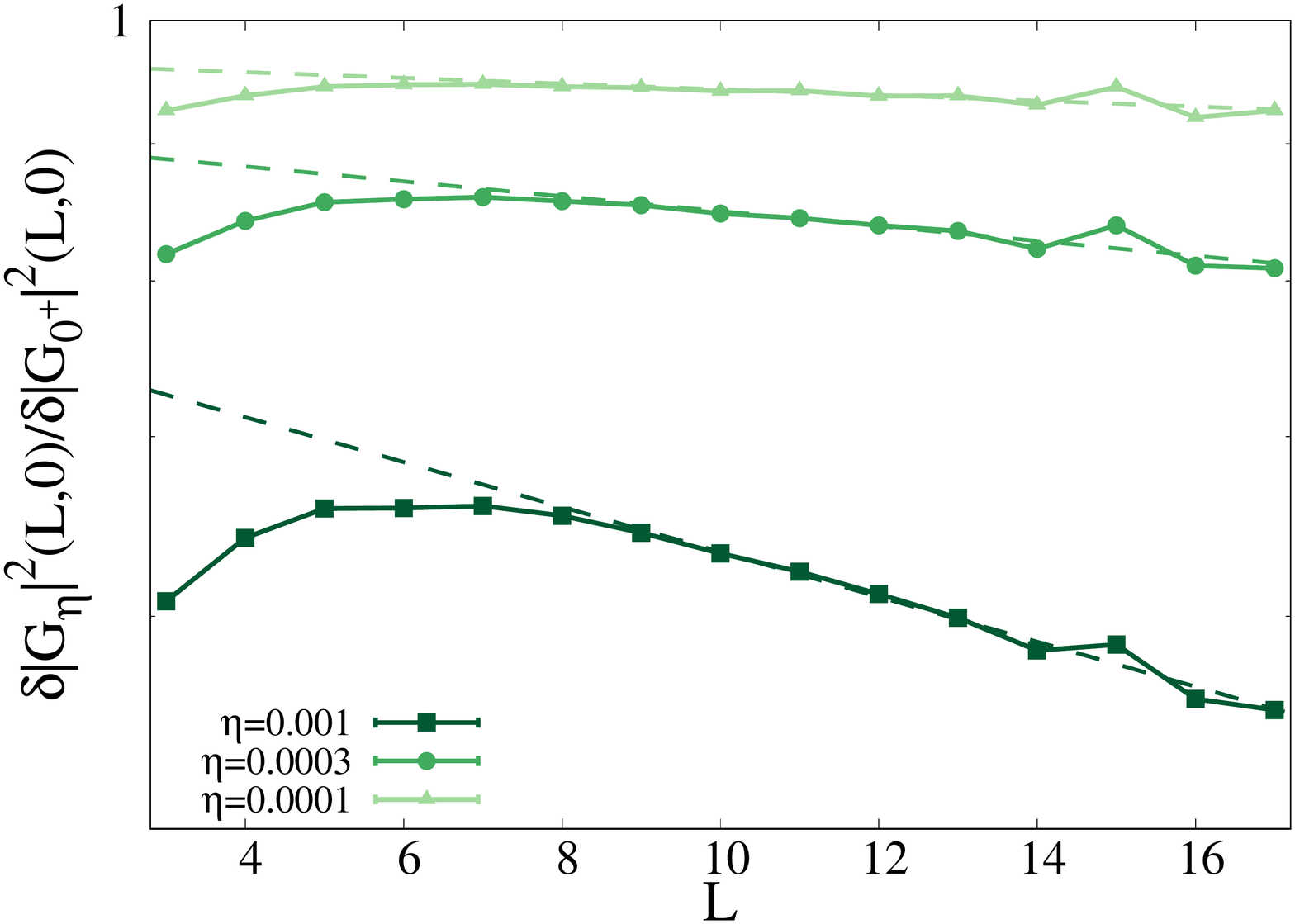} \hspace{-0.33cm} 
\includegraphics[width=0.338\textwidth]{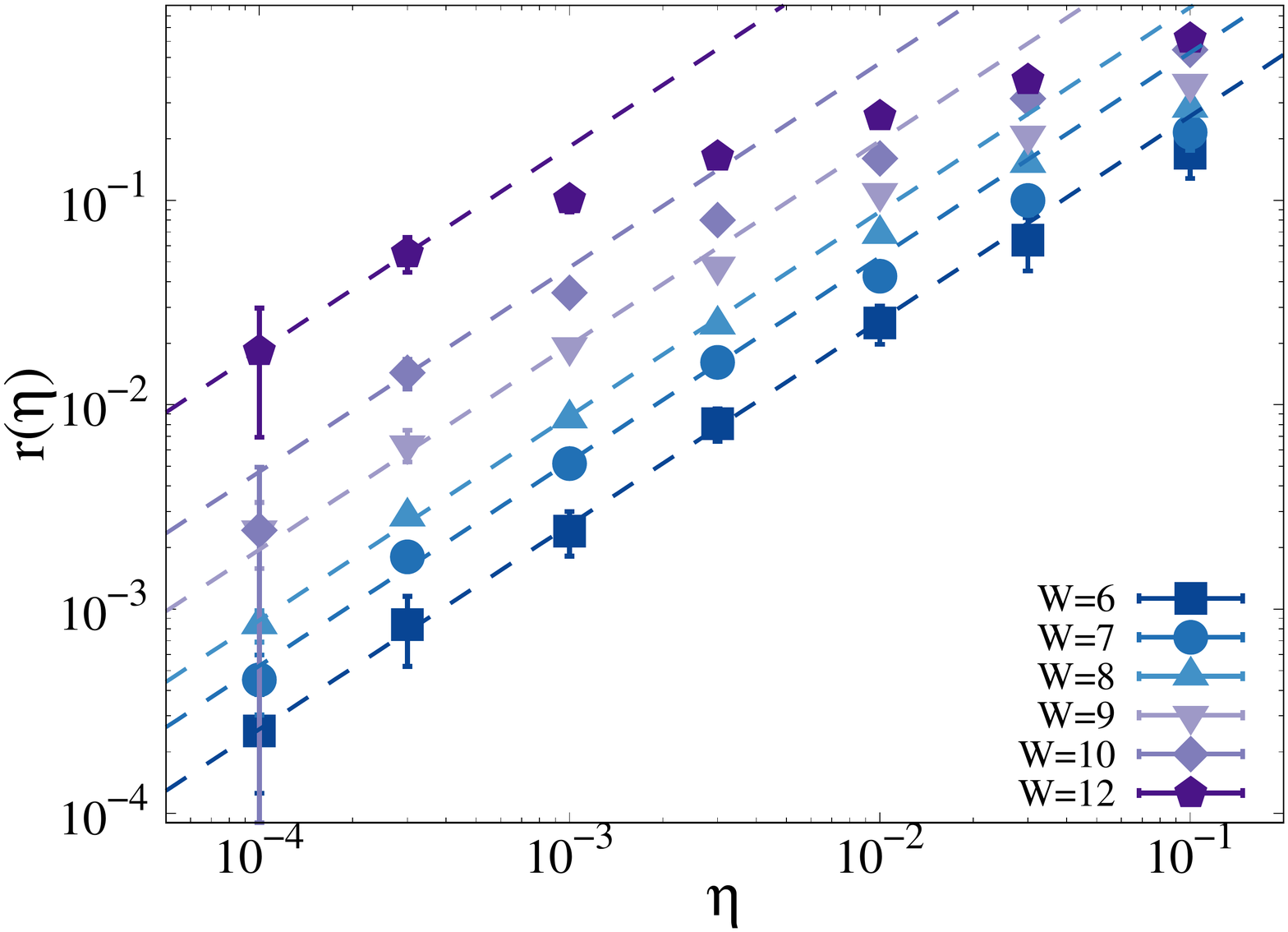} \hspace{-0.33cm} 
\includegraphics[width=0.338\textwidth]{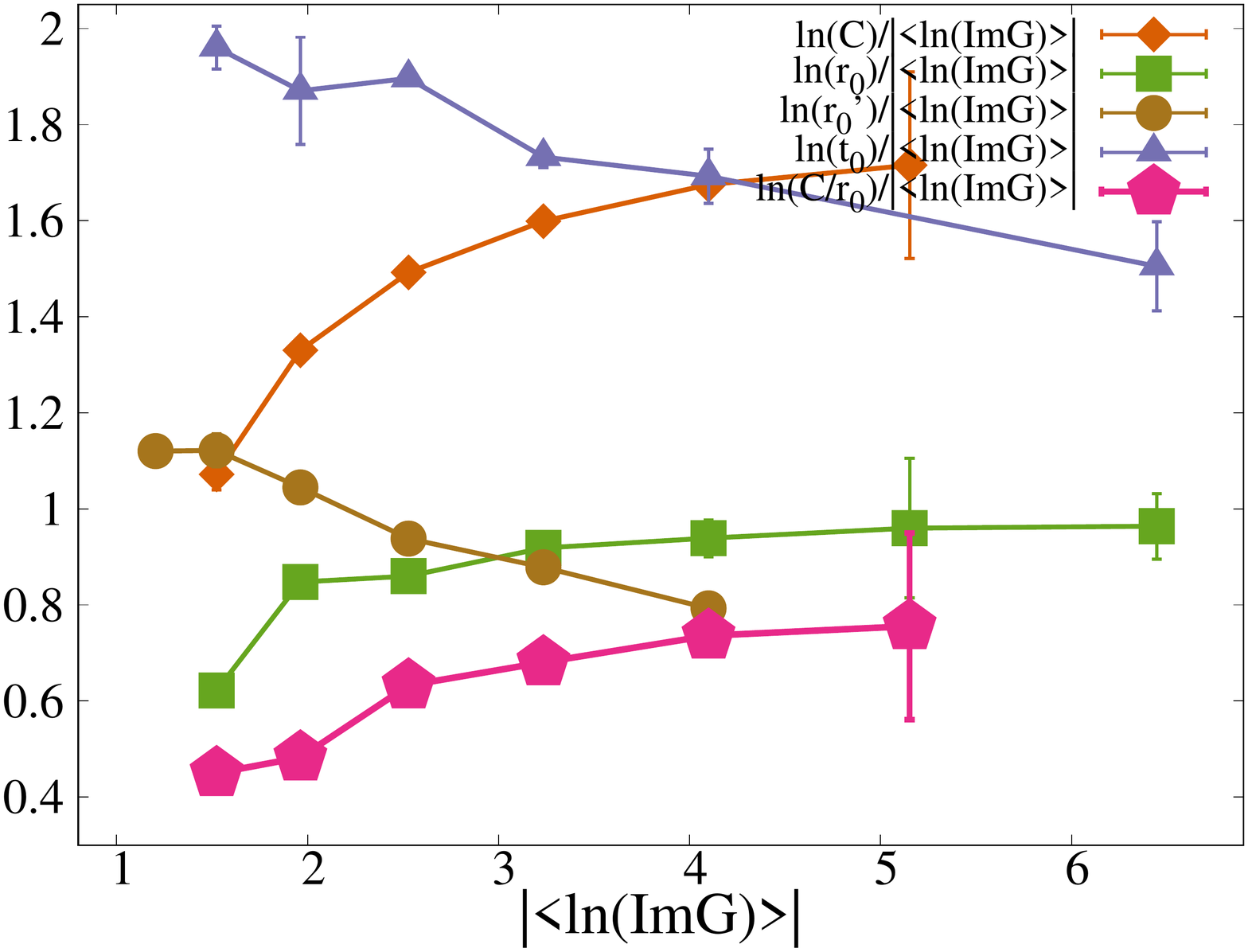}
\caption{Left panel:  $\Delta |G|^2 (L, \eta)/\Delta |G|^2 (L, \eta=0)$ as a function of $L$ for $L_1=0$ and $W=9$ and for three values of $\eta$ (and for $k=2$). The dashed lines correspond to fits of the data as $\Delta |G|^2 (L, \eta)/\Delta |G|^2 (L, \eta=0) = e^{-a(\eta,W)L - b(\eta,W)}$ for $L$ large enough. Middle panel: $\tilde{r}(\eta,W)$ as a function of $\eta$ for several values of the disorder. The dashed line correspond to linear fits of the data at small $\eta$ as $\tilde{r}(\eta,W) = \tilde{r}_0 \eta$. Right panel: Diamonds: $\ln C/|\avg{\ln {\rm Im}G}|$ obtained from fitting  $k^L \delta |G_{\eta=0^+}|^2 (L, 0)$ with a constant at large $L$ (see Fig.~\ref{fig:k2Iloop}) as a function of our estimator of the logarithm of the correlation volume $|\avg{\ln {\rm Im}G}|$. Squares, circles, and triangles: $\ln \tilde{r}_0/|\avg{\ln {\rm Im} G}|$, $\ln \tilde{r}_0^\prime/|\avg{\ln {\rm Im} G}|$, $\ln \tilde{t}_0/|\avg{\ln {\rm Im} G}|$ as a function of $|\avg{\ln {\rm Im} G}|$, where $\tilde{r}_0$, $\tilde{r}_0^\prime$, and $\tilde{t}_0$ are extracted from the fits of $\tilde{r}(\eta,W)$, $\tilde{r}^\prime(\eta,W)$, and $\tilde{t}(\eta,W)$ at small $\eta$. Pentagons: $\ln (C / \tilde{r}_0)$ {\it vs} $|\avg{\ln {\rm Im}G}|$.
\label{fig:k2Iratio}}
\end{figure*}

However, we find that $\delta |G_{0^+}|^2 (L, 0)$ behaves very differently from $\delta [\ln \rho_{0^+} (L,0)]$ (see Fig.~\ref{fig:loop}): Fig.~\ref{fig:k2Iloop_main} indicates that at fixed $L_1$ and for $L$ large enough (i.e. $L$ larger than a characteristic scale proportional to some power of $\ln \Lambda$) $\delta |G_{\eta=0^+}|^2$ behaves as $C k^{-L}$. The value of the disorder-dependent prefactor $C$ is obtained by fitting $k^L \delta |G|_{\eta=0^+}^2 (L, 0)$ with a constant at large $L$ (dashed lines). Since the asymptotic behavior is reached at large $L$ where an huge numerical precision is required to obtain reliable results, we are only able to estimate the prefactor $C$ for $W \le 11$. We find that $C$ grows very fast as the disorder is increased. In particular, in the right panel of Fig.~\ref{fig:k2Iratio} we plot the the ratio $\ln C / | \langle \ln {\rm Im} G \rangle |$ (diamonds), showing that at large enough disordered $\ln C$ is proportional to $\ln \Lambda$, $\ln C \approx c_4 \ln \Lambda$, with $c_4$ close to $2$. 

Fig.~\ref{fig:k2Iloop} shows that  $\delta |G_{\eta=0^+}|^2(L,L_1)$ behaves as $\lambda^{-L_1}$ for $L$ fixed, with $\lambda>k$. A numerical estimation of $\lambda$ is obtained by fitting $k^{L_1} \delta |G|_{\eta=0^+}^2 (3, L_1)$ with an exponential function at large $L_1$ (dashed lines). The values of $\lambda$ obtained from these fits are compatible, within errorbars, with those reported in the bottom left panel of Fig.~\ref{fig:loop1LA}, describing the exponential decay of $\delta [\ln \rho (L,L_1)]$ with $L_1$ at fixed $L$, implying that $\lambda$ approaches $k$ algebraically when $W$ approaches $W_c$ from below. 

Putting all these results together we obtain the asymptotic behavior at large $L$ and $L_1$ given in Eq.~\eqref{eq:dG2h0}, where $\lambda - k \propto (W_c-W)^\omega$ with $\omega \approx 3/2$, and $C \approx \Lambda^{c_4}$, with $c_4$ close to $2$. Yet, since the IPR is proportional to $\eta \avg{|G_\eta|^2}$ for $\eta \to 0$, in order to obtain the 1-loop corrections to the IPR we need to study the behavior of the  corrections to $\avg{|G_\eta|^2}$ at small but {\it finite} $\eta$. We start by focusing on the diagrams with $L_1=0$, since it is clear from the discussion above that the loop gives the most strongly divergent contribution. 

In the left panel of Fig.~\ref{fig:k2Iratio} we plot the ratio $\delta |G_\eta|^2 (L, 0)/\delta |G_{0^+}|^2 (L, 0)$ for $W=9$, and for three values of the imaginary regulator (similar results are found for other values of $W$). We observe that the regulator has the effect of introducing an exponential cutoff at large $L$ for the 1-loop corrections to $|G_\eta|^2$ very similar to the cutoff produced on $\delta [\ln \rho_\eta (L,L_1)]$ discussed in the previous section: For $L$ large enough (i.e. $L$ larger than a characteristic scale proportional to $\log \Lambda$) an exponential decay sets in, with a rate that decreases as $\eta$ is decreased. In order to characterize the dependence on $\eta$ and $W$ of such exponential cutoff we have fitted the numerical data at large $L$ as (dashed lines of the left panel):
\[
\frac{\delta |G_\eta|^2 (L, 0)}{\delta |G_{\eta = 0^+}|^2 (L, 0)} \simeq e^{- \tilde{r} (\eta,W)L - \tilde{t} (\eta,W)} \, .
\]
The results of these fits are shown in the middle panel of Fig.~\ref{fig:k2Iratio}, where $\tilde{r}(\eta,W)$ is plotted as a function of $\eta$ for several values of $W$. This plot indicates that when $\eta$ is small enough (i.e. when $\eta$ becomes smaller than the inverse of the correlation volume) $\tilde{r} (\eta,W)$ vanishes linearly with $\eta$. A similar behavior is found for $\tilde{t}(\eta,W)$:
\[
\begin{aligned}
\tilde{r}(\eta,W) &\approx \tilde{r}_0 (W) \, \eta \, , \\
\tilde{t} (\eta,W) & \approx \tilde{t}_0 (W) \, \eta \, .
\end{aligned}
\]
The values of the disorder dependent prefactors $\tilde{r}_0(W)$ and $\tilde{t}_0(W)$ can be estimated by performing a linear fit of $\tilde{r} (\eta,W)$ and $\tilde{t}(\eta,W)$ at small $\eta$ (dashed lines). The results of this procedure are given in the right panel of Fig.~\ref{fig:k2Iratio} where we plot the ratios $\ln \tilde{r}_0 /|\langle  \ln {\rm Im} G \rangle|$ (squares) and $\ln \tilde{t}_0/|\langle  \ln {\rm Im} G  \rangle|$ (triangles) for several values of $W$ across the delocalized phase, showing that at large enough disorder both $\ln \tilde{r}_0$ and $\ln \tilde{t}_0$ are essentially proportional to the logarithm of correlation volume $\ln \Lambda$, $\ln \tilde{r}_0 \approx c_5 |\langle  \ln {\rm Im} G \rangle|$, $\ln \tilde{t}_0 \approx c_6 |\langle  \ln {\rm Im} G \rangle|$, with $c_5$ and $c_6$ of order $1$.

For completeness, we have also studied the effect of the imaginary regulator on $\delta |G|^2_\eta (L, L_1)$ when $L$ is kept fixed and $L_1$ is varied, which yields an extra exponential cutoff of the form:
\[
\begin{aligned}
\delta |G_\eta|^2 (L, L_1) & \approx \delta |G_{0^+}|^2 (L, L_1) \\
& \qquad \times e^{- \tilde{r} (\eta,W)L - \tilde{r}^\prime (\eta,W) L_1 - \tilde{t} (\eta,W)} \, .
\end{aligned}
\]
The rate $\tilde{r}^\prime (\eta,W)$ behaves similarly to $\tilde{r} (\eta,W)$: For $\eta$ small enough (i.e. $\eta \ll \Lambda^{-1}$) $r^\prime (\eta,W)$ is proportional to $\eta$, $\tilde{r}^\prime(\eta,W) \approx  \tilde{r}_0^\prime (W) \eta$, through a constant $\tilde{r}_0^\prime$ which grows very fast as $W$ is increased. The values of $\tilde{r}_0^\prime$ obtained by fitting $\tilde{r}^\prime(\eta,W)$ at small $\eta$ for several values of $W$ are shown in the right panel of Fig.~\ref{fig:k2Iratio}, where we plot the ratio $\ln \tilde{r}_0^\prime / |\avg{\ln {\rm Im} G}|$ (circles). The plot indicates that  $\ln \tilde{r}_0^\prime \approx c_7 |\avg{\ln {\rm Im} G}|$, with $c_7$ of order $1$.  

\begin{figure}
\includegraphics[width=0.48\textwidth]{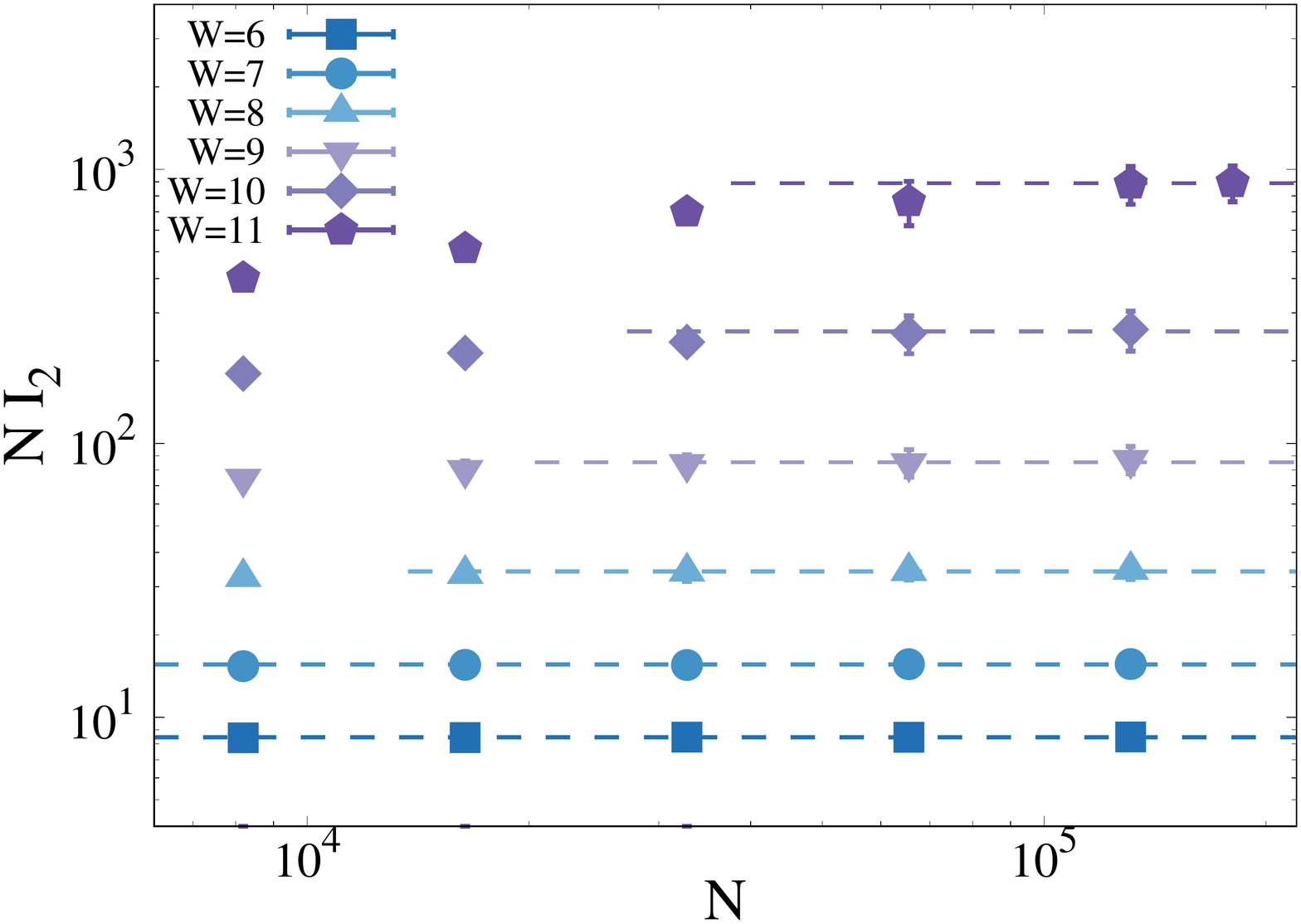} 
\caption{$N I_2$ {\it vs} $N$ for several values of the disorder $6 \le W \le 11$ obtained from EDs of finite RRGs of $N$ nodes and fixed connectivity $k+1=3$, with $N$ ranging from $8192$ to $180284$. The dashed lines are fits of the data at large $N$ with a constant. 
\label{fig:ipr}}
\end{figure}

Plugging these results into Eq.~\eqref{eq:iprRRG}, we can estimate the $1/N$ corrections on finite RRGs (in the $\eta \to 0$ limit) to $\eta \langle |G_\eta|^2 \rangle$:
\[
\begin{aligned}
I_2 & \simeq \frac{1}{\pi \avg{\rho}} \lim_{\eta \to 0^+}\sum_{L,L_1} \avg{{\cal N}_G (L,L_1)} \, \eta \delta |G|_\eta^2 (L,L_1) \\
& \propto   \lim_{\eta \to 0^+} \frac{\eta \, C \, e^{- \tilde{t}_0 \eta}  }{2 N} \sum_L e^{-\tilde{r}_0 \eta L} \sum_{L_1} \left( \frac{k}{\lambda} e^{- \tilde{r}_0^\prime} \right)^{L_1} \\
& \propto \frac{1}{N} \frac{1}{\pi \avg{\rho}} \, \frac{C}{\tilde{r}_0} \frac{1}{1 - \frac{k}{\lambda} } \, ,
\end{aligned}
\]
where we have used the expression of $\avg{{\cal N}_G (L,L_1)}$ given in Eq.~\eqref{eq:BLpaths}. As shown in the right panel of Fig.~\ref{fig:k2Iratio}  the ratio $C/\tilde{r}_0$ grows very fast as the disorder is increased proportionally to the correlation volume $\Lambda$. In particular, at sufficiently strong disorder the ratio $\ln (C/\tilde{r}_0)$ seems to approach a constant close to $1$. We thus finally obtain a simple expression of the value of the IPR on finite RRGs of $N$ nodes (provided that $\eta \ll \Lambda^{-1}$) given in Eq.~\eqref{eq:IPR} of the main text.

We can now compare the $1$-loop estimation above with the exact results obtained from EDs. In order to do this we have diagonalized the Anderson model on RRGs of $N$ nodes, with $N$ ranging from $8192$ to $180204$, focusing on few ($\sim 128$) eigenstates around the middle of the spectrum and averaging over several ($\sim 2^{26-\ln N/\ln 2}$) realizations of the disorder. The numerical results are shown in Fig.~\ref{fig:ipr} for several values of $W$. This plot shows that for large enough $N$ (i.e. $N \gg \Lambda$) the IPR behaves as:
\[
I_2 = \frac{Y_{\rm ED}}{N} \, .
\]
$Y_{\rm ED}$ can be estimated numerically by fitting $N I_2$ at large $N$ with a constant (dashed lines). Fig.~\ref{fig:IPRcorr} of the main text provides a comparison of the value of $Y$ obtained from ED and the one estimated from the $1$-loop corrections, Eq.~\eqref{eq:IPR}, showing an excellent agreement.

\subsection{Leading-order contribution of the correlation function $K_d(r)$} \label{app:ema}

We conclude this appendix by discussing the asymptotic behavior in the localized phase of the leading-order contribution on $d$-dimensional lattices of the correlation function $K_d(r)=  N \langle |\psi_\alpha (0) \psi_\alpha (r)|^2 \rangle = (\pi \langle \rho \rangle)^{-1} \lim_{\eta \to 0^+} \eta \langle |G_{0,r}|^2 \rangle$, which is directly related to the probability that a particle starting on site $0$ at $t=0$ is found on a  site at distance $r$ from it after an infinite time (note that $K(0)$ is the IPR). The asymptotic expression at large $L$ on the infinite BL can be computed analytically both in the insulating and in the extended phase~\cite{mirlin,zirn,verba,mirlintikhonov}: 
\begin{equation} \label{eq:corr}
\begin{aligned}
K_{\rm BL} (L) & \simeq
\left \{
\begin{array}{ll}
\eta \Lambda \, k^{-L}  L^{-3/2}  & \textrm{~for~} W<W_c \, ,\\
 k^{-L} e^{-L/\xi_{\rm loc}} L^{-3/2} & \textrm{~for~} W>W_c \, .
\end{array}
\right .
\end{aligned}
\end{equation}
On the insulating side of the transition the localization length diverges at the transition point as $\xi_{\rm loc} \propto (W-W_c)^{-1}$, with a critical exponent $\nu_{\rm loc} = 1$. Conversely, in the metallic phase, the localization length  is formally infinite, since the particles spread (although in a highly heterogeneous way) over the whole graph. The  critical behavior only manifests itself in the exponential divergence of the prefactor $\Lambda$, i.e. the correlation volume. 

Plugging the asymptotic expressions given above of $K_{\rm BL} (L)$ on the infinite BL in the localized phase into Eq.~\eqref{eq:cd}, using the asymptotic expressions of the number of paths between two points at distance $r$ on the euclidean lattice, Eq.~\eqref{eq:NrL}, and going to Fourier space, at the leading order of the $M$-layer expansion we obtain:
\[
\hat{K}_d (p) \propto \frac{1}{M} \sum_{L=1}^\infty \frac{e^{-L(p^2 + \xi_{\rm loc}^{-1})}}{L^{3/2}} =  \frac{1}{M}  \, {\rm Li}_{3/2} (e^{-p^2 - \xi_{\rm loc}^{-1}}) \, ,
\]
where ${\rm Li}_s (z) = \sum_{n=1}^\infty z^n/n^s$ is the polylogarithmic function of order $s$. 
Hence, similarly to the case of the percolation transition discussed in Sec.~\ref{app:GP},  at the leading order $K_d (p)$ can be expressed as a function of the variable $p / \sqrt{\xi_{\rm loc}}$. This implies that in the vicinity of the critical point  the correlation length associated to the Gaussian propagator in $d$ dimensions diverges as $\xi_{\rm loc}^{1/2} \propto (W-W_c)^{-1/2}$ and {\it not} as $\xi_{\rm loc}\propto (W-W_c)^{-1}$. 
In particular, in the small momentum, large distance region we can expand the expression above yielding:
\[
\hat{K}_d (p) \propto {\rm Li}_{3/2} (e^{- \xi_{\rm loc}^{-1}}) - {\rm Li}_{1/2} (e^{- \xi_{\rm loc}^{-1}}) \, p^2 + O(p^4) \, .
\]
Thus, in the vicinity of the critical point, $\xi_{\rm loc} \gg 1$,  the ``mass'' of the Gaussian propagator in $d$ dimensions at asymptotically small momentum vanishes as $[{\rm Li}_{1/2} (e^{- \xi_{\rm loc}^{-1}})]^{-1} \simeq 1/\sqrt{\pi \xi_{\rm loc}}$:
\[
\hat{K}_d (p) \propto \frac{1}{M} \, \frac{1}{1 + \sqrt{\xi_{\rm loc}} \, p^2 } \, .
\]
Note that ${K}_d (p=0) \sim 1$ to ensure the normalization of the wave-functions.

This result can be explained recalling  that a particle that diffuses freely on the BL is found at ({\it Hamming}) distance $L$ from the origin after $L$ steps, while it is found  at ({\it Euclidean}) distance $\sqrt{L}$ from the origin on a $d$-dimensional lattice. In this respect the leading order term of the $M$-layer construction allows to reproduce the results of Efetov’s effective medium approximation~\cite{ema} in a rigorous framework, and provides a simple and physically transparent way to reconcile the apparent discrepancy between $\nu_{\rm loc} = 1$, which describes the divergence of the localization length  when AL is approached from the insulating  phase, and $\nu_{\rm del} = 1/2$, which is associated to the exponential divergence of the logarithm of the correlation volume when the transition is approached from the metallic phase, without the need of resorting to the NL$\sigma$M.

\begin{acknowledgments}
We warmly thank G. Biroli, Y. V. Fyodorov, I. Khaymovich, A. D. Mirlin, and G. Semerjian for enlightening discussions.
\end{acknowledgments}

\end{document}